\documentclass[11pt]{article}
\usepackage{amsmath,inputenc}

\title{Stochastic Online Fisher Markets: \\ Static Pricing Limits and Adaptive Enhancements}

\author{Devansh Jalota$^\dagger$ \and Yinyu Ye$^\ddagger$
\thanks{$^\dagger$ Institute for Computational and Mathematical Engineering, Stanford University; {\tt djalota@stanford.edu}.}%
\thanks{$^\ddagger$ Department of Management Science and Engineering, Stanford University; {\tt yyye@stanford.edu}.}%
}

\newif\ifarxiv   
\arxivtrue 

\ifarxiv
\else
\fi

\usepackage{natbib}
 \bibpunct[, ]{(}{)}{,}{a}{}{,}%
 \def\bibsep{\smallskipamount}%

\usepackage{rotating}
\usepackage{fancyvrb}

\usepackage{graphicx}
\usepackage{csvsimple}
\usepackage{rotating}
\usepackage[letterpaper, portrait, margin=1in]{geometry}
\usepackage{hyperref}
\hypersetup{
    colorlinks=true,
    linkcolor=blue,
    citecolor=red,
    filecolor=magenta,      
    urlcolor=cyan
    }

\usepackage[nocomma]{optidef}
\usepackage{listings}
\usepackage{comment}
\usepackage{appendix}
\usepackage[ruled,vlined]{algorithm2e}
\usepackage{amsfonts} 
\usepackage{amssymb}
\usepackage{bm}

\usepackage{color} 
\definecolor{mygreen}{RGB}{28,172,0} 
\definecolor{mylilas}{RGB}{170,55,241}
\DeclareFixedFont{\ttb}{T1}{txtt}{bx}{n}{12} 
\DeclareFixedFont{\ttm}{T1}{txtt}{m}{n}{12}  
\usepackage{bbm}
\usepackage{amsmath,amsthm}

\newtheorem{theorem}{Theorem}
\newtheorem{corollary}{Corollary}

\newtheorem{lemma}{Lemma}

\newtheorem{assumption}{Assumption}

\usepackage{tikz,pgfplots}
\pgfplotsset{compat=1.14}
\pgfplotsset{scaled y ticks=false}

\usepackage{subcaption}

\pgfplotsset{scaled x ticks=false}

\theoremstyle{definition}
\newtheorem{definition}{Definition}

\newenvironment{hproof}{%
  \proof}{\endproof}

\usepackage{color}
\definecolor{deepblue}{rgb}{0,0,0.5}
\definecolor{deepred}{rgb}{0.6,0,0}
\definecolor{deepgreen}{rgb}{0,0.5,0}

\newcommand{\norm}[1]{\left\lVert#1\right\rVert}

\DeclareMathOperator*{\argmax}{arg\,max}
\DeclareMathOperator*{\argmin}{arg\,min}

\usepackage{accents}

\usepackage{mathtools}

\DeclarePairedDelimiter\floor{\lfloor}{\rfloor}

\usepackage{listings}

\renewcommand\footnotemark{}

\def\D{\mathcal{D}}
\def\Pp{\mathbb{P}}
\def\p{\mathbf{p}}
\def\c{\mathbf{c}}
\def\z{\mathbf{z}}

\def\H{\mathcal{H}}

\def\d{\mathbf{d}}

\def\v{\mathbf{v}}
\def\x{\mathbf{x}}

\def\u{\mathbf{u}}
\def\0{\mathbf{0}}
\def\1{\mathbf{1}}
\def\ppi{\boldsymbol{\pi}}
\def\alphaa{\boldsymbol{\alpha}}
\def\bbeta{\boldsymbol{\beta}}
\def\I{\mathcal{I}}

\begin{document}

\maketitle

\vspace{-5pt}

\begin{abstract}
Fisher markets are one of the most fundamental models for resource allocation. However, the problem of computing equilibrium prices in Fisher markets typically relies on complete knowledge of users' budgets and utility functions and requires transactions to happen in a static market where all users are present simultaneously. Motivated by these practical considerations, we study an online variant of Fisher markets, wherein users with privately known utility and budget parameters, drawn i.i.d. from a distribution, arrive sequentially. In this setting, we first study the limitations of static pricing algorithms, which set uniform prices for all users, along two performance metrics: (i) regret, i.e., the optimality gap in the objective of the Eisenberg-Gale program between an online algorithm and an oracle with complete information, and (ii) capacity violations, i.e., the over-consumption of goods relative to their capacities. Given the limitations of static pricing, we design adaptive posted-pricing algorithms, one with knowledge of the distribution of users’ budget and utility parameters and another that adjusts prices solely based on past observations of user consumption, i.e., revealed preference feedback, with improved performance guarantees. Finally, we present numerical experiments to compare our revealed preference algorithm’s performance to several benchmarks.
\end{abstract}

\section{Introduction}

The study of market \emph{equilibria} is central to economic theory and corresponds to a (uniform) price vector (for all agents) and an allocation of goods to agents (users) such that the \emph{market clears}~\citep{walras1954elements}, i.e., all goods are sold, while users obtain an affordable utility-maximizing bundle of goods under the set prices. 
One of the most fundamental equilibrium models for resource allocation is that of a Fisher market~\citep{Fisher-seminal}, wherein users spend a budget of (artificial) currency to purchase goods that maximize their utilities while a central planner sets prices on capacity-constrained goods. Since Fisher introduced his seminal framework, a focal point of the literature has been in developing methods to compute market equilibria. 
In a seminal work,~\cite{EisGale} developed a convex program that maximizes the (weighted) Nash social welfare objective to compute equilibrium prices and the corresponding allocations for a broad range of utility functions. 
In the setting when all buyers have the same budgets, the market equilibrium outcome corresponding to the solution of the Eisenberg-Gale program that maximizes the (unweighted) Nash social welfare objective is known as a competitive equilibrium from equal incomes (CEEI)~\citep{VARIAN197463}.

Despite the many desirable properties of the Eisenberg-Gale program~\citep{VARIAN197463,VARIAN1976249}, including it being polynomial time solvable~\citep{ye2008path,jain2007polynomial}, computing equilibrium prices via a centralized optimization problem relies on complete information on users' utilities and budgets, which are typically unavailable in practice. Thus, there has been a growing interest in developing distributed approaches for market equilibrium computation, e.g., tatonnement~\citep{Tatonnement,CHEUNG2019}, proportional response~\citep{zhang2011proportional,cheung2018dynamics}, primal-dual \citep{DevnaurPrimalDual}, alternating direction method~\citep{jalota2021fisher}, and auction-based \citep{vazirani_2007,nesterov-fisher-gale} approaches. In the distributed setting, the central planner typically updates the prices of goods until convergence to market equilibria (under certain, often mild, conditions). To enable the convergence of these algorithms to equilibrium prices, existing distributed approaches for Fisher markets, e.g., tatonnement, typically involve a simulated setting wherein all users repeatedly interact in a static market. In practice, however, users generally do not repeatedly interact in the market to enable the central planner to learn equilibrium prices and instead tend to arrive into the market sequentially, as users are often not all present at once.


Motivated by the aforementioned practical limitations of centralized and distributed approaches for Fisher markets, we study a generalization of Fisher markets to the setting of online user arrival wherein users with privately known utilities and budgets arrive sequentially. Such an online incomplete information setting of Fisher markets models several practical applications, which we elucidate through examples in Section~\ref{sec:examples}. In this online incomplete information variant of Fisher markets, we go beyond traditional centralized pricing mechanisms in classical Fisher markets through the development of \emph{posted-price} mechanisms, wherein each arriving user purchases their most favored bundle of goods given the set prices. Furthermore, in contrast to traditional distributed approaches for Fisher markets that involve a simulated setting with repeated user interactions, our work considers a real market setting wherein users arrive sequentially over time.
Notably, in this work, we develop a posted-price mechanism using a novel algorithmic approach that adjusts prices based solely on users' \emph{revealed preferences}, i.e., observations of the bundle of goods users purchase given the set prices. 
This algorithmic approach is conducive to deployment in practical contexts, as it involves posted prices and uses past user consumption behavior that is typically readily available with the proliferation of data on customer interactions enabled through advances in modern-day computing. We further note that these posted-price mechanisms are in contrast to prior work in online resource allocation~\citep{Balseiro2020DualMD,li2020simple} that typically involve \emph{hidden-price} mechanisms, wherein a central planner allocates resources to users after observing their attributes, e.g., user valuations revealed through bids in an auction.

Beyond developing posted-price mechanisms, 
we also establish the performance limitations of uniform (static) pricing, a primary focus of the equilibrium computation literature for Fisher markets, wherein a single price vector applies to all users.  
Furthermore, to analyse our developed algorithms, in addition to traditional tools in online learning, e.g., convex
programming duality, we leverage techniques from parametric optimization and also develop
a novel potential function argument that leverages the structural properties of Fisher markets. 
Thus, our work points toward developing novel methods to more deeply understand market equilibria in online variants of Fisher markets and, more generally, highlights the benefit of designing adaptive posted-price mechanisms for online resource allocation~\citep{arlotto2019uniformly}.




\vspace{-5pt}

\subsection{Examples} \label{sec:examples}

This section presents real-world applications to demonstrate the applicability of the online incomplete information variant of Fisher markets studied in this work. While we present two examples of artificial currency programs designed to improve outcomes for low-income communities, our framework is more generally applicable in settings where fairness is a primary concern for a central planner with incomplete information on the preferences (i.e., utilities) of users that arrive dynamically. Moreover, since we develop posted-price mechanisms, our algorithms have broader appeal in contexts such as online marketplaces, e.g., \emph{Amazon}, that dynamically adjust prices of the listed products over time to meet inventory capacities while customers arrive sequentially to purchase their preferred bundle of goods given the set prices and their budgets.

\vspace{-8pt}

\paragraph{Food Stamps:} Food stamp programs (e.g., the Supplemental Nutrition Assistance Program (SNAP) in the United States~\citep{snap-benefits}) are one of the most common government welfare programs and involve providing food stamps (or artificial currency) to enable low-income households to purchase food from designated locations that accept food stamps as payment for food~\citep{jalota2024catch}. To buy food, eligible food stamp holders go to designated grocery stores dynamically over time and buy their most preferred bundle of goods given the set prices and their available food stamp credits (i.e., budget). We note that this food stamp application, which considers a setting where users purchase their preferred bundle of resources given the set prices, is different from the food bank application considered in~\citet{sinclair2021sequential}, who study a setting where a central planner allocates food directly to users.

\vspace{-8pt}

\paragraph{Social Capital Credits:} Social capital credits (SoCCs) have been proposed as a poverty alleviation mechanism in many developing nation contexts, where low-income communities can redeem the credits to provide better education, healthcare, and up-skilling resources for community members~\citep{asia-initiatives}. Akin to the food stamp context, communities have heterogeneous preferences over the different critical resources (i.e., education, healthcare, and up-skilling) and enter the market dynamically to use their available social capital credits to avail the most preferred resources for their community given the market prices.

In both these applications, a central planner in charge of setting prices of the respective resources 
may not have complete information on user or community preferences (i.e., utilities); hence, the central planner seeks to dynamically adjust the prices of the goods in the market to achieve a desirable allocation of resources. Furthermore, both these applications correspond to settings where a central planner seeks to deliver benefits to low-income users in the population. Consequently, the Eisenberg-Gale objective studied in this work, with its desirable fairness properties (see Section~\ref{sec:offline}), is particularly apt for such applications.







\vspace{-5pt}

\subsection{Our Contributions}

We study an online variant of Fisher markets wherein budget-constrained users with privately known utility and budget parameters arrive sequentially. In particular, we focus on the setting when users have linear utilities and their budget and utility parameters are independently and identically (i.i.d.) distributed from a probability distribution $\D$. Since traditional methods, e.g., the Eisenberg-Gale program~\citep{EisGale}, are not amenable to computing equilibria in this setting, we study the problem of learning prices online to minimize two performance metrics: \emph{regret} and \emph{constraint violation}. We refer to regret as the optimality gap in the objective of the Eisenberg-Gale program between the online allocation and that of an offline oracle with complete information on users' budgets and utilities and constraint violation as the norm of the excess demand for goods beyond their capacity. For a detailed discussion on our problem setting and these performance metrics, we refer to Section~\ref{sec:moodelPerfMeasures}. While we focus on optimizing these performance metrics, we also develop feasible variants of our algorithms that respect resource capacities (see Section~\ref{sec:feasible-algo-design}). 

In this online incomplete information setting, we first study the limitations of static pricing algorithms (Section~\ref{sec:lb-result}), wherein the same price vector $\p$ applies to all users. For any static pricing algorithm, we establish that its expected regret or constraint violation must be $\Omega(\sqrt{n})$, where $n$ is the number of arriving users and the good capacities scale as $O(n)$. As an immediate consequence of this result, even an algorithm that sets expected equilibrium prices with knowledge of the distribution $\D$ must have a regret or constraint violation of $\Omega(\sqrt{n})$, which serves as a performance benchmark for an algorithm for online Fisher markets.

The limitations of static pricing motivate the design of adaptive (dynamic) pricing algorithms for online Fisher markets. To this end, we first present an adaptive variant of expected equilibrium pricing with $O(\log(n))$ regret and a constant constraint violation, i.e., independent of the number of users $n$, for discrete distributions with finite support (Section~\ref{sec:adaptiveExEq}). We establish these guarantees using techniques from parametric optimization, convex programming duality, and concentration inequalities. We note that while our adaptive expected equilibrium pricing algorithm achieves low regret and constraint violation guarantees, it requires knowledge of the distribution and does not naturally extend to continuous distributions.

Since the probability distribution $\D$, in general, may not be known (and the distribution may be continuous rather than discrete), in Section~\ref{sec:PPARMain}, we develop a simple yet effective approach to set prices that only relies on users' revealed preferences, i.e., past observations of user consumption. This algorithmic approach not only preserves user privacy as it requires no information on users' utility and budget parameters but also has a computationally efficient price update step, making it practically implementable. 

Notably, the revealed preference algorithm, under a mild assumption on the distribution $\D$, achieves an expected regret and constraint violation of $O(\sqrt{n})$ under a fixed step size of the price updates at each iteration. That is, our revealed preference algorithm achieves an expected regret and constraint violation (up to constants) that are no more than that of a static expected equilibrium pricing approach (and that of any static pricing algorithm) with complete knowledge of the distribution $\D$. Moreover, we show that the regret and constraint violation of the revealed preference algorithm can be reduced to $O(n^{2/5})$ for discrete distributions with finite support through a two-stage adjustment of the step size of the price updates, thereby strictly improving on the performance of static pricing algorithms for discrete distributions. Thus, our results highlight that a single static pricing rule is insufficient in achieving good performance in Fisher markets in the online incomplete information setting, and adaptive pricing methods, even with limited informational assumptions as in the revealed preference setting, can be developed with better performance guarantees. To establish these regret and constraint violation bounds, we develop a novel potential function argument that leverages the structural properties of Fisher markets and the price update step of our proposed algorithm. The key novelty in our regret analysis stems from the fact that our work considers a logarithmic objective that is both unbounded and negative, which is unlike the non-negative and bounded objectives typically considered in the online learning literature~\citep{Balseiro2020DualMD}. We also note that we term this algorithm as one based on revealed preferences (and thus one that preserves user privacy), as the \emph{only} information this algorithm relies on to make pricing decisions is users' revealed preferences, in contrast to the adaptive expected equilibrium pricing algorithm, which additionally requires knowledge of the distribution $\D$.

We also develop feasible variants of our adaptive pricing algorithms that respect resource capacities and present their corresponding regret bounds in Section~\ref{sec:feasible-algo-design}. In particular, we first introduce a general method of modifying any algorithm for online Fisher markets to guarantee feasibility, i.e., satisfy the capacity constraints. Then, we leverage this general framework to develop feasible variants of the adaptive expected equilibrium pricing approach with $O(\log(n))$ regret and the revealed preference algorithm with $O(\sqrt{n} \log(n))$ and $O(n^{2/5} \log(n))$ regret for the settings with a fixed step size and a two-stage adjustment of the step size of the price updates, respectively, under their respective distributional assumptions. These regret bounds indicate that our performance guarantees for the setting when violations of the capacity constraints are permissible extend to settings where resource capacities cannot be violated with minimal performance loss.

\ifarxiv
Finally, we evaluate the performance of the feasible variants of our revealed preference algorithm under both a fixed step size and a two-stage adjustment of the step size of the price updates relative to three benchmarks (Section~\ref{sec:experiments}). Our results validate our regret guarantees for the revealed preference algorithm under the chosen step sizes and demonstrate the efficacy of our revealed preference algorithm compared to the benchmarks, some of which have access to additional information on users' utility and budget parameters.

\fi

In the appendix, we review additional related literature, provide proofs omitted from the main text, and discuss additional numerical results to further demonstrate the efficacy of our developed algorithms.

\vspace{-5pt}
\section{Literature Review} \label{sec:literature}

Online resource allocation problems have been widely studied in operations research and computer science, and 
one of the most well-studied classes of such problems is online linear programming (OLP). 
While the traditional approach to OLP problems has been to develop worst-case guarantees~\citep{msvv}, these often pessimistic guarantees have prompted the study of beyond worst-case methods for such problems~\citep{beyond-comp-analysis}. As with several beyond worst-case approaches for OLP, we develop algorithms under the stochastic input model, where the input is drawn i.i.d. from some probability distribution. However, in contrast to the OLP literature, we study a non-linear concave objective.

Since non-linear objectives arise in many online resource allocation problems, there has been a growing interest in studying online convex optimization (OCO)~\citep{hazan2016introduction}. In this context,~\citet{AgrawalD15} study OCO problems with concave objectives and convex constraints and develop dual-based algorithms with low regret under both the random permutation and stochastic input models. More recently,~\citet{Balseiro2020DualMD} developed an online mirror descent algorithm with sub-linear regret for a general class of non-linear objectives. 
Akin to these works, we study an online resource allocation problem with a concave objective, which, in the Fisher market context, is the budget-weighted log utility objective~\citep{EisGale}, i.e., the sum of the logarithm of users' utilities weighted by their budgets. However, unlike these works that assume non-negativity and boundedness of the objective, we make no such assumption since the Fisher social welfare objective involves a logarithm, which can be both unbounded and negative. Thus, unlike classical online learning approaches, in the price update step of our revealed preference algorithm, we do not project the price vector onto the non-negative orthant. Consequently, to bound the regret and constraint violation of our revealed preference algorithm, we develop a novel potential function argument that leverages the structural properties of Fisher markets. Furthermore, we develop posted-price mechanisms, wherein users purchase their most favored bundle of goods given the set prices, which is in contrast to the OLP and OCO literature that have typically focused on \emph{hidden} price mechanisms, wherein the central planner decides on an allocation for each user after observing their attributes.

As in our work that studies an online variant of Fisher markets, several other online variants of Fisher markets have also been considered. While most works~\citep{azar2016allocate,banerjee2022online,gao2021online} study the setting when goods arrive sequentially and must be allocated upon their arrival to a fixed set of users, we consider the setting wherein users enter sequentially and purchase a fixed set of resources.


In the context of online user arrival in Fisher markets,~\cite{sinclair2021sequential} studies the problem of allocating a fixed set of resources to a random number of users that arrive over multiple rounds; however, our work differs from~\cite{sinclair2021sequential} in several ways. First, in~\cite{sinclair2021sequential}, the number of users arriving at each round is unknown (and is randomly drawn from some known distribution). In contrast, we consider a setting where the total number of users is known \emph{a priori} but consider a more general class of user preferences. In particular, in our setting, users' preferences can, in general, be drawn from an unknown continuous probability distribution, and users' budgets may not be equal, unlike in~\cite{sinclair2021sequential}, wherein users belong to a finite set of known types and have the same budgets. Moreover, unlike the theoretical guarantees in our work that apply to a more general class of utility and budget distributions, the theoretical guarantees in~\cite{sinclair2021sequential} require the setting of a single resource and a single user type (and their proof techniques extend to the multi-resource setting where each type desires a unique resource). 

In addition, unlike~\cite{sinclair2021sequential} who study a setting and design algorithms wherein a central planner determines the allocations for all users, we design posted-price mechanisms, wherein each arriving user purchases their most favored bundle of goods given the set prices. Consequently, given the difference in the settings between our work and that in~\cite{sinclair2021sequential}, we adopt different metrics to evaluate the performance of an algorithm (see Section~\ref{sec:moodelPerfMeasures}). While~\cite{sinclair2021sequential} study the notions of approximate Pareto inefficiency and approximate envy-freeness as their performance metrics, we study constraint violations and regret under the Eisenberg-Gale objective as our performance metrics, which serve as a proxy for a solution corresponding to an approximate market equilibrium (see Appendix~\ref{sec:apx-equilibria} for more details). We note here that our constraint violation metric resembles the Pareto inefficiency metric in~\cite{sinclair2021sequential}, which is related to the extent to which the capacity constraints are not satisfied. Moreover, unlike~\cite{sinclair2021sequential} who need to study the approximate envy-freeness as a central planner determines all users' allocations, in the revealed preference setting studied in this work where users observe the posted prices and freely choose which goods to purchase to obtain their most favored bundle of goods given the set prices, our proposed algorithms are envy-free by design (see Appendix~\ref{sec:apx-equilibria} for more details).

Our work is also closely related to prior works in revenue management on designing posted-price mechanisms in an incomplete information environment. In single-product settings,~\cite{kleinberg2003value,wang-ye-posted-price-singleproduct} study a revenue maximization problem for a seller having demand uncertainty and develop dynamic posted-price algorithms with asymptotically near-optimal performance. However, extending the proposed posted-price learning approaches to the multi-product setting, which we focus on in this work, has been noted as a significantly more challenging problem~\citep{wang-ye-posted-price-singleproduct}. In the multi-product setting,~\citet{NEURIPS2018_403ea2e8,JMLR:v20:17-357} study posted-price mechanisms in incomplete information settings with binary feedback, i.e., the seller observes whether a user purchased a product. However, in our setting, users purchase an optimal bundle of goods given the set prices, i.e., users always purchase resources, though the amount of resources purchased may differ across users. Moreover, compared to the revenue maximization objectives considered in these works, we consider a budget-weighted logarithmic utility objective that is both envy-free and Pareto optimal (see Section~\ref{sec:offline}).




Our work is also related to beyond worst-case analysis, revealed preference approaches, OCO with long-term constraints, and artificial currency mechanisms, which we review in Appendix~\ref{apdx:addnal-Related-work}.

\section{Model and Problem Formulation} \label{sec:moodelPerfMeasures}

In this section, we introduce our modeling assumptions and the individual optimization problem of users (Section~\ref{sec:preliminaries}), present the Eisenberg-Gale convex program used to compute equilibrium prices (Section~\ref{sec:offline}), and introduce the performance metrics used to evaluate the efficacy of an online algorithm (Section~\ref{sec:performancemeasures}).

\vspace{-10pt}

\subsection{Preliminaries and Individual Optimization Problem} \label{sec:preliminaries}

We study the problem of allocating $m$ divisible goods to $n$ users that arrive sequentially. Each good $j \in [m]$ has a capacity $c_j = n d_j$, where we denote $\c \in \mathbb{R}^m$ as the vector of good capacities, $\d \in \mathbb{R}^m$ as the vector of good capacities per user, and the set $[m] = \{1, \ldots, m \}$. Each user $t \in [n]$ has a budget $w_t$ of (artificial) currency, and to model users' preferences over the goods, we assume that each user's utility is linear in their allocations. In particular, for a vector of allocations $\x_t \in \mathbb{R}^m$, where $x_{tj}$ represents the consumption of good $j$ by user $t$, the utility function $u_t(\x_t): \mathbb{R}^m \rightarrow \mathbb{R}$ is given by $u_t(\x_t) = \u_t^\top \x_t = \sum_{j = 1}^m u_{tj} x_{tj}$, where $\u_t \in \mathbb{R}^m$ is a vector of utility coefficients and $u_{tj}$ is the utility received by user $t$ for consuming one unit of good $j$. Then, for a given price vector $\p \in \mathbb{R}^m$, the individual optimization problem for user $t$ can be described as \vspace{-10pt}
\begin{maxi!}|s|[2]                   
    {\mathbf{x}_t \in \mathbb{R}^m}                               
    { \u_t^\top \x_t = \sum_{j = 1}^m u_{tj} x_{tj},  \label{eq:Fisher1}}   
    {\label{eq:Eg001}}             
    {}                                
    \addConstraint{\mathbf{p}^\top \mathbf{x}_t}{\leq w_t, \label{eq:Fishercon1}}    
    \addConstraint{\mathbf{x}_t}{\geq \mathbf{0}, \label{eq:Fishercon3}}  
\end{maxi!}
where~\eqref{eq:Fishercon1} is a budget constraint. 
Since the utility function is linear, the optimal solution of Problem~\eqref{eq:Fisher1}-\eqref{eq:Fishercon3} given a price vector $\p$ is such that users purchase goods maximizing their \emph{bang-per-buck}, i.e., user $t$ purchases an affordable bundle of goods in the set $S_t^*(\p) = \{j: j \in \argmax_{j' \in [m]} \frac{u_{tj'}}{p_{j'}} \}$.

The prices of the goods in the market that users best respond to through the solution of Problem~\eqref{eq:Fisher1}-\eqref{eq:Fishercon3} are set by a central planner whose goal is to set equilibrium prices, defined as follows.

\begin{definition} (Equilibrium Price Vector) \label{def:eq-price}
A price vector $\p^* \in \mathbb{R}_{\geq 0}^m$ is an equilibrium price if there are allocations $\x_t^*(\p^*) \in \mathbb{R}^m$ for each user $t \in [n]$ such that:
\begin{enumerate}
    \item The allocation $\x_t^*(\p^*) \in \mathbb{R}^m$ is a solution of Problem~\eqref{eq:Fisher1}-\eqref{eq:Fishercon3} for all users $t \in [n]$ given $\p^*$;
    \item The prices of all goods are non-negative and the demand for all goods is no more than their capacity, i.e., $p_j^* \geq 0$ and $\sum_{t = 1}^n x_{tj}^*(\p^*) \leq c_j$ for all goods $j \in [m]$;
    \item If the price of a good $j$ is strictly positive, then the total demand for that good is equal to its capacity, i.e., if $p_j^*>0$ for some good $j \in [m]$, then $\sum_{t = 1}^n x_{tj}^*(\p^*) = c_j$.
\end{enumerate}
\end{definition}
The computation of equilibrium prices has been widely studied in the Fisher market literature and corresponds to one uniform price vector for all users at which the market clears. In classical (offline) Fisher markets, several methods, such as the Eisenberg-Gale convex program, have been developed to compute market equilibria (see Section~\ref{sec:offline}). However, these approaches assume that the central planner has complete knowledge of users' budget and utility parameters. 

Since information on users' utility and budget parameters are typically not known, and, in real markets, users tend to arrive sequentially over time, we study the online user arrival setting with incomplete information. In this context, we assume that users arrive sequentially with budget and utility parameters drawn i.i.d. from some distribution $\D$ with bounded and non-negative support. That is, the budget and utility parameters $(w_t, \u_t) \stackrel{i.i.d.}{\sim} \D$ for each user $t$, where the budget $w_t \in [\underline{w}, \Bar{w}]$ for some $\underline{w}> 0$ and the utility vector $\u_t \in [\underline{u}, \Bar{u}]^{m}$ for some $\underline{u} \geq \0$. We note that the stochastic input assumption has been used extensively in online Fisher markets~\citep{gao2021online,sinclair2021sequential} and in line with the Fisher market literature in the offline setting, we make the following assumption on the distribution $\D$.

\begin{assumption} [All Goods have Potential Buyers] \label{asmpn:mainRestriction}
The distribution $\D$ is such that for each good $j \in [m]$, there is a positive probability that users have a strictly positive utility for that good, i.e., $\mathbb{P}_{\D}((w, \u): u_{j}>0) > 0$.
\end{assumption}
Assumption~\ref{asmpn:mainRestriction} is akin to analogous assumptions in classical Fisher markets that require each good to have a potential buyer to guarantee the existence of equilibria~\citep{vazirani_2007}. In particular,~\cite{vazirani_2007} shows that if each good has a potential buyer and users have linear utilities, then equilibrium prices exist and are positive. Assumption~\ref{asmpn:mainRestriction} is mild since if no proportion of users had a positive utility for certain goods, those goods can be removed from the market as no user prefers to purchase them. 

A few comments about our modeling assumptions are in order. First, we assume that users' utilities are linear, a commonly used and well-studied utility function in classical Fisher markets~\citep{EisGale,vazirani_2007}. Next, 
in line with the online learning literature~\citep{li2020simple,chen2021linear}, we assume that the number of users $n$ is known. We defer the question of extending our results to the setting when $n$ is unknown~\citep{secretary-unknown} to future research. Finally, in the online setting, we study the arrival of users under the stochastic input model, which has been widely studied in the online resource allocation literature~\citep{li2020simple,li2021online,Balseiro2020DualMD}. We note that the stochastic input assumption naturally arises in several applications, including its common use in modeling customer arrivals~\citep{li2020simple}, and defer the exploration of online Fisher markets in non-i.i.d. settings to future research. In extending our work to non-i.i.d. settings, we do note that the arriving data process must be \emph{learnable}, i.e., past data provides insights into making better future decisions, as if past data does not inform pricing decisions for future users, e.g., under adversarial arrivals, then achieving a good performance may not be possible. To this end, we believe there is a significant scope to extend the online Fisher market setting studied in this work to settings including smoothed analysis, algorithm design with predictions, or the random permutation setting (where parameters are adversarially chosen but arrive in a random order).


\vspace{-10pt}

\subsection{Offline Allocations Using the Eisenberg-Gale Convex Program} \label{sec:offline}



In the offline setting, when complete information on the budgets and utilities of all users is known, the central planner can compute equilibrium prices through the dual variables of the capacity constraints of the following Eisenberg-Gale convex program~\citep{EisGale} \vspace{-8pt}
\begin{maxi!}|s|[2]                   
    {\mathbf{x}_t \in \mathbb{R}^m, \forall t \in [n]}                               
    {U(\mathbf{x}_1, ..., \mathbf{x}_n) = \sum_{t = 1}^{n} w_t \log \left( \sum_{j = 1}^m u_{tj} x_{tj} \right) , \label{eq:FisherSocOpt}}   
    {\label{eq:FisherExample2222}}             
    {}                                
    \addConstraint{\sum_{t = 1}^{n} x_{tj}}{ \leq c_j, \quad \forall j \in [m], \label{eq:FisherSocOpt1}}    
    \addConstraint{x_{tj}}{\geq 0, \quad \forall t \in [n], j \in [m], \label{eq:FisherSocOpt3}}  
\end{maxi!}
where~\eqref{eq:FisherSocOpt1} are capacity constraints 
and the Objective~\eqref{eq:FisherSocOpt} represents a budget-weighted geometric mean of buyer's utilities and is closely related to the Nash social welfare objective. 
If the market prices are set using the dual variables of the capacity Constraints~\eqref{eq:FisherSocOpt1}, then the optimal allocations of each user's individual optimization Problem~\eqref{eq:Fisher1}-\eqref{eq:Fishercon3} can be shown to be equal to that of Problem~\eqref{eq:FisherSocOpt}-\eqref{eq:FisherSocOpt3}~\citep{EisGale}. That is, the dual variables of the capacity Constraints~\eqref{eq:FisherSocOpt1} correspond to equilibrium prices.

The Eisenberg-Gale program has several desirable properties beyond its computational advantages, i.e., it can be solved in polynomial time~\citep{ye2008path,jain2007polynomial} with the same complexity as that of solving a linear program, that make it practically feasible. In particular, maximizing the weighted geometric mean results in an allocation satisfying both Pareto efficiency, i.e., no user can be made better off without making another user worse off, and envy-freeness, i.e., each user prefers their allocation compared to that of other users. In contrast, other welfare objectives often only satisfy one of these properties, e.g., the utilitarian welfare (sum of user's utilities) and egalitarian welfare (maximizing the minimum utility) objectives only achieve Pareto efficiency. Next, the Eisenberg-Gale objective achieves a natural compromise between the utilitarian and egalitarian objectives~\citep{BGM-2017}, thereby resulting in a simultaneously efficient and fair allocation. In particular, compared to maximizing utilitarian welfare, which may result in unfair allocations as some users may obtain zero utilities, under Objective~\eqref{eq:FisherSocOpt} all users receive a strictly positive utility. Further, compared to the egalitarian objective that may result in highly inefficient outcomes, optimizing the geometric mean of users' utilities is more robust as it provides a lower bound on the utilitarian welfare.

\vspace{-10pt}

\subsection{Algorithm Design and Performance Measures in Online Setting} \label{sec:performancemeasures}

While the offline allocations corresponding to the Eisenberg-Gale program have several desirable properties, achieving such allocations is generally not possible in the online setting when the central planner does not have access to information on users' utility and budget parameters. As a result, we focus on devising algorithms that achieve good performance relative to an offline oracle with complete information on users' utilities and budgets. In particular, we evaluate the efficacy of an online allocation policy through two metrics: (i) expected regret, i.e., the optimality gap in the social welfare Objective~\eqref{eq:FisherSocOpt} of this allocation policy relative to the optimal offline allocation, and (ii) expected constraint violation, i.e., the degree to which the goods are over-consumed relative to their capacities. Here the expectation is taken with respect to the distribution $\D$ from which users' budget and utility parameters are drawn. While we focus on jointly optimizing regret and constraint violation, we also develop feasible variants of our algorithms that respect resource capacities with little additional performance loss in terms of regret (see Section~\ref{sec:feasible-algo-design}). 


In this section, we first present the class of online policies (algorithms) we focus on in this work (Section~\ref{sec:algoDesignAmpns}) and then formally define our studied regret and constraint violation metrics (Section~\ref{sec:perfMeasures}). 
\vspace{-5pt}
\subsubsection{Online Algorithm Design} \label{sec:algoDesignAmpns}

In online Fisher markets, a central planner needs to make an allocation $\x_t$ (or a pricing decision $\p^t$) instantaneously upon the arrival of each user $t \in [n]$. The allocation or pricing decisions depend on the information set $\I_t$ available to the central planner at the time of arrival of each user. Examples of the information sets $\I_t$ include the distribution $\D$ from which users' budget and utility parameters are drawn, the history of user allocations, i.e., $\{ \x_{t'} \}_{t' = 1}^{t-1}$, and the history of budget and utility parameters, i.e., $(w_{t'}, \u_{t'})_{t' =1}^{t-1}$~\citep{li2020simple,online-agrawal}. Under a given information set $\I_t$, the allocations $\x_t$ (or pricing decisions $\p^t$) are specified by a policy $\ppi^A = (\pi_1^A, \ldots, \pi_n^A)$ (or $\ppi^P = (\pi_1^P, \ldots, \pi_n^P)$), where $\x_t = \pi_t^A(\I_t)$ (or $\p^t = \pi_t^P(\I_t)$). Here the superscript ``$A$'' refers to an allocation-based policy, while the superscript ``$P$'' refers to a pricing-based policy. Note that when the central planner sets prices $\p^t = \pi_t^P(\I_t)$ for each user, the corresponding allocations $\x_t$ are given by the optimal solution to Problem~\eqref{eq:Fisher1}-\eqref{eq:Fishercon1} given the price $\p^t$. For the remainder of this work, since we focus on designing pricing policies, we drop the superscript in the notation for conciseness and use $\ppi$ to refer to a pricing policy. However, we do note that our pricing policies have allocation-based analogues, as elucidated for one of our algorithms (Algorithm~\ref{alg:AlgoProbKnownDiscrete}) in Section~\ref{sec:adaptiveExEq}, and that both pricing and allocation policies are closely connected to each other by duality.





In this work, we focus on designing posted-price mechanisms in the \emph{revealed preference} setting, where each user's parameters are private information, and the prices are adjusted solely based on past user consumptions observable to the central planner (see Section~\ref{sec:PPARMain}). In particular, we devise pricing policies $\ppi = (\pi_1, \ldots, \pi_n)$ that set prices $\p^1, \ldots, \p^n$ such that $\p^t = \pi_t(\{ \x_{t'} \}_{t' = 1}^{t-1})$, where $\x_t$ is an optimal solution of Problem~\eqref{eq:Fisher1}-\eqref{eq:Fishercon1} given the price $\p^t$. Our focus on posted-price algorithms that use revealed preference information (e.g., see Section~\ref{sec:PPARMain}) is in contrast to traditional online resource allocation algorithms that require information on past user attributes to make subsequent allocations~\citep{li2021online}.

\vspace{-5pt}
\subsubsection{Performance Metrics} \label{sec:perfMeasures}

We now detail the regret and constraint violation performance metrics. 


\vspace{-8pt}

\paragraph{Regret:} We evaluate the regret of any online algorithm (pricing policy) $\ppi$ through the difference between the optimal objective of Problem~\eqref{eq:FisherSocOpt}-\eqref{eq:FisherSocOpt1} and that of the allocations resulting from the pricing policy $\ppi$. Let $U_n^*$ denote the optimal Objective~\eqref{eq:FisherSocOpt}, i.e., $U_n^* = U(\x_1^*, \ldots, \x_n^*)$, where $\x_1^*, \ldots, \x_n^*$ are the optimal allocations corresponding to the solution of Problem~\eqref{eq:FisherSocOpt}-\eqref{eq:FisherSocOpt1}, and let $\x_t$ be the solution to Problem~\eqref{eq:Fisher1}-\eqref{eq:Fishercon1} given the price vector $\p^t$ corresponding to the policy $\ppi$ for each user $t \in [n]$.
Then, the expected regret of an algorithm $\ppi$ is $R_n(\ppi) = \mathbb{E}_{\D} \left[ U_n^* - U(\x_1, \ldots, \x_n) \right]$, 
where the expectation is taken with respect to the budget and utility distribution $\D$. In the rest of this work, with a slight abuse of notation, we drop the subscript $\D$ in the expectation and assume all expectations are with respect to $\D$, unless stated otherwise.

While the regret measure is defined with respect to the objective of Problem~\eqref{eq:FisherSocOpt}-\eqref{eq:FisherSocOpt3}, regret guarantees derived for Objective~\eqref{eq:FisherSocOpt} directly translate into corresponding guarantees for the Nash social welfare objective, defined as $NSW(\x_1, \ldots, \x_n) = \left( \prod_{t = 1}^n u_t(\x_t^*) \right)^{\frac{1}{n}}$ in the setting when all users have the same budgets. In particular, if the regret of an algorithm $\ppi$ is $o(n)$, then the ratio of the Nash social welfare objective of the algorithm $\ppi$ approaches that of the optimal offline oracle as $n$ becomes large, i.e., if $U_n^* - U_n(\ppi) \leq o(n)$ for some algorithm $\ppi$, then $\frac{NSW(\x_1^*, \ldots, \x_n^*)}{NSW(\x_1, \ldots, \x_n)} \rightarrow 1$ as $n \rightarrow \infty$. We present a detailed discussion of this connection between the above-defined regret metric, which applies to Objective~\eqref{eq:FisherSocOpt}, and the ratio between the Nash social welfare objective of the optimal offline oracle and that of an online algorithm in Appendix~\ref{apdx:regretNSWConnection}.


\vspace{-8pt}

\paragraph{Constraint Violation:} We evaluate the constraint violation of an algorithm through the norm of the expected over-consumption of the goods beyond their capacity. In particular, for consumption bundles $\x_1, \ldots, \x_n$ corresponding to the pricing policy $\ppi$, the vector of excess demands is $\v(\x_1, \ldots, \x_n) = \left( \sum_{t = 1}^n \x_{t} - \c \right)_+$, and its constraint violation is $V_n(\ppi) = \mathbb{E} \left[\norm{\v(\x_1, \ldots, \x_n)}_2 \right].$

\vspace{-8pt}

\paragraph{}

A few comments about the regret and constraint violation metrics are in order. First, we define regret based on Objective~\eqref{eq:FisherSocOpt}, which is a natural choice as optimizing the (weighted) geometric mean is equivalent to finding the market equilibrium in offline Fisher markets, as elucidated in Section~\ref{sec:offline}. Our choice of this objective also stems from the fact that allocations resulting from the Eisenberg-Gale program satisfy many desirable properties, such as Pareto efficiency and envy-freeness, while allocations corresponding to other social welfare functions, e.g., utilitarian welfare, typically only satisfy one of these properties. Furthermore, in contrast to the utilitarian welfare objective where some users may receive no resources, optimizing the Eisenberg-Gale objective, with its logarithmic term, ensures that all users receive some resources. 

Moreover, since our performance metrics are intricately connected to the Eisenberg Gale program, obtaining sub-linear guarantees for our performance metrics serves as a proxy for a solution corresponding to an approximate market equilibrium. For further details on the connection between our performance metrics and an approximate equilibrium (and approximate Pareto efficiency and envy-freeness), see Appendix~\ref{sec:apx-equilibria}. However, we note that investigating algorithm design under other social welfare objectives, e.g., utilitarian welfare, and regret metrics for online Fisher markets is an interesting avenue for future research. 

Next, note that the budget-weighted geometric mean Objective~\eqref{eq:FisherSocOpt} is nonlinear and unbounded. Thus, our regret metric differs from that considered in the online linear programming and online convex optimization literature, which typically assume a linear or a concave objective that is bounded and non-negative. Further, while we defined our constraint violation metric with the $L_2$ norm, by norm-equivalence, any constraint violation guarantees obtained with the $L_2$ norm can be extended to any $p$-norm, e.g., the $L_{\infty}$ norm. 


In this work, we jointly optimize for regret and constraint violation, as in~\citet{li2020simple,yu2017online}, for two primary reasons. First, designing strictly feasible algorithms often necessitates a sudden change in the allocation or pricing decisions at the point when one of the goods is nearly exhausted, e.g., all users after a particular point in time receive no resources~\citep{Balseiro2020DualMD}. Thus, allowing for some constraint violation enables a more natural class of algorithms without abrupt changes in the allocation or pricing decisions across users, e.g., where users receive no resources. Nonetheless, we also present feasible variants of our algorithms that respect resource capacities and demonstrate that even these feasible algorithm variants achieve low regret. Second, we optimize for both regret and constraint violation as achieving good performance on either is typically easy. In particular, setting prices of all goods to be very low will result in low regret but potentially lead to constraint violations since users will purchase large quantities of goods at lower prices. On the other hand, setting exceedingly large prices will have the opposite effect. 
We note that such a fundamental trade-off between regret and constraint violation also persists when optimizing regret metrics corresponding to other social welfare functions, e.g., utilitarian welfare~\citep{li2020simple}.



Finally, allowing for some constraint violation is natural and aligns with the literature on online constrained convex optimization where the constraints are only required to be approximately satisfied in the long run~\citep{yu2017online,jenatton2016adaptive,mahdavi2012trading}. Further, in online resource allocation applications, resource capacities are often not hard constraints as there is often some flexibility in the number of goods that can be made available during the online allocation process. Thus, some constraint violation is acceptable in practical resource allocation contexts. Yet, we also develop a general method of modifying an algorithm for online Fisher markets to satisfy the resource capacity constraints in Section~\ref{sec:feasible-algo-design}.




\vspace{-2pt}

\section{Static Pricing Limits and Adaptive Pricing Enhancements} \label{sec:staticMainSec}

In classical Fisher markets, the central planner determines one uniform price vector that applies to all users, i.e., an equilibrium price vector, which can be computed through the solution of the Eisenberg-Gale program. Thus, we begin our study of online Fisher markets by establishing the performance limitations of static pricing in this online incomplete information setting. In particular, we develop a $\Omega(\sqrt{n})$ lower bound on the expected regret and constraint violation of static pricing algorithms, which includes an algorithm that sets expected equilibrium prices with knowledge of the distribution $\D$ (Section~\ref{sec:lb-result}). Given the limitations of static pricing, we present an adaptive (dynamic) variant of expected equilibrium pricing that achieves an $O(\log(n))$ regret and constant (i.e., independent of the number of users $n$) constraint violation (Section~\ref{sec:adaptiveExEq}) for discrete distributions with finite support. We also compare the static and adaptive expected equilibrium pricing algorithms through numerical experiments, which we present in Appendix~\ref{apdx:staticVdynamic}.

\vspace{-10pt}

\subsection{Lower Bound for Static Pricing Algorithms} \label{sec:lb-result}


This section establishes that the expected regret or constraint violation of any static pricing algorithm that sets prices $\p^t = \p^{t'}$ for all $t, t' \in [n]$ must be $\Omega(\sqrt{n})$ in the studied online variant of Fisher markets.


\begin{theorem} \label{thm:lbStatic}
Suppose that users' budget and utility parameters are drawn i.i.d. from a distribution $\D$. Then, there exists a market instance for which either the expected regret or expected constraint violation of any static pricing algorithm is $\Omega(\sqrt{n})$, where $n$ is the number of arriving users.
\end{theorem}

\hproof
To prove this claim, we consider a setting with $n$ users (with a budget of one for all users) and two goods, each with a capacity of $n$. Further, consider a distribution $\D$ where users have utility $(1, 0)$ or $(0, 1)$, each with probability $0.5$. For this instance, we first lower bound Objective~\eqref{eq:FisherSocOpt}, which we obtain by utilizing the property that users' utility distribution is binomial. Next, we consider two cases: (i) the price of either good is at most $0.5$ and (ii) the price of both goods is greater than $0.5$. In the first case, we use the central limit theorem to establish that the expected constraint violation is $\Omega(\sqrt{n})$. In the second case, we show that either the expected constraint violation or the expected regret is $\Omega(\sqrt{n})$ utilizing both the central limit theorem and the derived lower bound on Objective~\eqref{eq:FisherSocOpt}, which establishes our claim. \qed

\vspace{5pt}

For a complete proof of Theorem~\ref{thm:lbStatic}, see Appendix~\ref{apdx:pflb}. 
The counterexample to prove Theorem~\ref{thm:lbStatic} highlights a fundamental trade-off between the regret and constraint violation metrics and can be readily modified to the setting where the support of the utility distribution is strictly positive (see Section~\ref{sec:regret-ub}), i.e., the utility distribution is $(\epsilon, 1)$ and $(1, \epsilon)$ with probability $0.5$ each for some small $\epsilon>0$. Further, Theorem~\ref{thm:lbStatic} implies that any static pricing algorithm that respects the resource capacities will result in a regret of $\Omega(\sqrt{n})$.

Theorem~\ref{thm:lbStatic} establishes the limitations of static pricing in online Fisher markets and is in stark contrast to the efficacy of equilibrium pricing in classical Fisher markets, wherein one uniform price vector applies to all users. This result points toward developing novel methods for understanding market equilibria in online Fisher markets and augments the literature in online resource allocation where static pricing or allocation approaches have limited performance~\citep{arlotto2019uniformly}. We reiterate that the result in Theorem~\ref{thm:lbStatic} is not an algorithm-independent lower bound for online Fisher markets, which motivates the development of a more general lower bound for all adaptive pricing algorithms, a generally quite challenging task as evidenced by the literature on online linear programming (e.g., see~\cite{bray2019logarithmic}) under different informational assumptions, e.g., with and without information on the distribution $\D$, as a direction for future research. 

While the lower bound in Theorem~\ref{thm:lbStatic} does not apply to adaptive pricing algorithms, it provides a benchmark for the performance of any algorithm for online Fisher markets as static pricing includes an expected equilibrium pricing approach with knowledge of the distribution $\D$. Since Theorem~\ref{thm:lbStatic} establishes a lower bound on the expected regret and constraint violation of all static pricing algorithms, it, in particular, implies that even with complete information on the distribution $\D$, setting expected equilibrium prices will result in either an expected regret or constraint violation of $\Omega(\sqrt{n})$, as highlighted by the following corollary.


\begin{corollary} \label{cor:expectedEqPricing}
Suppose that users' budget and utility parameters are drawn i.i.d. from a known distribution $\D$. Then, there exists a market instance for which the expected constraint violation of an algorithm that sets equilibrium prices based on the expected number of user arrivals is $\Omega(\sqrt{n})$. 
\end{corollary}
We reiterate that the static expected equilibrium pricing policy $\ppi$ utilizes distributional information, and thus $\p^t = \pi_t(\D)$ for all users $t\in [n]$. As this price vector is uniform across all users, Corollary~\ref{cor:expectedEqPricing} follows as an immediate consequence of Theorem~\ref{thm:lbStatic}, and thus we omit its proof.

\vspace{-6pt}

\subsection{Adaptive Variant of Expected Equilibrium Pricing} \label{sec:adaptiveExEq}

Motivated by the limitations of static pricing, we now develop adaptive pricing algorithms for online Fisher markets. To this end, in this section, we introduce an adaptive variant of expected equilibrium pricing for discrete distributions $\D$ with finite support (Section~\ref{subsubsec:adaptiveExEq}), as in the counterexample to prove Theorem~\ref{thm:lbStatic}, and show that it achieves an $O(\log(n))$ regret and constant constraint violation (Section~\ref{subsubsec:RegretCapVioAlgo4}), thereby highlighting the benefit of adaptivity in algorithm design for online Fisher markets. We note that while the algorithm presented in this section applies to finitely supported discrete distributions, it does not naturally extend to continuous distributions, for which we present a revealed preference algorithm in Section~\ref{sec:PPARMain}.


\vspace{-1pt}

\subsubsection{Adaptive Expected Equilibrium Pricing Algorithm} \label{subsubsec:adaptiveExEq}


To present the adaptive expected equilibrium pricing algorithm, we first introduce some notation and the \emph{certainty equivalent} problem used to set static equilibrium prices. In particular, we assume that the utility and budget parameters $(w, \u)$ of users are drawn i.i.d. from a discrete probability distribution with finite support $(\Tilde{w}_k, \Tilde{\u}_k)_{k = 1}^K$, where the support size $K \in \mathbb{N}$, and the probability of a user having budget and utility parameters $(\Tilde{w}_k, \Tilde{\u}_k)$ is given by $q_k$. That is, $\mathbb{P}((w_t, \u_t) = (\Tilde{w}_k, \Tilde{\u}_k)) = q_k$ for all $k \in [K]$, where $q_k \geq 0$ and $\sum_{k \in [K]} q_k = 1$. Then, to set static expected equilibrium prices, we define the following certainty equivalent formulation of the Eisenberg-Gale program \vspace{-5pt}
\begin{maxi!}|s|[2]                   
    {\mathbf{z}_k \in \mathbb{R}^m, \forall k \in [K]}                               
    {U(\z_1, ..., \z_K) = \sum_{k = 1}^{K} q_k \Tilde{w}_k \log \left( \sum_{j = 1}^m  \Tilde{u}_{kj} z_{kj} \right) , \label{eq:FisherSocOptScenario}}   
    {\label{eq:FisherExample12222}}             
    {}                                
    \addConstraint{\sum_{k = 1}^{K} z_{kj} q_k }{ \leq d_{j}, \quad \forall j \in [m], \label{eq:FisherSocOpt1Scenario}}    
    \addConstraint{z_{kj}}{\geq 0, \quad \forall k \in [K], j \in [m], \label{eq:FisherSocOpt3Scenario}}  
\end{maxi!}
where the stochastic or uncertain parameters of the problem, i.e., the proportion of users with a utility and budget of $(\Tilde{w}_k, \Tilde{\u}_k)$ for all $k \in [K]$, are assumed to be equal to their expected values $q_k$. 
Here~\eqref{eq:FisherSocOptScenario} is the objective of the Eisenberg-Gale program weighted by the probability of occurrence of the corresponding budget and utility parameters, and~\eqref{eq:FisherSocOpt1Scenario} are the capacity constraints wherein the allocations are weighted by their corresponding probabilities. For brevity, we denote the certainty equivalent Problem~\eqref{eq:FisherSocOptScenario}-\eqref{eq:FisherSocOpt3Scenario} as $CE(\d)$ for a vector of average resource capacities $\d$. 

Observe that the optimal dual variables of the capacity constraints of this certainty equivalent problem correspond to the static expected equilibrium prices. As an example, for the two-good counterexample used to prove Theorem~\ref{thm:lbStatic}, the distribution support size $K = 2$, where the budget and utility parameters $(w, \u)$ are $(1, (1, 0))$ and $(1, (0, 1))$, each with probability $q_1 = q_2 = 0.5$, and the average resource capacity of each good per user is one, i.e., $d_1 = d_2 = 1$. For these parameters, the optimal dual variables of the capacity constraints of the certainty equivalent Problem~\eqref{eq:FisherSocOptScenario}-\eqref{eq:FisherSocOpt3Scenario} correspond to a price vector of $(0.5, 0.5)$, which are the static expected equilibrium prices for the market instance described in the proof of Theorem~\ref{thm:lbStatic}.

However, as observed in Corollary~\ref{cor:expectedEqPricing}, an issue with static expected equilibrium prices is that the probability $q_k$ of a user belonging to type $k$ generally differs from the empirically observed fraction of users in each type, which can result in large constraint violations as some goods may be consumed too early, i.e., well before the arrival of the last user, due to the stochasticity in user arrivals. To circumvent this issue, we design an adaptive variant of expected equilibrium pricing (see Algorithm~\ref{alg:AlgoProbKnownDiscrete}) that increases the prices (relative to the static expected equilibrium prices) of goods that have been over-consumed and vice-versa. In particular, Algorithm~\ref{alg:AlgoProbKnownDiscrete} tracks the average ``remaining'' capacity of all goods at the arrival of each user $t$, i.e., a vector $\d_t$, where $d_{tj} = \frac{c_j - \sum_{t'=1}^{t-1}x_{t'j}}{n-t+1}$ for all users $t \in [n]$, and sets prices based on the certainty equivalent problem $CE(\d_t)$ with the updated average remaining good capacities. Such a dynamic adjustment of the average remaining good capacities ensures that over-consumed goods have higher dual prices (relative to the static expected equilibrium prices), guaranteeing that no resource is consumed too quickly in Algorithm~\ref{alg:AlgoProbKnownDiscrete}.

Finally, since the certainty equivalent problem $CE(\d_t)$ is only well-defined for an average capacity vector $\d_t > 0$, in Algorithm~\ref{alg:AlgoProbKnownDiscrete} we adopt two pricing mechanisms depending on the difference between the average remaining capacity $\d_t$ and the initial average capacity $\d_1 = \frac{\c}{n}$. In particular, if the average remaining capacity $\d_t$ does not deviate too far from the initial average capacity $\d_1 = \frac{\c}{n}$, i.e., $\d_t \in [\d - \Delta, \d + \Delta]$ for some constant vector $\0<\Delta < \d$, e.g., $\Delta = \frac{\d}{2}$, then prices are set based on the dual variables of the above certainty equivalent problem $CE(\d_t)$, which is well defined for any such $\d_t \in [\d - \Delta, \d + \Delta]$. After the first time $\tau$ that $\d_t \notin [\d - \Delta, \d + \Delta]$, static expected equilibrium prices are set in the market. We note that following period $\tau$, any strictly positive prices can be set, e.g., the price corresponding to the dual variables of $CE(\d_{\tau-1})$, and we choose the static expected equilibrium prices for simplicity. Furthermore, we note that the second part of this algorithm, following period $\tau$ can be modified to guarantee feasibility with respect to the resource capacity constraints (see Section~\ref{sec:specific-bds-noCapVio-algos1-2}). Finally, we note that the pricing policy $\ppi$ in Algorithm~\ref{alg:AlgoProbKnownDiscrete} depends on the distribution $\D$ and the history of past allocations, i.e., $\p^t = \pi_t(\D, \{ \x_{t'} \}_{t' = 1}^{t-1})$. In response to the set prices, users consume their optimal bundle of goods given by the solution of Problem~\eqref{eq:Fisher1}-\eqref{eq:Fishercon3}. 

\vspace{-5pt}

\begin{algorithm} 
\SetAlgoLined
\SetKwInOut{Input}{Input}\SetKwInOut{Output}{Output}
\Input{Initial Good Capacities $\c$, Number of Users $n$, Threshold Parameter Vector $\Delta$, \\ 
Support of Distribution $\{ w_k, \u_k \}_{k = 1}^K$, Occurrence Probabilities $\{q_k\}_{k = 1}^K$}
Initialize $\c_1 = \c$ and the average remaining good capacity to $\d_1 = \frac{\c}{n}$  \; 
 \For{$t = 1, 2, ..., n$}{
 \textbf{Phase I: Set Price} \\ 
 \eIf{$\d_{t'} \in [\d - \Delta, \d+\Delta] \text{ for all } t' \leq t$}{
   Set price $\p^t$ as the dual variables of the capacity constraints of $CE(\d_t)$ \;}
   {
   Set price $\p^t$ using the dual variables of the capacity constraints of $CE(\d)$ with $\d = \d_1$  \; 
   }
 \textbf{Phase II: Observe User Consumption and Update Available Good Capacities} \\  
   User purchases optimal bundle $\x_t$ by solving Problem~\eqref{eq:Fisher1}-\eqref{eq:Fishercon3} given price $\p^t$ \; 
 Update the available good capacities $\c_{t+1} = \c_{t} - \x_{t}$ \; 
 Compute the average remaining good capacities $\d_{t+1} = \frac{\c_{t+1}}{n-t}$ \; 
  }
\caption{Adaptive Expected Equilibrium Pricing}
\label{alg:AlgoProbKnownDiscrete}
\end{algorithm}

\vspace{-5pt}

Algorithm~\ref{alg:AlgoProbKnownDiscrete} is similar in spirit to the adaptive allocation algorithm in~\cite{chen2021linear}; however, in contrast to~\cite{chen2021linear}, who study online linear programming, our algorithm applies for online Fisher markets with a non-linear objective. Furthermore, while the algorithm in~\cite{chen2021linear} involves solving a sampling based linear program using observed parameters of users that have previously arrived, Algorithm~\ref{alg:AlgoProbKnownDiscrete} solves a certainty equivalent problem at each step when a user arrives. We also note that there is an allocation-based variant of Algorithm~\ref{alg:AlgoProbKnownDiscrete}, as with the allocation-based algorithm in~\cite{chen2021linear}. In particular, for a remaining average good capacity $\d_t$, where $\z_1^*, \ldots, \z_K^*$ are the optimal solutions of the certainty equivalent problem $CE(\d_t)$, the allocation made to user $t$ can be given by $\x_t = \z_k^*$ if user $t$ has the budget and utility parameters $(\Tilde{w}_k, \Tilde{\u}_k)$. Note here that $\z_k^*$ is one of the optimal consumption vectors given the price $\p^t$ for a user $t$ of type $k \in [K]$. Furthermore, note that as compared to the adaptive expected equilibrium pricing, which only requires information on the distribution $\D$ and the history of past allocations, this allocation-based policy additionally requires information on the budget and utility parameters of the user for which an allocation decision needs to be made, i.e., $\x_t = \pi_t^A(\D, \{ \x_{t'} \}_{t' = 1}^{t-1}, (w_t, \u_t))$.

Finally, Algorithm~\ref{alg:AlgoProbKnownDiscrete} crucially relies on the assumption that the distribution $\D$ is discrete with finite support and does not naturally extend to the continuous distribution setting. Observe that Algorithm~\ref{alg:AlgoProbKnownDiscrete} involves solving a certainty equivalent problem at each step, which can be done in polynomial time for discrete distributions with finite support. In contrast, for continuous distributions, a stochastic program would need to be solved at each step, which is computationally challenging. Thus, extending Algorithm~\ref{alg:AlgoProbKnownDiscrete} to the continuous distribution setting would require resorting to approximations, e.g., approximating the dual prices of the associated stochastic programs through sample average approximations or discretizing the continuous distribution $\D$ and applying Algorithm~\ref{alg:AlgoProbKnownDiscrete} to the discretized distribution. We defer a deeper exploration of extending Algorithm~\ref{alg:AlgoProbKnownDiscrete} and its analysis to the continuous distribution setting to future research.


\vspace{-3pt}

\subsubsection{Regret and Constraint Violation Guarantee of Algorithm~\ref{alg:AlgoProbKnownDiscrete}} \label{subsubsec:RegretCapVioAlgo4}

We now show that Algorithm~\ref{alg:AlgoProbKnownDiscrete} achieves an $O(\log(n))$ regret and a constant constraint violation. 

\begin{theorem} [Regret and Constraint Violation Bounds for Algorithm~\ref{alg:AlgoProbKnownDiscrete}] \label{thm:RegretCapVioAlgo4}
Suppose that users' budget and utility parameters are drawn i.i.d. from a discrete distribution $\D$ satisfying Assumption~\ref{asmpn:mainRestriction}, and let $\ppi$ denote the online pricing policy described by Algorithm~\ref{alg:AlgoProbKnownDiscrete}, with $\underline{p}, \bar{p}>0$ as the lower and upper bounds, respectively, for the prices $p_j^t$ for all goods $j$ and for all users $t \in [n]$. Further, let $\x_1, \ldots, \x_n$ be the allocations for the $n$ users, where $\x_t$ is an optimal solution for that user corresponding to the certainty equivalent problem $CE(\d_t)$ for $t \leq \tau$, where $\tau$ is the first time at which $\d_t \notin [\d-\Delta, \d+\Delta]$, and $\x_t$ is an optimal solution to $CE(\d)$ for $t > \tau$. Then, the constraint violation $V_n(\ppi) \leq O(1)$ and the regret $R_n(\ppi) \leq O(\log(n))$.
\end{theorem}

\hproof
We prove this claim using four lemmas presented in Appendix~\ref{apdx:pfAlgo4}. Our first lemma establishes a generic upper bound on the regret of an algorithm for online Fisher markets using convex programming duality. This bound is composed of two terms: (i) the first term is akin to the constraint violation of
the algorithm and, in particular, accounts for the loss of over (or under-consuming)
goods, and (ii) the second term accounts for the loss of setting prices that deviate far from the static expected equilibrium prices. 

To bound the first term of the generic regret bound (and the constraint violation of Algorithm~\ref{alg:AlgoProbKnownDiscrete}), we first note that the constraint violation is upper bounded by $O(\mathbb{E}[n-\tau])$, where $\tau$ is the first time at which the average remaining resource capacity vector $\d_t \notin [\d - \Delta, \d+\Delta]$. Then, using concentration inequalities and arguments analogous to those in~\cite{chen2021linear}, we show that $O(\mathbb{E}[n-\tau])$ is a constant, establishing our desired constraint violation bound. To upper bound the second term in the generic regret bound, we first apply a variable transformation to the dual of the certainty equivalent problem and leverage techniques from parametric optimization to establish a lipschitzness relation between the optimal price of the certainty equivalent Problem $CE(\d_t)$ and the vector $\d_t$. 
We then apply this Lipschitzness relation and use induction on the optimal dual prices of $CE(\d_t)$ to bound the difference between the adaptive and static expected equilibrium prices for each user, which gives the desired $O(\log(n))$ regret upper bound. \qed

For a complete proof of Theorem~\ref{thm:RegretCapVioAlgo4}, see Appendix~\ref{apdx:pfAlgo4}. We note that in addition to the tools used to analyse the algorithm in~\cite{chen2021linear} for linear programs, our regret analysis leverages the structural properties of Fisher markets and combines that with techniques from parametric optimization to establish necessary sensitivity relations for the non-linear Eisenberg-Gale program. Furthermore, as the counterexample used to prove Theorem~\ref{thm:lbStatic} involved a discrete distribution with finite support, Theorem~\ref{thm:RegretCapVioAlgo4} implies that Algorithm~\ref{alg:AlgoProbKnownDiscrete} achieves a constant constraint violation and logarithmic regret on the distribution $\D$ in that counterexample. 

As mentioned in the statement of Theorem~\ref{thm:RegretCapVioAlgo4}, the prices $p_j^t$ are strictly positive and bounded throughout Algorithm~\ref{alg:AlgoProbKnownDiscrete}. To see this, first note that the boundedness of the optimal dual prices of the certainty equivalent problem $CE(\d_t)$ at each step follows directly from the restriction that the vector of average remaining capacities $\d_t \in [\d-\Delta, \d+\Delta]$ for some $\Delta < \d$. In particular, the optimal dual prices of $CE(\d_t)$ for any vector $\d_t>0$ remain bounded as users' budgets and utilities are bounded. Next, these optimal dual prices are guaranteed to be positive under Assumption~\ref{asmpn:mainRestriction}~\citep{vazirani_2007}, i.e., as long as the distribution $\D$ is such that for each good $j$, there is at least one type $k \in [K]$ with probability $q_k>0$ such that the utility $u_{kj}>0$. 

We also note in the statement of Theorem~\ref{thm:RegretCapVioAlgo4} that the allocation $\x_t$ is a solution to the associated certainty equivalent problem at each step. That is, if user $t$ is of type $k$, then the allocation $\x_t = \z_k^*$ for the allocation-based analogue of Algorithm~\ref{alg:AlgoProbKnownDiscrete} presented in Section~\ref{subsubsec:adaptiveExEq}, where $\z_k^*$ is a solution to the certainty equivalent problem $CE(\d_t)$. Observe that such an allocation corresponds to one of the optimal consumption vectors for each user given the price $\p^t$, which is the dual price of that certainty equivalent problem. Further, in some special cases, e.g., the counter-example in the proof of Theorem~\ref{thm:lbStatic} where users only have utility for one good, there is only one optimal consumption vector, characterized by the solution of the certainty equivalent problem, for each user for any price $\p^t>\0$. Given this observation and noting that the utility of a user remains the same at all optimal consumption vectors given a price $\p^t$, focusing on the optimal solution to the corresponding certainty equivalent problem is without loss of generality. Our purpose of doing so is that it guarantees that the expected consumption at each step is equal to the average remaining good capacity, i.e., $\mathbb{E}[\x_t] = \d_t$ for all $t \leq \tau$, which we require to bound the constraint violation of Algorithm~\ref{alg:AlgoProbKnownDiscrete}.






Theorem~\ref{thm:RegretCapVioAlgo4} implies that Algorithm~\ref{alg:AlgoProbKnownDiscrete} satisfies the capacity constraints, up to constants, and achieves an $O(\log(n))$ regret, significantly improving upon the $\Omega(\sqrt{n})$ lower bound on either the expected regret or constraint violation of static pricing obtained in Theorem~\ref{thm:lbStatic}. Thus, Theorem~\ref{thm:RegretCapVioAlgo4} highlights the benefit of adaptivity in online Fisher markets and motivates the further development of adaptive pricing algorithms for this problem setting. We also note that the logarithmic regret guarantee achieved by Algorithm~\ref{alg:AlgoProbKnownDiscrete} matches known lower bounds in the online linear programming literature~\citep{bray2019logarithmic}, further highlighting the efficacy of Algorithm~\ref{alg:AlgoProbKnownDiscrete}. To further underscore the advantages of Algorithm~\ref{alg:AlgoProbKnownDiscrete} compared to static pricing, we present numerical experiments in Appendix~\ref{apdx:staticVdynamic}, which show that Algorithm~\ref{alg:AlgoProbKnownDiscrete} achieves a low regret with almost no constraint violation even for large problem instances with $n=10,000$ users. Moreover, in Section~\ref{sec:specific-bds-noCapVio-algos1-2}, we present a variant of Algorithm~\ref{alg:AlgoProbKnownDiscrete} that respects resource capacities while still achieving a regret of $O(\log(n))$.


Despite the significant advantages of Algorithm~\ref{alg:AlgoProbKnownDiscrete} compared to static pricing, it has its limitations in applications when knowledge of the distribution $\D$ is not readily available. Moreover, as noted in Section~\ref{subsubsec:adaptiveExEq}, Algorithm~\ref{alg:AlgoProbKnownDiscrete} requires a discrete distribution $\D$ with finite support (in which case the certainty equivalent problem $CE(\d_t)$ can be tractably solved at each step) and does not naturally extend to continuous distributions. To address these concerns, in the next section, we develop a revealed preference algorithm that does not use any distributional information and is applicable for general (non-discrete) distributions.


\vspace{-5pt}

\section{Revealed Preference Algorithm and Regret Guarantees} \label{sec:PPARMain}

\vspace{-1pt}

This section presents a revealed preference algorithm for online Fisher markets and its corresponding regret and constraint violation guarantees. In particular, this algorithm solely utilizes observations of past user consumption to inform pricing decisions for future arriving users without requiring any
information on users’ utility and budget parameters, thereby preserving user privacy. We show that this algorithm achieves a regret and constraint violation of $O(\sqrt{n})$ under a fixed step size of the price updates at each iteration. Moreover, for discrete distributions with finite support, we show that the revealed preference algorithm's regret and constraint violation can be improved to $O(n^{2/5})$ through an appropriate two-stage adjustment of the step size of the price updates. 
Thus, our obtained regret and constraint violation guarantees highlight that adaptive pricing methods, even with the limited informational assumptions of the revealed preference setting, can be developed with improved performance relative to static pricing approaches with complete distributional information on users' budget and utility parameters (see Theorem~\ref{thm:lbStatic} and Corollary~\ref{cor:expectedEqPricing}).



In this section, we first present the dual of the Eisenberg-Gale Program~\eqref{eq:FisherSocOpt}-\eqref{eq:FisherSocOpt3} (Section~\ref{sec:dualProblem}) and the revealed preference algorithm (Section~\ref{sec:algoMain}), which follows from performing sub-gradient descent on this dual. Then, we establish upper bounds on the regret and constraint violation of this algorithm with a fixed step size and a two-stage adjustment of the step size of the price updates in Sections~\ref{sec:regret-ub} and~\ref{sec:variable-step-size}, respectively.



\vspace{-5pt}

\subsection{Dual Formulation of Eisenberg-Gale Program} \label{sec:dualProblem}

Letting the price $p_j$ be the dual variable of the capacity constraint for good $j$, Problem~\eqref{eq:FisherSocOpt}-\eqref{eq:FisherSocOpt3}'s dual is \vspace{-5pt}
\begin{equation} \label{eq:dual}
\begin{aligned}
\min_{\mathbf{p}} \quad & \sum_{t = 1}^n w_t \log(w_t) - \sum_{t = 1}^n w_t \log \left( \min_{j \in [m]} \frac{p_j}{u_{tj}} \right) + \sum_{j = 1}^m p_j c_j - \sum_{t = 1}^n w_t.
\end{aligned}
\end{equation}
For a derivation of the above dual using the Lagrangian of Problem~\eqref{eq:FisherSocOpt}-\eqref{eq:FisherSocOpt3}, we refer to Appendix~\ref{apdx:dual-pf}. We note that the above dual problem is the unconstrained version of the dual problem presented in~\cite{duality2017CDGJMVY} with the additional terms $\sum_{t = 1}^n w_t \log(w_t)$ and $- \sum_{t = 1}^n w_t$ in the objective. Observe that these terms in the objective are independent of the prices and thus do not influence the optimal solution of the dual problem but are necessary to analyze the regret of the algorithm we develop.

Since users' budget and utility parameters are drawn i.i.d. from the same distribution, this dual problem can be re-formulated as the following sample average approximation (SAA) problem \vspace{-7pt}
\begin{equation} \label{eq:SAA2}
\begin{aligned}
\min_{\mathbf{p}} \quad & D_n(\p) = \sum_{j = 1}^m p_j d_j + \frac{1}{n} \sum_{t = 1}^n \left( w_t \log(w_t) - w_t \log \left(\min_{j \in [m]} \frac{p_j}{u_{tj}} \right) - w_t \right)
\end{aligned}
\end{equation}
by dividing the dual objective in Problem~\eqref{eq:dual} by the number of users $n$, where, recall that $d_j = \frac{c_j}{n}$. Note that each term in the second summation of the objective of the above problem is independent of each other under the i.i.d. assumption on the utility and budget parameters of users. 


\vspace{-5pt}

\subsection{Revealed Preference Algorithm} \label{sec:algoMain}


In this section, we present an algorithm to dynamically update the prices of the goods solely based on observations of user consumption. We term this algorithm as one based on revealed preferences, as the \emph{only} information this algorithm relies on to make pricing decisions is users' revealed preferences, in contrast to Algorithm~\ref{alg:AlgoProbKnownDiscrete}, which additionally requires knowledge of the distribution $\D$. 
In particular, we devise a pricing policy $\ppi = (\pi_1, \ldots, \pi_n)$ that sets a sequence of prices $\p^1, \ldots, \p^n$ such that the pricing decision at each step only depends on the observed history of user consumption at the previous steps, i.e., $\p^t = \pi_t(\{ \x_{t'} \}_{t' = 1}^{t-1})$, where the allocations $\x_t$ are given by the solutions to Problem~\eqref{eq:Fisher1}-\eqref{eq:Fishercon3} given the price $\p^t$. Our algorithm adjusts market prices when a user arrives based on whether the previous user consumed more or less than their market share of each good. In particular, the price of a good $j$ is increased (decreased) if the previous user consumed more (less) than the average good capacity $d_j = \frac{c_j}{n}$ units of good $j$. The prices are updated using a step size $\gamma_t$. This process of updating the prices is presented formally in Algorithm~\ref{alg:PrivacyPreserving}. 

\vspace{-5pt}

\begin{algorithm}
\SetAlgoLined
\SetKwInOut{Input}{Input}\SetKwInOut{Output}{Output}
\Input{Number of users $n$, Vector of good capacities per user $\d = \frac{\c}{n}$}
Initialize $\p^1 > \0$ \; 
 \For{$t = 1, 2, ..., n$}{
    \textbf{Phase I:} User Consumes Optimal Bundle $\x_t$ by solving Problem~\eqref{eq:Fisher1}-\eqref{eq:Fishercon3} given the price $\p^t$ \; 
   \textbf{Phase II (Price Update):} $\p^{t+1} \leftarrow \p^t - \gamma_t \left( \d - \x_{t} \right) $  \;
  }
\caption{Revealed Preference Algorithm for Online Fisher Markets}
\label{alg:PrivacyPreserving}
\end{algorithm}

\vspace{-5pt}

A few comments about Algorithm~\ref{alg:PrivacyPreserving} are in order. First, Algorithm~\ref{alg:PrivacyPreserving} is akin to several revealed preference approaches in the literature~\citep{roth2016watch,ji2018social}. However, unlike prior approaches focusing on quasi-linear utilities, Algorithm~\ref{alg:PrivacyPreserving} applies when users are budget-constrained, as in Fisher markets. Further, unlike much of the literature in online resource allocation that involves hidden-price mechanisms, wherein a central planner decides an allocation for each user after observing their attributes, Algorithm~\ref{alg:PrivacyPreserving} involves a posted-price mechanism. Consequently, since Algorithm~\ref{alg:PrivacyPreserving} relies on users' revealed preferences, the price update step does not require any information on users' budgets and utilities and thus preserves user privacy.


Next, while Algorithm~\ref{alg:PrivacyPreserving} is akin to the dual mirror descent algorithm in~\cite{Balseiro2020DualMD}, our regret analysis differs from that in~\cite{Balseiro2020DualMD}, as we consider a logarithmic objective that can be both unbounded and negative. We reiterate that prior works in online learning (e.g.,~\cite{Balseiro2020DualMD}) typically consider a non-negative and bounded objective to establish sub-linear regret. Hence, unlike standard dual sub-gradient descent approaches~\citep{Balseiro2020DualMD}, we do not project the price vector $\p^t$ to the non-negative orthant in the price update step of Algorithm~\ref{alg:PrivacyPreserving}. Instead, we develop a novel potential function argument to show that the prices always remain positive and bounded throughout the operation of Algorithm~\ref{alg:PrivacyPreserving} under a mild assumption on the distribution $\D$ (see Section~\ref{sec:regret-ub}). We reiterate that establishing the strict positivity of prices is not required for non-negative and bounded objectives, as considered in~\cite{Balseiro2020DualMD}. Moreover, while we obtain an $O(\sqrt{n})$ regret and constraint violation guarantee for Algorithm~\ref{alg:PrivacyPreserving} with a fixed step size of the price updates, akin to the corresponding guarantee in~\cite{Balseiro2020DualMD}, we additionally present a method to adjust the step size of the price updates of Algorithm~\ref{alg:PrivacyPreserving} to obtain $O(n^{2/5})$ regret and constraint violation guarantees for discrete distributions with finite support (Section~\ref{sec:variable-step-size}).

Algorithm~\ref{alg:PrivacyPreserving} is also practically implementable with low computational overhead since the computational complexity of the price updates is only $O(m)$ each time a user arrives. Note here that Phase I of Algorithm~\ref{alg:PrivacyPreserving}, wherein each arriving user solves their individual optimization problem, also has an $O(m)$ complexity as users purchase the good with the maximum bang-per-buck ratio. However, since Phase I of Algorithm~\ref{alg:PrivacyPreserving} is a distributed step, the central planner only incurs a cost when performing the price updates in Phase II. Finally, for each user $t$, the price update step follows from performing sub-gradient descent on the $t$'th term of the dual Problem~\eqref{eq:SAA2}. In particular, if the optimal consumption set $S_t^*$ for user $t$, given the price vector $\p^t$, consists of one good, then the sub-gradient of the $t$'th term of the dual Problem~\eqref{eq:SAA2} is given by
$\left.\partial_{\p}\left( \sum_{j = 1}^m p_{j} d_j + w_t \log \left(w_t\right)-w_t \log \left(\min _{j \in[m]} \frac{p_{j}}{u_{tj}}\right) - w_t \right)\right|_{\p=\p^t} = \d - \x_t,$
where $\x_t$ is an optimal bundle corresponding to the solution of Problem~\eqref{eq:Fisher1}-\eqref{eq:Fishercon3} of agent $t$. Note here that $x_{tj^*} = \frac{w_t}{p_{tj^*}}$ for the good $j^*$ in its optimal consumption set $S_t^*$, which is of cardinality one, and $x_{tj} = 0$ for all goods $j \neq j^*$. 

\ifarxiv

Given the connection between gradient descent and the price updates in Algorithm~\ref{alg:PrivacyPreserving}, we note that other price update steps could also have been used in Algorithm~\ref{alg:PrivacyPreserving} based on mirror descent. For instance, instead of adjusting the prices through an additive update, as in Algorithm~\ref{alg:PrivacyPreserving}, prices can be modified through the widely studied multiplicative update rule~\citep{CHEUNG2019,Balseiro2020DualMD}: $\p^{t+1} \leftarrow \p^t e^{- \gamma_t \left( \d - \x_{t} \right)}$. 

In Appendix~\ref{apdx:AddVMult}, we compare the regret and constraint violation of Algorithm~\ref{alg:PrivacyPreserving} with the additive price update step to the corresponding algorithm with a multiplicative price update step through numerical experiments. For our theoretical analysis, we focus on the additive price update step in Algorithm~\ref{alg:PrivacyPreserving} and defer an exploration of the regret and constraint violation guarantees under the multiplicative price update step to future research. To this end, we do mention that this mirror descent-based multiplicative price update step achieves $O(\sqrt{n})$ regret in~\cite{Balseiro2020DualMD} for bounded and non-negative objectives and believe that some of their techniques can be extended to the budget-weighted log utility objective studied in this work.

\fi


\subsection{Regret and Constraint Violation Bound with Fixed Step Size of Price Updates} \label{sec:regret-ub}


This section establishes an $O(\sqrt{n})$ bound on the expected regret and constraint violation of Algorithm~\ref{alg:PrivacyPreserving} when the step sizes of its price updates are fixed to $\gamma = \gamma_t = O(\frac{1}{\sqrt{n}})$ for all users $t$. To establish the regret and constraint violation bounds, we make the following regularity assumption on the utility and budget parameters of users, which ensures that all users' utilities and budgets are strictly positive and bounded.
\begin{assumption} [Support of $\D$] \label{asmpn:regularity}
The support of the distribution $\D$ is such that the utilities and budgets are strictly positive and bounded, i.e., $\u \in [\underline{u}, \Bar{u}]$ and $w \in [\underline{w}, \Bar{w}]$, where $\underline{u}, \underline{w}>0$.  
\end{assumption}
Assumption~\ref{asmpn:regularity} imposes a mild restriction on the set of allowable distributions from which the budget and utility parameters of users are drawn. In particular, the boundedness of the utilities and user budgets are standard assumptions in the Fisher market literature, and the positivity of the budgets is a natural condition as users with no budgets have no buying power and can thus be removed from consideration of the set of users in the market. Furthermore, the condition on the positivity of the utilities is mild and introduced mainly for simplicity as it aligns with practical contexts where users typically receive some positive utility (given by any small constant $\underline{u}>0$) for obtaining resources. Note that changing the utility parameters from zero to some small positive constant also generally does not influence the optimal choice set of users. 

We now establish an $O(\sqrt{n})$ bound on both the expected regret and constraint violation of Algorithm~\ref{alg:PrivacyPreserving}.
\begin{theorem} [Regret and Constraint Violation Bounds for Algorithm~\ref{alg:PrivacyPreserving}] \label{thm:PrivacyPreserving}
Suppose that the budget and utility parameters of users are drawn i.i.d. from a distribution $\D$ satisfying Assumptions~\ref{asmpn:mainRestriction} and~\ref{asmpn:regularity}. Furthermore, let $\ppi$ denote the online pricing policy described by Algorithm~\ref{alg:PrivacyPreserving}, $\x_1, \ldots, \x_n$ be the corresponding allocations for the $n$ users. Then, for a step size $\gamma = \gamma_t = \frac{\Bar{D}}{\sqrt{n}}$ for some constant $\Bar{D}>0$ for all users $t \in [n]$, the regret $R_n(\ppi) \leq O(\sqrt{n})$ and the constraint violation $V_n(\ppi) \leq O(\sqrt{n})$.
\end{theorem}

\hproof
The proof of Theorem~\ref{thm:PrivacyPreserving} relies on two intermediate lemmas. First, we show that if the price vector at every step of Algorithm~\ref{alg:PrivacyPreserving} is bounded above and below by some positive constant, then the $O(\sqrt{n})$ bounds on both the regret and constraint violation hold. The proof of this claim uses convex programming duality and the stochastic assumption on the budget and utility parameters of users. We then show that the price vector $\p^t$ in Algorithm~\ref{alg:PrivacyPreserving} remains strictly positive and bounded for all users $t$ if the distribution $\D$ satisfies Assumption~\ref{asmpn:regularity}. We prove this result in two steps. First, we show that if the price vector $\p^t$ at each iteration of Algorithm~\ref{alg:PrivacyPreserving} is bounded below by some vector $\underline{\p} > \0$, then the price vector is also bounded from above. 
Next, we show that the prices will always remain positive under Assumption~\ref{asmpn:regularity}. To show this, we develop a novel potential function argument that leverages the structural properties of Fisher markets and the price update rule in Algorithm~\ref{alg:PrivacyPreserving}. In particular, we define a potential $V_t = (\p^t)^T \d$, and show that this potential is non-decreasing, i.e., $V_{t+1} \geq V_t$, if the prices of all goods are below a specified threshold for user $t$. We then combine this potential function argument with Assumption~\ref{asmpn:regularity} to show that the prices of all goods are lower bounded by some price $\underline{p}>0$, which establishes our claim. \qed

\vspace{5pt}

For a more detailed proof sketch of Theorem~\ref{thm:PrivacyPreserving}, see Appendix~\ref{sec:pfThm3}, and for a complete proof of the lemmas presented in Appendix~\ref{sec:pfThm3} required to prove Theorem~\ref{thm:PrivacyPreserving}, see Appendices~\ref{apdx:ubpf} and~\ref{apdx:positivePricesPf}. We reiterate here that, as opposed to prior literature on online learning, our regret analysis enables us to establish the regret and constraint violation bounds for Algorithm~\ref{alg:PrivacyPreserving} without projecting the price vector $\p^t$ to the non-negative orthant. Note that if the price of some good became non-positive during Algorithm~\ref{alg:PrivacyPreserving}, then users would purchase an infinite amount of that good by the solution of Problem~\eqref{eq:Fisher1}-\eqref{eq:Fishercon3}, resulting in an unbounded constraint violation. Thus, our potential function argument to show that the positivity of prices during Algorithm~\ref{alg:PrivacyPreserving} is crucial to establishing the regret and constraint violation bounds in Theorem~\ref{thm:PrivacyPreserving}. 

Theorem~\ref{thm:PrivacyPreserving} establishes that the expected regret and constraint violation of Algorithm~\ref{alg:PrivacyPreserving} are sub-linear in the number of users $n$, which suggests that the obtained solution corresponding to Algorithm~\ref{alg:PrivacyPreserving} serves as a proxy for an approximate market equilibrium. For further details on the connection between our performance metrics and an approximate market equilibrium (and corresponding approximate Pareto efficiency and envy-freeness), see Appendix~\ref{sec:apx-equilibria}. As with Theorem~\ref{thm:RegretCapVioAlgo4}, the obtained regret and constraint violation bounds depend on the specific problem instance and the support of the distribution $\D$ (see Appendices~\ref{apdx:ubpf} and~\ref{apdx:positivePricesPf}), but we focus our attention here on the dependency of the regret bound on the number of users $n$. Note that Theorems~\ref{thm:lbStatic} and~\ref{thm:PrivacyPreserving} jointly imply that Algorithm~\ref{alg:PrivacyPreserving}, while preserving user privacy, achieves expected regret and constraint violation guarantees, up to constants, that are no more than that of an expected equilibrium pricing approach (see Corollary~\ref{cor:expectedEqPricing}) with complete information on the distribution $\D$. 


Further, compared to Algorithm~\ref{alg:AlgoProbKnownDiscrete} that achieves a constant constraint violation and an $O(\log(n))$ regret (see Theorem~\ref{thm:RegretCapVioAlgo4}), Algorithm~\ref{alg:PrivacyPreserving} with a fixed step size of the price updates achieves a higher regret and constraint violation of $O(\sqrt{n})$. For discrete distributions with finite support, we further close the performance gap between these two algorithms 
in Section~\ref{sec:variable-step-size}. 

While the regret and constraint violation bounds of Algorithms~\ref{alg:AlgoProbKnownDiscrete} and~\ref{alg:PrivacyPreserving} highlight a performance loss in the absence of distributional information, Algorithm~\ref{alg:PrivacyPreserving} has several advantages to Algorithm~\ref{alg:AlgoProbKnownDiscrete}. First, Algorithm~\ref{alg:PrivacyPreserving} is applicable for a broader range of distributions compared to Algorithm~\ref{alg:AlgoProbKnownDiscrete}, which only applies for discrete distributions $\D$. Next, the price update step in Algorithm~\ref{alg:PrivacyPreserving} has a low computational overhead, while Algorithm~\ref{alg:AlgoProbKnownDiscrete} involves solving a convex program at each step. Finally, Algorithm~\ref{alg:PrivacyPreserving} is more practically viable as it only relies on users' revealed preferences, while Algorithm~\ref{alg:AlgoProbKnownDiscrete} requires knowledge of the distribution $\D$.

While Assumption~\ref{asmpn:regularity} is crucial to establishing Theorem~\ref{thm:PrivacyPreserving}, we can also extend this result to distributions such as in the counterexample used to prove Theorem~\ref{thm:lbStatic}. In particular, in Appendix~\ref{apdx:ubPrices}, we show for distributions such as in the counterexample to prove Theorem~\ref{thm:lbStatic} that the prices remain strictly positive through the operation of Algorithm~\ref{alg:PrivacyPreserving} with high probability using concentration inequalities. We also validate the positivity of prices through the operation of Algorithm~\ref{alg:PrivacyPreserving} using numerical experiments in Appendix~\ref{apdx:numericalValidationStrictPositivity}.

\vspace{-8pt}

\subsection{Regret and Constraint Violation Bound with Two-Stage Step Size Adjustment} \label{sec:variable-step-size}

\vspace{-3pt}

In this section, we show that the $O(\sqrt{n})$ bound on the expected regret and constraint violation of our revealed preference algorithm can be improved to $O(n^{2/5})$ for discrete distributions $\D$ with finite support (see Section~\ref{sec:adaptiveExEq}) through a more nuanced selection of the step-size of the price update step in Algorithm~\ref{alg:PrivacyPreserving}. Rather than considering a fixed step-size $\gamma = \gamma_t = O(\frac{1}{\sqrt{n}})$ for all users $t \in [n]$, we consider two different step sizes, corresponding to \emph{exploration} and \emph{exploitation} stages of the algorithm. In particular, we set the step size in the exploration stage, which corresponds to the first $n^{4/5}$ periods, to $\gamma_t = O(\frac{1}{n^{2/5}})$ and the step size in the exploitation stage, which corresponds to the remaining $n - n^{4/5}$ periods, to $\gamma_t = O(\frac{1}{n^{3/5}})$. Note that the step size in the exploration stage is larger than $O(\frac{1}{\sqrt{n}})$ to enable sufficient exploration, while the step size in the exploitation stage is smaller than $O(\frac{1}{\sqrt{n}})$, which enables the revealed preference algorithm to exploit by selecting prices in a small neighborhood of the price vector learned in the exploration stage.

We now establish that with the above choice of the step-sizes, Algorithm~\ref{alg:PrivacyPreserving} achieves an $O(n^{2/5})$ bound on both the expected regret and constraint violation metrics for discrete distributions with finite support.

\begin{theorem} [Regret and Constraint Violation for Two-Stage Revealed Preference Algorithm] \label{thm:PrivacyPreservingImproved}
Suppose that users' budget and utility parameters are drawn i.i.d. from a discrete distribution $\D$ (with finite support) satisfying Assumptions~\ref{asmpn:mainRestriction} and~\ref{asmpn:regularity}. Further, let $\ppi$ denote the pricing policy described by Algorithm~\ref{alg:PrivacyPreserving}, where the step-size $\gamma_t = O(\frac{1}{n^{2/5}})$ for the first $n^{4/5}$ periods and $\gamma_t = O(\frac{1}{n^{3/5}})$ for the remaining periods. 
Then, the regret $R_n(\ppi) \leq O(n^{2/5})$ and the constraint violation $V_n(\ppi) \leq O(n^{2/5})$.
\end{theorem}

Theorem~\ref{thm:PrivacyPreservingImproved} establishes that the regret and constraint violation of Algorithm~\ref{alg:PrivacyPreserving} can be improved from $O(\sqrt{n})$ (when the step-size of the price updates is fixed to $O(\frac{1}{\sqrt{n}})$) to $O(n^{2/5})$ for discrete distributions with finite support through a two-stage adjustment of the step-size of the price updates. 
Moreover, Theorem~\ref{thm:PrivacyPreservingImproved} implies that Algorithm~\ref{alg:PrivacyPreserving} with a two-stage adjustment of the step-size of the price updates, while preserving user privacy, achieves lower regret and constraint violation than that achievable through any static pricing approach (see Theorem~\ref{thm:lbStatic}), including one with knowledge of the distribution $\D$ (see Corollary~\ref{cor:expectedEqPricing}), for discrete distributions with finite support. In other words, Theorems~\ref{thm:lbStatic} and~\ref{thm:PrivacyPreservingImproved} highlight that a single static pricing rule is insufficient in achieving good performance in Fisher markets in the online incomplete information setting, and adaptive pricing methods, even under the limited informational assumptions of the revealed preference setting, can be developed with \emph{strictly} improved performance guarantees for certain classes of distributions.

While an appropriate two-stage adjustment of the step size of the price updates of Algorithm~\ref{alg:PrivacyPreserving} closes the performance gap between Algorithms~\ref{alg:AlgoProbKnownDiscrete} and~\ref{alg:PrivacyPreserving} for discrete distributions with finite support relative to the revealed preference algorithm with a fixed step size, we defer the problem of closing the performance gap between these algorithms further to future research. Furthermore, the performance gap of these algorithms motivates the development of a more general lower bound for all adaptive pricing algorithms, a very challenging problem as evidenced by the literature on online linear programming (e.g., see~\cite{bray2019logarithmic}), in the revealed preference setting without information on the distribution $\D$ as a direction for future research.

We omit the details of the proof of Theorem~\ref{thm:PrivacyPreservingImproved} for brevity and present its complete proof in Appendix~\ref{apdx:pf-PrivacyPreservingImproved}. In the following, we note that while the proof technique for Theorem~\ref{thm:PrivacyPreservingImproved} is inspired by the analysis of a similar approach in online linear programming~\citep{gao2024decoupling}, there are key differences between our revealed preference algorithm (with a two-stage adjustment of the step size of the price updates) and its analysis compared to the corresponding algorithm and analysis in~\cite{gao2024decoupling}. First, as noted in Section~\ref{sec:regret-ub}, our regret analysis, through a novel potential function argument, enables us to establish the regret and constraint violation bound for Algorithm~\ref{alg:PrivacyPreserving} with the two-stage adjustment of the step-size of the price updates without projecting the price vector to the non-negative orthant. While the prices are allowed to drop to zero in~\cite{gao2024decoupling} as the resource consumption at each iteration is bounded, in a Fisher market setting, if a good's price became non-positive, users would purchase an infinite amount of that good, resulting in an unbounded constraint violation. Hence, our potential function argument plays a critical role in proving our regret and constraint violation bounds by establishing the positivity of prices throughout the operation of Algorithm~\ref{alg:PrivacyPreserving}.

Next, unlike~\cite{gao2024decoupling}, who require an assumption that is analogous to a strong convexity condition for the linear program's dual objective to establish their regret and constraint violation bounds, we leverage our analysis to derive the Lipshitzness relation between the optimal prices and the corresponding average resource capacities established in the proof of Theorem~\ref{thm:RegretCapVioAlgo4} to show that this required condition is a property of the dual of the Eisenberg-Gale program when the distribution is discrete with finite support. We note that extending our analysis to general continuous distributions is challenging as our variable transformation used to establish the Lipshitzness relation in the proof of Theorem~\ref{thm:RegretCapVioAlgo4} would be infinite-dimensional in the setting with continuous distributions. Consequently, evaluating the norm of this infinite-dimensional variable would involve introducing additional regularity conditions, and we defer the question of extending our analysis of Algorithm~\ref{alg:PrivacyPreserving} with a two-stage adjustment of the step size of the price updates to general continuous distributions as a direction for future research. Moreover, unlike~\cite{gao2024decoupling} who study online linear programming, we tailor our analysis to the Fisher market setting with a logarithmic objective. Finally, unlike~\cite{gao2024decoupling} who only study a setting where constraint violations are allowed, we also analyse the feasible variant of Algorithm~\ref{alg:PrivacyPreserving} with the two-stage adjustment of the step-size in Section~\ref{sec:specific-bds-noCapVio-algos1-2}.

We also note that~\cite{gao2024decoupling} develop an approach with $O(n^{1/3})$ regret and constraint violation with an \emph{exploration} phase running two algorithms in parallel, a \emph{learning} algorithm and a \emph{decision} algorithm, and an \emph{exploitation} phase where the price learned via the learning algorithm is used to ``re-start'' the decision algorithm. While such a decoupling of learning and decision-making is possible in the setting of~\cite{gao2024decoupling}, where they observe each user's parameters, enabling the learning algorithm to update prices, doing so is not possible in our revealed preference setting, where users' utility and budgets are private information.

\vspace{-4pt}

\section{Designing Feasible Algorithms for Online Fisher Markets} \label{sec:feasible-algo-design}


Thus far, we have studied the setting when some constraint violation is acceptable to achieve low regret. This section considers the setting when exceeding resource constraints is not permissible and shows that Algorithms~\ref{alg:AlgoProbKnownDiscrete} and~\ref{alg:PrivacyPreserving} can be adapted to satisfy the resource capacities while still achieving low regret. To that end, we first highlight a method of modifying any algorithm for online Fisher markets to guarantee feasibility and present a corresponding generic regret bound of such an algorithm (Section~\ref{sec:general-algo-framework-bd}). Then, we use this generic bound to obtain regret bounds for the feasible variants of Algorithms~\ref{alg:AlgoProbKnownDiscrete} and~\ref{alg:PrivacyPreserving} (Section~\ref{sec:specific-bds-noCapVio-algos1-2}).



\vspace{-10pt}
\subsection{General Framework for Feasible Algorithm Design} \label{sec:general-algo-framework-bd}
\vspace{-2pt}


To introduce our general purpose method to modify any algorithm for online Fisher markets to guarantee feasibility, we first introduce some notation. In particular, let $\ppi$ denote any online algorithm (that may result in an infeasible outcome violating the capacity constraints) for online Fisher markets that sets strictly positive prices with a price lower bound of $\underline{p}>0$. Next, let $\epsilon>0$ be some constant and define a stopping time $\tau^{\ppi}$ as the first time less than $n$ at which there exists a resource $j$ with $\sum_{t' = 1}^{\tau^{\ppi}} x_{tj} + \frac{\Bar{w}}{\underline{p}} \geq c_j - \epsilon$. 

Then, we define the feasible variant ($\ppi^f$) of algorithm $\ppi$ as follows. First, run algorithm $\ppi$ until the stopping time $\tau^{\ppi}$, following which all remaining users receive $\frac{\epsilon}{n}$ of the remaining resources. 

A few comments about algorithm $\ppi^f$ are in order. First, notice that $\ppi^f$ is feasible by design as all users until $\tau^{\ppi}$ cannot use up more than $c_j-\epsilon$ units of any good $j$, following which all remaining users are allocated at most $\epsilon$ units of each good. Next, while $\ppi^f$ is a natural choice to design a feasible variant of an algorithm $\ppi$, we note that many other feasible variants of an online algorithm $\ppi$ can be developed.

Furthermore, our procedure to define a feasible variant of an algorithm $\ppi$ is unlike prior literature in online learning~\citep{Balseiro2020DualMD}. In particular, unlike~\cite{Balseiro2020DualMD}, where users arriving after the stopping time $\tau^{\ppi}$ are given no resources, in our setting, we terminate the algorithm before at least $\epsilon$ units of each resource are remaining and provide all subsequent users $\frac{\epsilon}{n}$ units of the remaining resources. Our rationale behind doing so is that unlike prior objectives in the online learning literature~\citep{Balseiro2020DualMD}, where the objective is assumed to be bounded and non-negative, our regret measure involves a budget-weighted logarithmic utility objective that can be both unbounded and negative. 

We now establish an upper bound on the regret of the feasible algorithm $\ppi^f$.

\begin{theorem} [Generic Regret Bound for Feasible Algorithms] \label{thm:general-regret-stoppingtime}
Suppose that the budget and utility parameters of users are drawn i.i.d. from a distribution $\D$ satisfying Assumption~\ref{asmpn:mainRestriction}. Let $\ppi$ denote an online pricing policy with stopping time $\tau^{\ppi}$ and let $\ppi^f$ denote its feasible variant where all users after $\tau^{\ppi}$ are given $\frac{\epsilon}{n}$ units of each of the remaining resources. Then, the regret
$R_n(\ppi^f) \leq R_{\tau^{\ppi}}(\ppi) + O(\mathbb{E}[(n - \tau^{\ppi})] \log(n))$.
\end{theorem}
For a proof of Theorem~\ref{thm:general-regret-stoppingtime}, see Appendix~\ref{apdx:pf-thm-no-capvio-stoppintime}. Theorem~\ref{thm:general-regret-stoppingtime} states that the regret of the feasible algorithm $\ppi^f$ is upper bounded by the sum of the regret of the algorithm $\ppi$ accumulated until period $\tau^{\ppi}$ and the difference between the total number of users $n$ and the stopping time $\tau^{\ppi}$ up to a logarithmic factor.

\vspace{-5pt}

\subsection{Regret Bounds for Feasible Variants of Algorithms~\ref{alg:AlgoProbKnownDiscrete} and~\ref{alg:PrivacyPreserving}} \label{sec:specific-bds-noCapVio-algos1-2}

This section applies the general algorithmic framework and the regret bound (Theorem~\ref{thm:general-regret-stoppingtime}) in Section~\ref{sec:general-algo-framework-bd} to obtain regret bounds for the feasible variants of Algorithm~\ref{alg:PrivacyPreserving} under both a fixed step size and a two-stage adjustment of the step size of the price updates. We also employ a similar analysis to develop a regret bound for the feasible variant of Algorithm~\ref{alg:AlgoProbKnownDiscrete}. Our results highlight that our earlier obtained regret guarantees naturally extend to the setting where the capacity violations are not permissible with little additional regret.


We first present an upper bound on the regret of the feasible variant of Algorithm~\ref{alg:PrivacyPreserving} with a fixed step size (Corollary~\ref{cor:noCapVioAlgo2}) and a two-stage adjustment of the step size of the price updates (Corollary~\ref{cor:noCapVioAlgo2-variable}), where all users after the corresponding stopping time of Algorithm~\ref{alg:PrivacyPreserving} are given $\frac{\epsilon}{n}$ units of the remaining resources. 


\begin{corollary}[Regret of Feasible Variant of Algorithm~\ref{alg:PrivacyPreserving} with a Fixed Step Size] \label{cor:noCapVioAlgo2}
Suppose users' budget and utility parameters are drawn i.i.d. from a distribution $\D$ satisfying Assumptions~\ref{asmpn:mainRestriction} and~\ref{asmpn:regularity}. Further, let $\ppi$ denote the online pricing policy described by Algorithm~\ref{alg:PrivacyPreserving} with a fixed step size of $\gamma_t = O(\frac{1}{\sqrt{n}})$ for all users $t$ with stopping time $\tau^{\ppi}$ and let $\ppi^{f}$ denote its feasible variant where all users after the stopping time are given $\frac{\epsilon}{n}$ units of each of the remaining resources. Then, the regret $R_n(\ppi^{f}) \leq O(\sqrt{n} \log(n))$.
\end{corollary}
\begin{corollary}[Regret of Feasible Variant of Algorithm~\ref{alg:PrivacyPreserving} with a Two-Stage Step Size Adjustment] \label{cor:noCapVioAlgo2-variable}
Suppose users' budget and utility parameters are drawn i.i.d. from a discrete distribution $\D$ (with finite support) satisfying Assumptions~\ref{asmpn:mainRestriction} and~\ref{asmpn:regularity}. Furthermore, let $\ppi$ denote the online pricing policy described by Algorithm~\ref{alg:PrivacyPreserving} with a two-stage adjustment of the step-size of the price updates, where $\gamma_t = O(\frac{1}{n^{2/5}})$ for the first $n^{4/5}$ users and $\gamma_t = O(\frac{1}{n^{3/5}})$ for the remaining users. Moreover, let $\tau^{\ppi}$ denote the stopping time of this algorithm and let $\ppi^{f}$ denote its feasible variant where all users after the stopping time are given $\frac{\epsilon}{n}$ units of each of the remaining resources. Finally, let $\underline{p}>0$ be the minimum price at any point during this algorithm. Then, for all $n \geq \max \{ 1, \left( \frac{2 \Bar{w}}{\underline{p} \min_{j} d_j} \right)^5 \}$, the regret $R_n(\ppi^{f}) \leq O(n^{2/5} \log(n))$.
\end{corollary}
For a proof of Corollaries~\ref{cor:noCapVioAlgo2} and~\ref{cor:noCapVioAlgo2-variable}, see Appendices~\ref{apdx:pf-cor-no-capVioAlgo2} and~\ref{apdx:pf-cor-no-capVioAlgo2-variable}, respectively. While no condition on the number of users $n$ is required to establish Corollary~\ref{cor:noCapVioAlgo2}, which corresponds to Algorithm~\ref{alg:PrivacyPreserving} with a fixed step size, we require the number of users $n$ to be sufficiently large in Corollary~\ref{cor:noCapVioAlgo2-variable} to ensure that the stopping time occurs after the change in the step size of Algorithm~\ref{alg:PrivacyPreserving}. Corollaries~\ref{cor:noCapVioAlgo2} and~\ref{cor:noCapVioAlgo2-variable} establish that designing a strictly feasible variant of Algorithm~\ref{alg:PrivacyPreserving} that satisfies resource capacities results in only an additional $O(\log(n))$ factor loss in the regret compared to when capacity violations are permissible (Theorems~\ref{thm:PrivacyPreserving} and~\ref{thm:PrivacyPreservingImproved}). Our obtained results, where we incur an additional $\log(n)$ factor loss in the regret, are in contrast to prior online resource allocation literature~\citep{Balseiro2020DualMD}. In particular, due to the non-negative and boundedness assumption on the rewards, prior work~\citep{Balseiro2020DualMD} obtains no additional logarithmic factor loss in the regret for an algorithm satisfying capacity constraints, where no user after the stopping time is given any resources. In contrast, due to the logarithmic objective that may be negative and unbounded studied in this work, all users must receive some resources; thus, the feasible variant of Algorithm~\ref{alg:PrivacyPreserving} incurs an additional $\log(n)$ factor in the regret as all users after the stopping time are given $\frac{\epsilon}{n}$ units of the remaining resources.

Next, to design a feasible variant of Algorithm~\ref{alg:AlgoProbKnownDiscrete}, we first recall that this algorithm already has a stopping time $\tau$ that represents the first period at which $\d_t \notin [\d - \Delta, \d+\Delta]$, where $\0 < \Delta < \d$. For ease of exposition, we define $d - \Delta = \epsilon$ and note that the feasible variant $\ppi^f$ of Algorithm~\ref{alg:AlgoProbKnownDiscrete} can be defined as follows. First, we run Algorithm~\ref{alg:AlgoProbKnownDiscrete} until its stopping time $\tau$ following which all remaining users receive $\frac{\epsilon}{n}$ of the remaining resources. We obtain the following regret upper bound for this feasible variant of Algorithm~\ref{alg:AlgoProbKnownDiscrete}.
\begin{corollary}[Regret Guarantee of Feasible Variant of Algorithm~\ref{alg:AlgoProbKnownDiscrete}] \label{cor:noCapVioAlgo1}
Suppose users' budget and utility parameters are drawn i.i.d. from a discrete distribution $\D$ (with finite support) satisfying Assumption~\ref{asmpn:mainRestriction}. Further, let $\ppi$ denote an online pricing policy described by Algorithm~\ref{alg:AlgoProbKnownDiscrete} with stopping time $\tau^{\ppi}$ representing the first period at which $\d_t \notin [\d - \Delta, \d+\Delta]$. In addition, let $\ppi^f$ denote its feasible variant where all users after $\tau^{\ppi}$ are given $\frac{\epsilon}{n}$ units of each of the remaining resources. Then, the regret $R_n(\ppi^f) \leq O(\log(n))$.
\end{corollary}
For a proof of Corollary~\ref{cor:noCapVioAlgo1}, see Appendix~\ref{apdx:pf-cocapVioAlgo1}. Note that Corollary~\ref{cor:noCapVioAlgo1} implies that designing a feasible variant of Algorithm~\ref{alg:AlgoProbKnownDiscrete} comes at no additional loss, up to constants, in the regret compared to the setting when the capacity constraints can be violated (Theorem~\ref{thm:RegretCapVioAlgo4}). Overall, Corollaries~\ref{cor:noCapVioAlgo2}-\ref{cor:noCapVioAlgo1} imply that our obtained regret guarantees for Algorithms~\ref{alg:AlgoProbKnownDiscrete} and~\ref{alg:PrivacyPreserving} in the setting when some violation of the capacity constraints is acceptable extend to the setting where capacity constraints cannot be violated with little performance loss.





\vspace{-5pt}

\section{Numerical Experiments} \label{sec:experiments}


This section compares the performance of the feasible variants of our revealed preference algorithm under our chosen step sizes (see Sections~\ref{sec:regret-ub} and~\ref{sec:variable-step-size}) to several benchmarks. Our results not only validate the regret bounds in Corollaries~\ref{cor:noCapVioAlgo2} and~\ref{cor:noCapVioAlgo2-variable} but also demonstrate the efficacy of our revealed preference algorithms compared to three benchmarks, some of which have access to additional information on users' utilities and budgets. In the following, we present an overview of the benchmarks (Section~\ref{sec:benchmark-overview}), the experimental setup and implementation details of the benchmarks and our revealed preference algorithms (Section~\ref{sec:experimental-setup-details}), and results comparing the performance of our revealed preference algorithms to these benchmarks (Section~\ref{sec:experimental-results-all-algos}).


\vspace{-5pt}

\subsection{Overview of Benchmarks} \label{sec:benchmark-overview}


Our first benchmark (\emph{Stochastic Program}) assumes knowledge of the distribution $\D$ of users' budget and utility parameters, as with an algorithm that sets static expected equilibrium prices, and involves solving a stochastic program to set prices. The second benchmark (\emph{Dynamic Learning SAA}) assumes that users' utility and budget parameters are revealed to the central planner when they enter the market. In this benchmark, a sampled version of the Eisenberg-Gale program is solved with the observed budget and utility parameters of users that have previously arrived to set prices for subsequent users. Our final benchmark (\emph{Decaying step-size RP}) involves applying our revealed preference algorithm (Algorithm~\ref{alg:PrivacyPreserving}) with a decaying step-size of $\gamma_t = O(\frac{1}{t})$ for each arriving user $t$. For a more detailed discussion of these benchmarks, see Appendix~\ref{apdx:numericalBenchmarks}.

\vspace{-5pt}

\subsection{Experimental Setup and Implementation Details} \label{sec:experimental-setup-details}
We consider the following two market instances for our numerical study.


\emph{Instance 1:} Our first instance corresponds to the counterexample in the proof of Theorem~\ref{thm:lbStatic}. In particular, we consider a setting of $n$ users, where all users have a fixed budget of one, and $m = 2$ goods, each with a capacity of $n$. The utility parameters of users are drawn i.i.d. from a distribution $\D$, where users have an equal $0.5$ probability of having the utility vector $(1, 0)$ or $(0, 1)$.

\emph{Instance 2:} We consider an instance of $m = 5$ goods, each with a capacity of $n$ when the market has $n$ users. Users' budget and utility parameters are generated i.i.d. from the following distribution. Each user's budget can take on one of three values -- $2$, $5$, or $10$ -- with equal $\frac{1}{3}$ probability. Further, user utilities are independent of their budget, and their utility for each good is drawn uniformly at random in the range $[5, 10]$.

We choose a continuous utility distribution in instance 2 to validate Corollary~\ref{cor:noCapVioAlgo2}, as Algorithm~\ref{alg:PrivacyPreserving} applies for general (non-discrete) probability distributions. Since this utility distribution is continuous, the adaptive expected equilibrium pricing algorithm (Algorithm~\ref{alg:AlgoProbKnownDiscrete}) does not apply in this setting. For numerical experiments comparing Algorithm~\ref{alg:AlgoProbKnownDiscrete} to a static expected equilibrium pricing approach, see Appendix~\ref{apdx:staticVdynamic}.

For the two market instances, we let the number of users $n$ range between 100 to 7500. To implement the stochastic programming benchmark, we compute the solution to the associated stochastic program using a sample average approximation with $5000$ samples of budget and utility parameters generated from the above-described distributions $\D$ to evaluate the expectation. For the revealed preference algorithm with a fixed step size (henceforth, termed as Fixed step-size RP), we select $\gamma = \gamma_t = \frac{1}{5 \sqrt{n}}$ for all users $t \in [n]$, and for the revealed preference algorithm with the two-stage adjustment in the step size of the price updates (henceforth, termed as Two-Stage RP), we select a step size of $\gamma_t = \frac{1}{5 n^{2/5}}$ for the first $n^{4/5}$ users and a step size of $\gamma_t = \frac{1}{5 n^{3/5}}$ for the remaining users. Moreover, for the benchmark revealed preference algorithm, i.e., Decaying step-size RP, we select a step size of $\gamma_t = \frac{1}{10 t}$ for all $t \in [n]$. Finally, we implement the feasible variants of these algorithms using the procedure in Section~\ref{sec:feasible-algo-design} (with $\epsilon = 10$) to compare their efficacy.

\vspace{-5pt}

\subsection{Results} \label{sec:experimental-results-all-algos}



This section compares the performance of our revealed preference algorithm under our chosen step sizes to the three benchmarks in Section~\ref{sec:benchmark-overview}. Figure~\ref{fig:regret_comp_feasible_algos} depicts the log-log plots of the regret with the number of users $n$ on instances 1 (left) and 2 (right) for the five algorithms, i.e., the three benchmarks (labeled A1-A3), Fixed step-size RP (A4), and Two-Stage RP (A5). From this figure, we observe that our revealed preference algorithms (A4 and A5) achieve significantly lower regret than the three benchmarks without relying on the additional assumptions on users' budget and utility parameters that the first two benchmarks require. Figure~\ref{fig:regret_comp_feasible_algos} also highlights the performance improvement of our revealed preference algorithms (A4 and A5) compared to static pricing, as even static pricing with knowledge of the distribution $\D$, which corresponds to the stochastic program benchmark (A1), achieves higher regret than our revealed preference algorithms on both instances. We also note that our proposed revealed preference algorithms (A4 and A5) achieve lower regret than the revealed preference benchmark with a decaying step size (A3) for both instances. 


Our results in Figure~\ref{fig:regret_comp_feasible_algos} also validate the regret bounds in Corollaries~\ref{cor:noCapVioAlgo2} and~\ref{cor:noCapVioAlgo2-variable} for the revealed preference algorithms under both the choices of the step sizes of the price updates. In particular, the solid black line representing the empirically observed regret of Fixed step-size RP (A4) on the two instances is very close to the theoretical $O(\sqrt{n} \log(n))$ regret bound, represented by a dotted black line of slope $0.5$ in the log-log plots. Analogously, the solid gray line representing the empirical regret of Two-Stage RP (A5) on the two instances is very close to the theoretical $O(n^{2/5} \log(n))$ regret bound, represented by a dashed gray line of slope $0.4$. We note that Two-Stage RP (A5) achieves a slope of $0.4$ on the log-log plot even on instance two corresponding to a continuous distribution, despite the guarantee in Corollary~\ref{cor:noCapVioAlgo2-variable} only applying for discrete distributions with finite support. Such a result points towards the possibility of achieving a $O(n^{2/5} \log(n))$ regret bound of the revealed preference algorithm for general distributions, which we defer to future research.

\begin{figure}[tbh!]
    \centering  \hspace{-90pt}
    \begin{subfigure}[t]{0.45\columnwidth}
        \centering \vspace{-4pt}
%
%
\definecolor{mycolor3}{rgb}{0.00000,0.44700,0.74100}%
\definecolor{mycolor2}{rgb}{1.0, 0.88, 0.21}
\definecolor{mycolor1}{rgb}{0.89, 0.0, 0.13}
\definecolor{mycolor4}{rgb}{0.20, 0.63, 0.17} 
\definecolor{mycolor5}{rgb}{0.58, 0.44, 0.86} 
\definecolor{mycolor6}{rgb}{0.90, 0.60, 0.00} 
\definecolor{mycolor7}{rgb}{0.67, 0.85, 0.90} 

\begin{tikzpicture}

\begin{axis}[%
width=2in,
height=0.9in,
scale only axis,
xmin=4.5,
xmax=9,
xlabel style={font=\color{white!15!black}, font = \footnotesize, yshift=0.5em},
xlabel={$\log(\text{Number of Users})$},
xticklabel style= {font = \scriptsize},
yticklabel style= {font = \scriptsize},
ymin=3,
ymax=8.2,
ylabel style={font=\color{white!15!black}, align=center, font = \footnotesize},
ylabel={$\log(\text{Regret})$},
axis background/.style={fill=white},
]
\addplot [solid, color=black!70, line width=1.5pt, dashdotted]
  table[row sep=crcr]{%
4.60517019	4.17434207\\
5.52146092	4.50335809\\
6.2146081	4.99843463\\
6.62007321	5.16427591\\
6.90775528	4.81947977\\
7.13089883	5.27827087\\
7.31322039	5.59670926\\
7.60090246	5.65593287\\
7.82404601	5.44212832\\
8.00636757	6.15541014\\
8.51719319  5.97155053\\
8.9226583   6.43399546\\
};

\addplot [dotted, color=black!30, line width=1.5pt, densely dashed]
  table[row sep=crcr]{%
4.60517019	4.16359857\\
5.52146092	4.78526554\\
6.2146081	4.99243054\\
6.62007321	5.31511308\\
6.90775528	5.49029381\\
7.13089883	5.88680379\\
7.31322039	6.31746906\\
7.60090246	6.03613686\\
7.82404601	6.34034009\\
8.00636757	6.36509065\\
8.51719319  6.75599494\\
8.9226583   6.95619458\\
};

\addplot [dash pattern=on 4pt off 2pt on 1pt off 2pt, color=black!55, line width=1.5pt, densely dotted]
  table[row sep=crcr]{%
4.60517019	3.95877953\\
5.52146092	4.80138642\\
6.2146081	5.42945479\\
6.62007321	5.79832918\\
6.90775528	6.05631691\\
7.13089883	6.25210337\\
7.31322039	6.41144149\\
7.60090246	6.67343402\\
7.82404601	6.87189862\\
8.00636757	7.03099634\\
8.51719319  7.4866245\\
8.9226583   7.84511642\\
};

\addplot [solid, color=black!90, line width=1.5pt]
  table[row sep=crcr]{%
4.60517019	3.3570144\\
5.52146092	3.86585761\\
6.2146081	4.22356239\\
6.62007321	4.42726558\\
6.90775528	4.57663677\\
7.13089883	4.69440857\\
7.31322039	4.79120949\\
7.60090246	4.92456188\\
7.82404601	5.04038877\\
8.00636757	5.13750365\\
8.51719319  5.39466191\\
8.9226583   5.59223301\\
};

\addplot [solid, color=black!40, line width=1.5pt]
  table[row sep=crcr]{%
4.60517019	3.30380668\\
5.52146092	3.66361016\\
6.2146081	3.67130816\\
6.62007321	3.83930571\\
6.90775528	3.94536602\\
7.13089883	4.06370595\\
7.31322039	4.15607832\\
7.60090246	4.2011347\\
7.82404601	4.30508793\\
8.00636757	4.3879296\\
8.51719319  4.743432\\
8.9226583   4.74682723\\
};

\addplot [dash dot, color=black!90, line width=1.5pt, domain=4.5:9, dotted] {0.5*x + 1.1} node[pos=0.3, sloped, above, yshift = -4pt, font = \scriptsize] {Slope 0.5};;

\addplot [densely dashed, color=black!40, line width=1.5pt, domain=4.5:9, dotted] {0.4*x + 1.3}
    node[pos=0.5, sloped, below, yshift = 1pt, font = \scriptsize] {Slope 0.4};

\end{axis} 

\begin{axis}[%
width=0in,
height=0in,
at={(0in,0in)},
scale only axis,
xmin=0,
xmax=1,
ymin=0,
ymax=1,
axis line style={draw=none},
ticks=none,
axis x line*=bottom,
axis y line*=left
]
\end{axis}

\end{tikzpicture}%
    \end{subfigure} \hspace{-25pt}
    \begin{subfigure}[t]{0.45\columnwidth}
        \centering
%
%
\definecolor{mycolor3}{rgb}{0.00000,0.44700,0.74100}%
\definecolor{mycolor2}{rgb}{1.0, 0.88, 0.21}
\definecolor{mycolor1}{rgb}{0.89, 0.0, 0.13}
\definecolor{mycolor4}{rgb}{0.20, 0.63, 0.17} 
\definecolor{mycolor5}{rgb}{0.58, 0.44, 0.86} 
\definecolor{mycolor6}{rgb}{0.90, 0.60, 0.00} 
\definecolor{mycolor7}{rgb}{0.67, 0.85, 0.90} 

\begin{tikzpicture}

\begin{axis}[%
width=2in,
height=0.9in,
legend style={
legend cell align=left, align=left,
  fill opacity=0.8,
  draw opacity=1,
  text opacity=1,
  at={(1.02,0.9)},
  anchor=north west,
  draw=white!80!black
},
scale only axis,
xmin=4.5,
xmax=9,
xlabel style={font=\color{white!15!black}, font = \footnotesize, yshift=0.5em},
yticklabel style= {font = \scriptsize},
xticklabel style= {font = \scriptsize},
xlabel={$\log(\text{Number of Users})$},
ymin=6,
ymax=10.5,
axis background/.style={fill=white},
legend style = {font = \scriptsize}
]
\addplot [solid, color=black!70, line width=1.5pt, dashdotted]
  table[row sep=crcr]{%
4.60517019	6.526882\\
5.52146092	7.02008089\\
6.2146081	7.58511781\\
6.62007321	8.09861169\\
6.90775528	7.87786183\\
7.13089883	8.33013976\\
7.31322039	8.45469033\\
7.60090246	8.57390693\\
7.82404601	8.92316792\\
8.00636757	8.62312393\\
8.51719319  9.17692765\\
8.9226583   9.26020055\\
};
\addlegendentry{$A1$: Stochastic Program}

\addplot [dotted, color=black!30, line width=1.5pt, densely dashed]
  table[row sep=crcr]{%
4.60517019	6.76697065\\
5.52146092	7.23562502\\
6.2146081	7.84255336\\
6.62007321	8.03922352\\
6.90775528	8.28098756\\
7.13089883	8.5117879\\
7.31322039	8.52187341\\
7.60090246	8.85680472\\
7.82404601	8.27596007\\
8.00636757	9.34983768\\
8.51719319  9.1378408\\
8.9226583   9.68810978\\
};
\addlegendentry{$A2$: Dynamic Learning SAA}

\addplot [dash pattern=on 4pt off 2pt on 1pt off 2pt, color=black!55, line width=1.5pt, densely dotted]
  table[row sep=crcr]{%
4.60517019	6.42573667\\
5.52146092	7.33785691\\
6.2146081	8.00175493\\
6.62007321	8.11834341\\
6.90775528	8.36422155\\
7.13089883	8.62041519\\
7.31322039	8.55693306\\
7.60090246	8.84800383\\
7.82404601	9.07084242\\
8.00636757	9.38862651\\
8.51719319  9.66181112\\
8.9226583   10.13528635\\
};
\addlegendentry{$A3$: Decaying step-size RP}

\addplot [solid, color=black!90, line width=1.5pt]
  table[row sep=crcr]{%
4.60517019	6.05981788\\
5.52146092	6.45033799\\
6.2146081	7.03944381\\
6.62007321	7.16177218\\
6.90775528	7.27278607\\
7.13089883	7.25460784\\
7.31322039	7.35482247\\
7.60090246	7.50409955\\
7.82404601	7.54799023\\
8.00636757	7.77715049\\
8.51719319  8.02500518\\
8.9226583   8.13715561\\
};
\addlegendentry{$A4$: Fixed step-size RP}

\addplot [solid, color=black!40, line width=1.5pt]
  table[row sep=crcr]{%
4.60517019	6.21083146\\
5.52146092	6.56885442\\
6.2146081	7.10007146\\
6.62007321	7.22543583\\
6.90775528	7.19807465\\
7.13089883	7.30450409\\
7.31322039	7.33992819\\
7.60090246	7.33164707\\
7.82404601	7.44531057\\
8.00636757	7.85815503\\
8.51719319  7.65307587\\
8.9226583   8.01859472\\
};
\addlegendentry{$A5$: Two-Stage RP}

\addplot [dash dot, color=black!40, line width=1.5pt, domain=4.5:9, dotted] {0.4*x + 4.4} node[pos=0.83, sloped, below, yshift = 1pt, font = \scriptsize] {Slope 0.4};

\addplot [densely dashed, color=black!90, line width=1.5pt, domain=4.5:9, dotted] {0.5*x + 3.7} node[pos=0.4, sloped, below, yshift = 2pt, font = \scriptsize] {Slope 0.5};

\end{axis} 

\begin{axis}[%
width=0in,
height=0in,
at={(0in,0in)},
scale only axis,
xmin=0,
xmax=1,
ymin=0,
ymax=1,
axis line style={draw=none},
ticks=none,
axis x line*=bottom,
axis y line*=left
]
\end{axis}

\end{tikzpicture}%
    \end{subfigure}
    \vspace{-35pt}
    \caption{{\small Log-log plots comparing the regret of the feasible variants of the revealed preference algorithms, one with a fixed step size and another with a two-stage adjustment in the step size of the price updates, and the three benchmarks presented in Section~\ref{sec:benchmark-overview} for instances 1 (left) and 2 (right).
    }} 
    \label{fig:regret_comp_feasible_algos}
\end{figure}

\vspace{-15pt}

\section{Conclusion and Future Work} \label{sec:conclusion}

\ifarxiv
In this work, we studied an online variant of Fisher markets wherein users with linear utilities arrive sequentially and have privately known budget and utility parameters drawn i.i.d. from some distribution $\D$. 
In this setting, we first established that no static pricing algorithm 
can achieve a regret and constraint violation of less than $\Omega(\sqrt{n})$ (where $n$ is the number of users). Given the limitations of static pricing, we developed adaptive posted-price algorithms, one with knowledge of the distribution $\D$ and another that adjusts prices solely based on past observations of user consumption, i.e., revealed preference feedback, with improved performance guarantees. We further developed feasible variants of our two adaptive pricing algorithms that respect resource capacities with overall low regret comparable to that in the setting where violations of the resource capacities are permissible. Finally, we presented numerical experiments highlighting the efficacy of our revealed preference algorithm under our chosen step sizes relative to several benchmarks.

\fi

There are several future research directions. First, as we obtained a lower bound on the regret and constraint violation of static pricing (Theorem~\ref{thm:lbStatic}), it would be worthwhile to develop algorithm-independent lower bounds to characterize the performance limits of adaptive pricing algorithms for online Fisher markets. Next, while Algorithm~\ref{alg:AlgoProbKnownDiscrete} achieved an $O(\log(n))$ regret and constant constraint violation for discrete distributions, it would be interesting to study whether adaptive pricing algorithms can achieve a performance better than the $O(\sqrt{n})$ regret and constraint violation of the revealed preference algorithm (Algorithm~\ref{alg:PrivacyPreserving}) for general distributions. Moreover, it would be interesting to investigate the possibility of achieving a regret better than $O(n^{2/5})$ in the revealed preference setting for discrete distributions with finite support. There is also scope to generalize the studied model to settings with more general concave utilities and study settings beyond the stochastic input model of user arrival, e.g., the random permutation model. 

\vspace{-8pt}

\section*{Acknowledgements} \vspace{-5pt}
This research was supported by the Stanford Interdisciplinary Graduate Fellowship (SIGF). We thank Yale Wang and Vladimir Gonzalez Migal for their assistance with the simulation experiments, Chunlin Sun and Xiaocheng Li for proof-reading the paper, and anonymous reviewers for their valuable comments.

\bibliographystyle{unsrtnat}
\bibliography{main}



\clearpage

\appendix

\section{Additional Related Work} \label{apdx:addnal-Related-work}

This section surveys additional works beyond those covered in Section~\ref{sec:literature} that are also related to this work.

\paragraph{Beyond worst-case Analysis for OLP:} Beyond worst-case approaches for OLP problems have focused on designing algorithms under (i) the random permutation and (ii) the stochastic input models. In the random permutation model, the constraints and objective coefficients arrive according to a random permutation of an adversarially chosen input sequence. In this context,~\citep{devanur-adwords,online-agrawal} develop a two-phase algorithm, which includes training the model on a small fraction of the input sequence and then using the learned parameters to make online decisions on the remaining input sequence. Contrastingly, in the stochastic input model, the input sequence is drawn i.i.d. from some potentially unknown distribution. In this setting,~\cite{li2020simple} investigate the convergence of the dual price vector and design algorithms using LP duality to obtain logarithmic regret bounds. Since the algorithms developed in~\cite{li2020simple} involve solving an LP at specified intervals,~\cite{li2020simple,gao2024decoupling} developed gradient descent-based algorithms wherein the dual prices are adjusted solely based on the allocation to users at each time step. Furthermore,~\cite{chen2021linear} devised an adaptive allocation algorithm with constant regret when the samples are drawn from a discrete distribution. As with some of these works, we develop algorithms for online Fisher markets under the stochastic input model; however, in contrast to these works that assume a linear objective, we develop regret guarantees for a non-linear concave objective function.

\paragraph{Revealed Preferences:} Our approach of adjusting prices using users' revealed preferences, i.e., observed user consumption information, is analogous to price update mechanisms that use information from interactions with earlier buyers to inform pricing decisions for future buyers~\citep{kleinberg2003value}. While our dual-based price update mechanism is akin to those used in prior work on revealed preferences~\citep{roth2016watch,ji2018social}, our work considers a setting with budget-constrained users, unlike the quasi-linear utility setting studied in these works. Prior literature on revealed preference has also considered the setting of budget-constrained users~\citep{zadimoghaddam2012efficiently,bei2016learning,balcan2014learning,beigman2006learning} as in this work. However, these works focus on the problem of learning the budgets and valuation functions of users that rationalize their observed buying behavior rather than designing algorithms with low regret, which is one of the main focuses of this work.

\paragraph{Online Constraint Convex Optimization with Long-term Constraints:} Furthermore, since we focus on jointly optimizing regret and constraint violation, our work closely relates to the literature on online constrained convex optimization with long-term constraints~\citep{yi2021regret,liakopoulos2019cautious,jenatton2016adaptive,mahdavi2012trading,valls2020online}. However, compared to these works that focus on a regret measure defined based on the sub-optimality of an optimal static action in hindsight, we adopt a more powerful oracle model, wherein the oracle can vary its actions across time steps as in~\cite{yu2017online,oco-clg-2017,oco-tac-2019}. Even though our chosen regret metric is akin to the dynamic regret notions in these works, our work differs from~\cite{yu2017online,oco-clg-2017,oco-tac-2019} in several ways. First, unlike these works, which consider a setting wherein the central planner observes a convex cost function after each user arrival, we study a revealed preference setting, wherein users' utility and budget parameters are private information. Next, as opposed to the gradient descent projection step used in the algorithms developed in~\cite{yu2017online,oco-clg-2017,oco-tac-2019}, we establish regret and constraint violation bounds for our revealed preference algorithm without projecting the price vector to the non-negative orthant. We do so by developing a novel potential function argument that relies on the structural properties of Fisher markets (see Section~\ref{sec:regret-ub} and Appendix~\ref{sec:pfThm3}). Finally, compared to~\cite{yu2017online,oco-clg-2017,oco-tac-2019}, we also consider the informational setting when the distribution $\D$ is discrete and known to the central planner and develop an adaptive expected equilibrium pricing algorithm in this setting with constant constraint violation and logarithmic regret.

\paragraph{Artificial Currency Mechanisms:} Our work is also closely related to the design and analysis of artificial currency mechanisms~\citep{kash2007optimizing,gorokh2020nonmonetary}. Such mechanisms have found applications in various resource allocation settings, including the allocation of food to food banks~\citep{prendergast2016allocation}, the allocation of students to courses~\citep{Budish}, and the allocation of public goods to people~\citep{Jalota2020MarketsFE}. Mechanisms that involve artificial currencies have also been designed for repeated allocation settings~\citep{gorokh2016near}, as is the main focus of this paper. However, unlike~\cite{gorokh2016near} that studies the repeated allocation of goods that arrive online, we investigate the setting of online user arrival.

\section{Regret and Nash Social Welfare} \label{apdx:regretNSWConnection}

We establish a fundamental connection between the regret measure studied in this work and the ratio between the Nash social welfare objective of the optimum offline oracle and that corresponding to an online algorithm. In particular, we show that if the regret $U_n^* - U_n(\ppi) \leq o(n)$ for some algorithm $\ppi$, then $\frac{NSW(\x_1^*, \ldots, \x_n^*)}{NSW(\x_1, \ldots, \x_n)} \rightarrow 1$ as $n \rightarrow \infty$. Here, $\x_1^*, \ldots, \x_n^*$ are the optimal offline allocations, and $\x_1, \ldots, \x_n$ are the optimal consumption vectors given by the solution of Problem~\eqref{eq:Fisher1}-\eqref{eq:Fishercon3} under the prices corresponding to the online pricing policy $\ppi$. Without loss of generality, consider the setting when the budgets of all users are equal. Note that if the budgets are not equal, then we can just re-scale the utilities of each user based on their budget. In this setting, it holds that
\begin{align*}
    \frac{1}{n} U_n^* = \frac{1}{n} \sum_{t = 1}^n \log(u_t(\x_t^*)) = \frac{1}{n} \log \left( \prod_{t = 1}^n u_t(\x_t^*)) \right) = \log \left( \left( \prod_{t = 1}^n u_t(\x_t^*) \right)^{\frac{1}{n}} \right) = \log(NSW(\x_1^*, \ldots, \x_n^*)),
\end{align*}
and $\frac{1}{n} U_n(\ppi) = \log \left( \left( \prod_{t = 1}^n u_t(\x_t) \right)^{\frac{1}{n}} \right) = \log(NSW(\x_1, \ldots, \x_n))$. Then, it follows that
\begin{align*}
    \frac{NSW(\x_1^*, \ldots, \x_n^*)}{NSW(\x_1, \ldots, \x_n)} = \frac{e^{\frac{1}{n} U_n^*}}{e^{\frac{1}{n} U_n(\ppi)}} = e^{\frac{1}{n} (U_n^* - U_n(\ppi))} \leq e^{\frac{o(n)}{n}}.
\end{align*}
Observe that as $n \rightarrow \infty$, the term $e^{\frac{o(n)}{n}} \rightarrow 1$. That is, if the regret of an algorithm $\ppi$ is $o(n)$, then the ratio of the Nash social welfare of algorithm $\ppi$ approaches that of the optimal offline oracle as $n$ becomes large.

\section{Proof of Theorem~\ref{thm:lbStatic}} \label{apdx:pflb}

Consider a setting with $n$ users with a fixed budget of one and two goods, each with a capacity of $n$. Further, let the utility parameters of users be drawn i.i.d. from a distribution, where the users have utility $(1, 0)$ with probability $0.5$ and a utility of $(0, 1)$ with probability $0.5$. That is, users only have utility for good one or good two, each with equal probability. For this instance, we first derive a tight bound for the expected optimal social welfare objective, i.e., Objective~\eqref{eq:FisherSocOpt}. Then, to establish the desired lower bound, we consider two cases: (i) the price of either of the two goods is at most $0.5$, and (ii) the price of both goods is strictly greater than $0.5$. In the first case, we establish that the expected constraint violation is $\Omega(\sqrt{n})$ while in the second case, we establish that either the expected constraint violation or the expected regret is $\Omega(\sqrt{n})$.

\subsection{Tight Bound on Expected Optimal Social Welfare Objective}

To obtain a bound on the expected optimal social welfare objective, we first find an expression for the objective given the number of arrivals $s$ of users with the utility $(1, 0)$. To this end, for the defined problem instance, given $s$ arrivals of users with the utility $(1, 0)$ (for ease of exposition, let the first $s$ indexed users have a utility of $(1, 0)$), we have the following offline social optimization problem
\begin{maxi!}|s|[2]                   
    {\substack{\x_{t} \in \mathbb{R}^2, \\ \forall t \in [n]}}                               
    {\sum_{t = 1}^s \log \left( x_{t1} \right) + \sum_{t = s+1}^n \log(x_{t2}), \label{eq:FisherSocOptLB11}}   
    {\label{eq:FisherExample111}}             
    {U^*(s) = }                                
    \addConstraint{\sum_{t = 1}^{n} x_{t1}}{ \leq n, \label{eq:FisherSocOpt1LB11}}    
    \addConstraint{\sum_{t = 1}^{n} x_{t2}}{ \leq n, \label{eq:FisherSocOpt2LB11}}
    \addConstraint{x_{tj}}{\geq 0, \quad \forall t \in [n], j \in [2]. \label{eq:FisherSocOpt3LB11}}  
\end{maxi!}
If $0<s<n$, then the optimal solution of the above problem is to allocate $\x_t = (\frac{n}{s}, 0)$ to each user $t$ with a utility of $(1, 0)$ and to allocate $\x_t = (0, \frac{n}{n-s})$ to each user $t$ with a utility of $(0, 1)$. In this case, the optimal objective value is given by
\begin{align*}
    U^*(s) = s \log \left(\frac{n}{s} \right) + (n-s) \log \left( \frac{n}{n-s} \right) = n \log(n) - s \log(s) - (n-s) \log(n-s).
\end{align*}

We now develop a tight bound on the expected optimal objective $U^*(s)$ using the fact that the number of arrivals $s$ of users with utility $(0, 1)$ is binomially distributed with a probability of $0.5$. That is, we seek to develop a tight bound for
\begin{align*}
    \mathbb{E}[U^*(s)] &= \mathbb{E}[n \log(n) - s \log(s) - (n-s) \log(n-s)], \\ &= n \log(n) - \mathbb{E}[s \log(s)] - \mathbb{E}[(n-s) \log(n-s)].
\end{align*}
To this end, we present an upper bound for $s \log(s)$ and $(n-s) \log(n-s)$, which will yield a lower bound for $\mathbb{E}[U^*(s)]$. 

We begin by observing that the expectation of the binomial random variable is given by $\mathbb{E}[s] = \frac{n}{2}$ and its variance is $\mathbb{E}[(s - \frac{n}{2})^2] = \frac{n}{4}$. Next, letting $\sigma = \frac{2}{n} (s - \frac{n}{2})$, which has zero mean and a standard deviation of $\frac{1}{\sqrt{n}}$, we obtain the following upper bound on the term $s \log(s)$
\begin{align}
    s \log(s) &= s \log \left(\frac{n}{2} + s - \frac{n}{2} \right), \nonumber \\
    &= s \log \left(\frac{n}{2} \left(1 + \frac{2}{n} \left(s - \frac{n}{2} \right) \right) \right), \nonumber \\
    &= s \log \left(\frac{n}{2} \right) + s \log \left(1 + \frac{2}{n} \left(s - \frac{n}{2} \right) \right), \nonumber \\
    &= s \log \left(\frac{n}{2} \right) + s \log(1 + \sigma), \nonumber \\
    &\leq s \log \left(\frac{n}{2} \right) + s \sigma. \label{eq:slns}
\end{align}

Similarly, we obtain the following upper bound for $(n-s) \log(n-s)$:
\begin{align}
    (n-s) \log(n-s) = (n-s) \log \left(\frac{n}{2} \right) + (n-s) \log(1-\sigma) \leq (n-s) \log \left(\frac{n}{2} \right) - (n-s) \sigma \label{eq:nslnns}
\end{align}
Adding Equations~\eqref{eq:slns} and~\eqref{eq:nslnns}, we have that
\begin{align*}
    s \log(s) + (n-s) \log(n-s) \leq n \log \left(\frac{n}{2} \right) + (2s-n) \sigma = n \log \left(\frac{n}{2} \right) + n \sigma^2.
\end{align*}
As a result, it holds that
\begin{align} \label{eq:u_star_s_lb}
    U^*(s) = n \log(n) - s \log(s) - (n-s) \log(n-s) \geq n \log(n) - n \log \left(\frac{n}{2} \right) - n \sigma^2 = n \log(2) - n \sigma^2
\end{align}
for all $0<s<n$. Next, letting $q_s$ be the probability of observing $s$ users with utility $(1, 0)$, it follows that
\begin{align*}
    \mathbb{E}[U^*(s)] &= \sum_{s = 0}^n q_s U^*(s) \stackrel{(a)}{=} \sum_{s = 1}^{n-1} q_s U^*(s) \stackrel{(b)}{\geq} \sum_{s = 1}^{n-1} q_s ( n \log(2) - n \sigma^2 ), \\ 
    &\stackrel{(c)}{\geq} \left( 1 - \frac{1}{2^{n-1}} \right) n \log(2) - n \mathbb{E}[\sigma^2], \\
    &= \left( 1 - \frac{1}{2^{n-1}} \right) n \log(2) - n \mathbb{E} \left[ \left(\frac{2}{n} \left(s - \frac{n}{2} \right) \right)^2 \right], \\
    &= \left( 1 - \frac{1}{2^{n-1}} \right) n \log(2) - 1
\end{align*}
where (a) follows as $U^*(0) = 0$ and $U^*(n) = 0$, (b) follows by Equation~\eqref{eq:u_star_s_lb}, (c) follows as $\sum_{s = 1}^{n-1} q_s = 1 - \frac{1}{2^{n-1}}$ and $\sum_{s = 1}^{n-1} q_s \sigma^2 \leq \sum_{s = 0}^{n} q_s \sigma^2 = \mathbb{E}[\sigma^2]$.

Finally, using Jensen's inequality for a concave function, we obtain the following upper bound on the expected optimal social welfare objective:
\begin{align*}
    \mathbb{E}[U^*(s)] \leq U^*(\mathbb{E}(s)) \leq n \log(2).
\end{align*}
As a result, we have shown the following tight bound on the expected optimal social welfare objective for the earlier defined instance:
\begin{align*}
    \left( 1 - \frac{1}{2^{n-1}} \right) n \log(2) - 1 \leq \mathbb{E}[U^*(s)] \leq n \log(2).
\end{align*}

\subsection{Lower bound on Expected Regret and Constraint Violation}

\paragraph{Case (i):}
We first consider the case when the price of either of the two goods is at most $0.5$. Without loss of generality, let $p_1 \leq 0.5$. Then, with $s$ arrivals of users with utility $(1, 0)$, the expected constraint violation of good one is given by
\begin{align*}
    v_1 = \mathbb{E} \left[\left(\frac{s}{p_1}  - n \right)_+ \right] \geq  \mathbb{E} \left[ (2s - n)_+ \right],
\end{align*}
which is $O(\sqrt{n})$ by the central limit theorem as $\frac{n}{2}$ users of each type arrive in expectation. As a result, the norm of the constraint violation $\Omega(\sqrt{n})$. This establishes that if the price of either of the goods is below $0.5$, the expected constraint violation is $\Omega(\sqrt{n})$.

\paragraph{Case (ii):}
Next, we consider the case when the price of both goods is strictly greater than $0.5$. In particular, suppose that $\p = (p_1, p_2) = (\frac{1}{2-\epsilon_1(n)}, \frac{1}{2-\epsilon_2(n)})$, where $\epsilon_1(n), \epsilon_2(n) > 0$ can depend on the number of users $n$ and are constants for any fixed value of $n$. We now show that for any choice of $\epsilon_1(n), \epsilon_2(n)>0$ that either the expected regret or the expected constraint violation is $\Omega(\sqrt{n})$.

To this end, first note by the central limit theorem that the expected constraint violation for good one for $s$ arrivals of users with utility $(1, 0)$ is given by
\begin{align}
    v_1 = \mathbb{E} \left[\left(\frac{s}{p_1}  - n \right)_+ \right] = \mathbb{E} \left[\left(s (2 - \epsilon_1(n))  - n \right)_+ \right] \geq  \mathbb{E} \left[\left(2s  - n \right)_+ \right]  - \epsilon_1(n) \mathbb{E}[s] = \Omega(\sqrt{n}) - \epsilon_1(n) \frac{n}{2}.
\end{align}
Similarly, the expected constraint violation of good two is lower bounded by $\Omega(\sqrt{n}) - \epsilon_2(n) \frac{n}{2}$.

Next, using the lower bound on the expected optimal social welfare objective we obtain the following lower bound on the regret of any static pricing policy with $\p = (\frac{1}{2-\epsilon_1(n)}, \frac{1}{2-\epsilon_2(n)})$:
\begin{align*}
    \text{Regret} &\geq \left( 1 - \frac{1}{2^{n-1}} \right) n \log(2) - 1 - \mathbb{E} \left[ \sum_{t = 1}^n \log\left(\frac{1}{\frac{1}{2-\epsilon(n)}} \right) \right], \\
    &= \left( 1 - \frac{1}{2^{n-1}} \right) n \log(2) - 1 - \mathbb{E} \left[ \sum_{t = 1}^n \log(2-\epsilon(n)) \right], \\
    &= \left( 1 - \frac{1}{2^{n-1}} \right) n \log(2) - 1 - \mathbb{E} \left[ \sum_{t = 1}^n \log \left(2 \left(1-\frac{\epsilon(n)}{2} \right) \right) \right], \\
    &= - \frac{1}{2^{n-1}} n \log(2) - 1 - \mathbb{E} \left[ \sum_{t = 1}^n \log \left(1-\frac{\epsilon(n)}{2} \right) \right], \\
    &\geq - \frac{1}{2^{n-1}} n \log(2) -1 + \frac{n \epsilon(n)}{2},
\end{align*}
where $\epsilon(n) = \min \{\epsilon_1(n), \epsilon_2(n) \}$ and $0< \epsilon(n) < 2$.

Finally, to simultaneously achieve the lowest regret and constraint violation, we set $\Omega(\sqrt{n}) - \epsilon(n) \frac{n}{2} = -1 + \frac{n \epsilon(n)}{2} - \frac{1}{2^{n-1}} n \log(2)$. Solving for $\epsilon(n)$, we get that $\epsilon(n) = O(\frac{1}{\sqrt{n}})$ as $n$ becomes large. This relation implies that to minimize both regret and constraint violation, $\epsilon(n)$ needs to be set on the order of $\frac{1}{\sqrt{n}}$, which will result in a corresponding expected regret and constraint violation of $\Omega(\sqrt{n})$. Observe that for any other choice of $\epsilon(n)$, either the regret or the constraint violation must be $\Omega(\sqrt{n})$ since setting $\epsilon(n) = O(\frac{1}{\sqrt{n}})$ guarantees that both the regret and constraint violation are minimized. This establishes our claim that either the regret or the constraint violation must be $\Omega(\sqrt{n})$ when the price of both goods is strictly greater than $0.5$, which proves our claim.

\section{Proof of Theorem~\ref{thm:RegretCapVioAlgo4}} \label{apdx:pfAlgo4}

We prove Theorem~\ref{thm:RegretCapVioAlgo4} using four intermediate lemmas, which we elucidate below. After presenting the statements of these lemmas, we then present their proofs.

Our first lemma establishes a generic upper bound on the regret of an algorithm for the online Fisher market setting considered in this work. To define this generic regret bound, we first introduce the following stochastic program
\begin{equation} \label{eq:stochastic22}
\begin{aligned}
\min_{\mathbf{p}} \quad & D(\p) = \sum_{j = 1}^m p_j d_j + \mathbb{E}\left[\left(w \log \left(w\right)-w \log \left(\min _{j \in [m]} \frac{p_{j}}{u_{j}}\right)-w\right)\right],
\end{aligned}
\end{equation}
which is the stochastic programming formulation of the dual of the Eisenberg-Gale program (see Equation~\eqref{eq:SAA2}) presented in Section~\ref{sec:dualProblem}. Letting $\p^*$ be the optimal solution to this stochastic program, we obtain the following generic bound on the regret of any algorithm for online Fisher markets.


\begin{lemma} [Generic Regret Bound] \label{lem:genericRegret2}
Suppose that the budget and utility parameters of users are drawn i.i.d. from a probability distribution $\D$. Furthermore, let $\ppi$ denote an online pricing policy, $\x_1, \ldots, \x_n$ be the corresponding allocations for the $n$ users, and $\underline{p}, \bar{p}>0$ be the lower and upper bounds, respectively, for the prices $p_j^t$ for all goods $j$ and for all users $t \in [n]$, where the price upper bound $\Bar{p} \geq \max_{j \in [m]} p_j^*$. Then, the regret $R_n(\ppi) \leq \frac{2 \sqrt{m} \Bar{w}}{\underline{p}} \sum_{t = 1}^n \mathbb{E} \left[ \norm{\p^* - \p^t}_2 \right] + \mathbb{E} \left[ \Bar{p} \left| \sum_{j = 1}^m \left(\sum_{t=1}^n x_{tj} - c_j  \right) \right| \right]$.
\end{lemma}
A few comments about Lemma~\ref{lem:genericRegret2} are in order. First, observe that the generic regret bound obtained in Lemma~\ref{lem:genericRegret2} applies to general (non-discrete) probability distributions $\D$. Next, the generic regret bound is composed of two terms: (i) the first term accounts for the loss from setting prices that deviate from the optimal expected prices $\p^*$, and (ii) the second term is akin to the constraint violation of the algorithm and, in particular, accounts for the loss corresponding to over (or under-consuming) certain goods. 

As a result, to upper bound the regret of Algorithm~\ref{alg:AlgoProbKnownDiscrete}, we now present lemmas that upper bound both the terms in the generic regret upper bound. To this, end, we first show that the upper bound on the expected constraint violation is constant in the number of arriving users. This result not only establishes the desired constraint violation bound in the statement of Theorem~\ref{thm:RegretCapVioAlgo4} but its analysis also provides a bound on the second term of the generic regret upper bound in Lemma~\ref{lem:genericRegret2}.

\begin{lemma} [Constraint Violation Bound of Algorithm~\ref{alg:AlgoProbKnownDiscrete}] \label{lem:capVioAlgo4}
Suppose that the budget and utility parameters of users are drawn i.i.d. from a discrete probability distribution $\D$ and let $\ppi$ denote the online pricing policy described by Algorithm~\ref{alg:AlgoProbKnownDiscrete}. Furthermore, let $\x_1, \ldots, \x_n$ be the corresponding allocations for the $n$ users, where $\x_t$ is an optimal solution for that user corresponding to the certainty equivalent problem $CE(\d_t)$ for $t \leq \tau$, where $\tau$ is the first time at which $\d_t \notin [\d-\Delta, \d+\Delta]$, and $\x_t$ is an optimal solution to $CE(\d)$ for $t > \tau$. Then, the constraint violation $V_n(\ppi) \leq O(1)$.
\end{lemma}
The proof of Lemma~\ref{lem:capVioAlgo4} follows from an application of similar techniques to that used in~\cite{chen2021linear}. In this proof, we leverage the fact that the allocations $\x_t$ are given by the optimal solution of the certainty equivalent problem $CE(\d_t)$ for $t \leq \tau$, which is one of the optimal consumption vectors corresponding to the price $\p^t$. Note that doing so is without loss of generality, since the utility of the users is unchanged for any optimal consumption bundle. Furthermore, recall from Section~\ref{subsubsec:adaptiveExEq} that the allocations corresponding to the optimal solution of the certainty equivalent problem $CE(\d_t)$ at each step can be implemented in Algorithm~\ref{alg:AlgoProbKnownDiscrete} using an allocation-based algorithm, wherein users are given allocations based on their observed type $k \in [K]$.


Having obtained a bound on the constraint violation, we next upper bound the first term in the generic regret upper bound. To do so, we proceed in two steps. First, we establish a Lipschitzness relation between the optimal price vector of the certainty equivalent problem $CE(\d_t)$ and the average remaining resource capacity vector $\d_t$, as is elucidated through the following lemma.

\begin{lemma} [Lipschitz Relation Between Prices and Average Remaining Resource Capacities] \label{lem:lipshitzness}
Suppose $\d, \d' > \0$ are two average remaining resource capacity vectors and $\p^*(\d), \p^*(\d')$ are the optimal price vectors corresponding to the certainty equivalent problems $CE(\d), CE(\d')$, respectively. Then, $\norm{\p^*(\d) - \p^*(\d')}_2 \leq L \norm{\d - \d'}_2$ for some constant $L>0$. 
\end{lemma}
Lemma~\ref{lem:lipshitzness} establishes that small changes in the average remaining capacity vector will only result in small changes in the corresponding optimal price vector of the certainty equivalent Problem $CE(\d_t)$. In particular, Lemma~\ref{lem:lipshitzness} implies that if $\norm{\d - \d'}_2 \leq O(\frac{1}{n-t})$ for a given $t \in [n-1]$, then, the optimal price vectors $\p, \p'$ of the certainty equivalent problems $CE(\d)$ and $CE(\d')$, respectively, satisfy $\norm{\p - \p'}_2 \leq O(\frac{1}{n-t})$. We also numerically validate this obtained Lipschitz relation in Appendix~\ref{apdx:priceStabilization}.

We then leverage Lemma~\ref{lem:lipshitzness} to establish an $O(\log(n))$ upper bound on the first term of the generic regret bound, as is elucidated through the following lemma.

\begin{lemma} [Bound on Difference in Prices] \label{lem:firstTermUB}
Suppose that the budget and utility parameters of users are drawn i.i.d. from a discrete probability distribution $\D$. Furthermore, let $\ppi$ denote the online pricing policy described by Algorithm~\ref{alg:AlgoProbKnownDiscrete} and let $\x_1, \ldots, \x_n$ be the corresponding allocations for the $n$ users. Then, $\frac{2 \sqrt{m} \Bar{w}}{\underline{p}} \sum_{t = 1}^n \mathbb{E} \left[ \norm{\p^* - \p^t}_2 \right] \leq O(\log(n))$.
\end{lemma}

Finally, we combine the results obtained in Lemmas~\ref{lem:capVioAlgo4} and~\ref{lem:firstTermUB} to obtain the $O(\log(n))$ upper bound on the regret of Algorithm~\ref{alg:AlgoProbKnownDiscrete}.
\begin{corollary} [Regret Upper Bound of Algorithm~\ref{alg:AlgoProbKnownDiscrete}] \label{cor:regretAlgo4}
Suppose that the budget and utility parameters of users are drawn i.i.d. from a discrete probability distribution $\D$ and let $\ppi$ denote the online pricing policy described by Algorithm~\ref{alg:AlgoProbKnownDiscrete}. Furthermore, let $\x_1, \ldots, \x_n$ be the corresponding allocations for the $n$ users, where $\x_t$ is an optimal solution for that user corresponding to the certainty equivalent problem $CE(\d_t)$ for $t \leq \tau$, where $\tau$ is the first time at which $\d_t \notin [\d-\Delta, \d+\Delta]$, and $\x_t$ is an optimal solution to $CE(\d)$ for $t > \tau$. Then, the regret $R_n(\ppi) \leq O(\log(n))$.
\end{corollary}
Note that Lemma~\ref{lem:capVioAlgo4} and Corollary~\ref{cor:regretAlgo4} jointly imply Theorem~\ref{thm:RegretCapVioAlgo4}, which thus proves our claim.

\subsection{Proof of Lemma~\ref{lem:genericRegret2}}

We now establish a generic bound on the regret of any online algorithm as long as the prices $\p^t$ are strictly positive and bounded, i.e., $0< \underline{p} \leq p^t_j \leq \Bar{p}$ for all goods $j$ and for all users $t \in [n]$. To establish a generic upper bound on the regret, we first obtain a bound on the expected value of the optimal objective, i.e., Objective~\eqref{eq:FisherSocOpt}, and a relation for the expected value of the objective for any online allocation policy $\ppi$. We finally combine both these relations to obtain an upper bound on the regret.

To perform our analysis, we define the function $g(\p) = \mathbb{E} [ w_t \log(\u_t^T \x_t) + \sum_{j = 1}^m (d_j - x_{tj}(\p)) p_j^* ]$, where $\p^*$ is the optimal price vector of the stochastic Program~\eqref{eq:stochastic22}. Then, by duality we have that the expected primal objective value $\mathbb{E}[U_n^*]$ is no more than the dual objective value with $\p = \p^*$, which gives the following upper bound on the optimal objective
\begin{align}
    \mathbb{E}[U_n^*] &\leq \mathbb{E} \left[ \sum_{j = 1}^m p_j^{*} c_j + \sum_{t = 1}^n \left( w_t \log(w_t) - w_t \log \bigg(\min_{j \in [m]} \frac{p_j^{*}}{u_{tj}} \bigg)  -  w_t \right) \right], \nonumber \\
    &= nD(\p^*), \label{eq:ThmProbKnown11}
\end{align}
by the definition of $D(\p)$ in Problem~\eqref{eq:stochastic22}. Next, we establish a relation between the function $g(\p)$ and the above obtained bound on the expected value of the optimal objective value by noting that
\begin{align}
    g(\p^*) &= \mathbb{E} [ w_t \log(\u_t^T \x_t^*) + \sum_{j = 1}^m (d_j - x_{tj}^*(\p^*)) p_j^* ], \nonumber \\ 
    &\stackrel{(a)}{=} \mathbb{E} \left[ w_t \log(w_t) - w_t \log \left(\min_{j \in [m]} \frac{p_j^{*}}{u_{tj}} \right) + \sum_{j \in [m]} p_j^{*} d_j -  w_t \right], \nonumber \\ 
    &= nD(\p^*), \label{eq:ThmProbKnown12}
\end{align}
where (a) follows by the definition of $g(\p)$ and noting that for each agent $t \in [n]$ it holds that $\u_t^T \x_t = u_{tj'} \frac{w_t}{p_{j'}^*}$ for some good $j'$ in the optimal bundle for the user $t$, and that $\sum_{j \in [m]} x_{tj}(\p^*) p_j^* = w_t$ since each user spends their entire budget when consuming its optimal bundle of goods given the price vector $\p^*$. Combining the relations obtained in Equations~\eqref{eq:ThmProbKnown11} and~\eqref{eq:ThmProbKnown12}, we obtain the following upper bound on the expected value of the optimal objective:
\begin{align}
    \mathbb{E}[U_n^*] \leq n g(\p^*). \label{eq:UBHelper1}
\end{align}

Having obtained an upper bound on the expected optimal objective, we now obtain the following relationship for the true accumulated social welfare objective, i.e., Objective~\eqref{eq:FisherSocOpt}, accrued by any online policy $\ppi$ that sets prices $\p^1, \ldots, \p^n$ with corresponding allocations $\x_1, \ldots, \x_n$:
\begin{align}
     \mathbb{E} \left[U_n(\ppi)\right] &= \mathbb{E} \left[ \sum_{t = 1}^{n} w_t \log(\u_t^T \x_t) \right], \nonumber \\
     &= \mathbb{E} \left[ \sum_{t = 1}^{n} w_t \log(\u_t^T \x_t) + \sum_{j = 1}^m p_j^* \left(c_j - \sum_{t =1}^n x_{tj} \right) - \sum_{j =1}^n p_j^* \left( c_j - \sum_{t =1}^n x_{tj} \right) \right], \nonumber \\
     &= \mathbb{E} \left[ \sum_{t = 1}^{n} \left( w_t \log(\u_t^T \x_t) + \sum_{j=1}^m p_j^* \left( d_j - x_{tj} \right) \right) \right] + \mathbb{E} \left[ \sum_{j =1}^m p_j^* \left(\sum_{t=1}^m x_{tj} - c_j \right) \right] \label{eq:helperBothTerms}
\end{align}
We can analyse the first term on the right hand side of Equation~\eqref{eq:helperBothTerms} as follows:
\begin{align}
    \mathbb{E} \left[ \sum_{t = 1}^{n} \left( w_t \log(\u_t^T \x_t) + \sum_{j=1}^m p_j^* \left( d_j - x_{tj} \right) \right) \right] &\stackrel{(a)}{=} \sum_{t = 1}^{n} \mathbb{E} \left[  w_t \log(\u_t^T \x_t) + \sum_{j=1}^m p_j^* \left( d_j - x_{tj} \right) \right], \nonumber \\
    &\stackrel{(b)}{=} \sum_{t = 1}^{n} \mathbb{E} \left[ \mathbb{E} \left[ w_t \log(\u_t^T \x_t) + \sum_{j=1}^m p_j^* \left( d_j - x_{tj} \right) | \H_{t-1} \right]  \right], \nonumber \\
    &\stackrel{(c)}{=} \sum_{t = 1}^{n} \mathbb{E} \left[ g(\p^t) \right] = \mathbb{E} \left[ \sum_{t = 1}^{n} g(\p^t) \right], \label{eq:helperFirstTerm}
\end{align}
where (a) follows by the linearity of expectation, (b) follows from nesting conditional expectations, where the history $\H_{t-1} = \{w_i, \u_i, \x_i \}_{i = 1}^{t-1}$, and (c) follows from the definition of $g(\p)$ and the fact that the allocation $x_{tj}$ depends on the vector of prices $\p_t$.

Finally, combining the above analysis in Equations~\eqref{eq:UBHelper1},~\eqref{eq:helperBothTerms}, and~\eqref{eq:helperFirstTerm} for $\mathbb{E}[U_n^*]$ and $\mathbb{E}[U_n(\ppi)]$, we obtain the following bound on the regret of any online allocation policy $\ppi$ for $\Bar{p} \geq \max_{j \in [m]} p_j^*$:
\begin{align}
    \mathbb{E}[U_n^* - U_n(\ppi)] &\leq n g(\p^*) -  \mathbb{E} \left[ \sum_{t = 1}^{n} g(\p^t) \right] - \mathbb{E} \left[ \sum_{j =1}^m p_j^* \left(\sum_{t=1}^m x_{tj} - c_j \right) \right], \nonumber \\
    &\leq \mathbb{E} \left[ \sum_{t = 1}^{n} (g(\p^*) - g(\p^t)) \right] + \mathbb{E} \left[ \Bar{p} \left| \sum_{j = 1}^m \left(\sum_{t=1}^n x_{tj} - c_j  \right) \right| \right]. \label{eq:genericRegretBound}
\end{align}

Finally, to obtain the desired generic regret bound, we establish that $\mathbb{E} \left[ g(\p^*) - g(\p^t) \right] \leq O(\mathbb{E} \left[ \norm{\p^* - \p^t}_2 \right])$. To this end, first observe from the definition of the function $g$ that for the optimal solution $\x_t(\p)$ of the individual optimization Problem~\eqref{eq:Fisher1}-\eqref{eq:Fishercon3} given a price vector $\p$ that
\begin{align*}
    g(\p^*) - g(\p^t) &= \mathbb{E} \left[ w_t \log(\u_t^T \x_t(\p^*)) + \sum_{j=1}^m (d_j - x_{tj}(\p^*)) p_j^* \right] \\ 
    &- \mathbb{E} \left[ w_t \log(\u_t^T \x_t(\p^t)) + \sum_{j=1}^m (d_j - x_{tj}(\p_t)) p_j^* \right], \\
    &= \mathbb{E}\left[w_{t} \log \left(\min _{j \in [m]}\left\{\frac{p_{j}^t}{u_{t j}}\right\} \frac{1}{\min_{j \in [m]}\left\{\frac{p_{j}^{*}}{u_{t j}}\right\}}\right) \right] + \mathbb{E} \left[ \sum_{j = 1}^m (x_{tj}(\p^t) - x_{tj}(\p^*)) p_j^* \right].
\end{align*}
Then, letting the good $j' \in \argmin_{j \in [m]} \{ \frac{p_{j}^{*}}{u_{t j}} \}$ and $j^*(\p)$ be a good in the optimal consumption set of user $t$ given the price $\p$, we observe that
\begin{align}
    g(\p^*) - g(\p_t) & \stackrel{(a)}{\leq} \mathbb{E} \left[ w_t \log \left( \frac{p_{j'}^t}{p_{j'}^*} \right) \right] + \mathbb{E} \left[ \sum_{j = 1}^m \left( \mathbbm{1}_{j = j^*(\p^t)} \frac{w_t}{p_{j}^t} - \mathbbm{1}_{j = j^*(\p^*)} \frac{w_t}{p_{j}^*} \right) p_j^* \right], \nonumber \\
    &\stackrel{(b)}{=} \mathbb{E} \left[ w_t \log \left( 1+  \frac{p_{j'}^t - p_{j'}^*}{p_{j'}^*} \right) \right] + \mathbb{E} \left[ \sum_{j = 1}^m \frac{w_t (p_{j}^* - p_{j}^t)}{p_j^* p_{j}^t} \left( \mathbbm{1}_{j = j^*(\p^t)}  - \mathbbm{1}_{j = j^*(\p^*)} \right) p_j^* \right], \nonumber \\
    &\stackrel{(c)}{\leq} \mathbb{E} \left[ w_t \frac{p_{j'}^t - p_{j'}^*}{p_{j'}^*} \right] + \mathbb{E} \left[ \sum_{j = 1}^m \frac{w_t(p_{j}^* - p_{j}^t)}{p_{j}^t} \right], \nonumber \\
    &\stackrel{(d)}{\leq} \frac{2 \Bar{w}}{\underline{p}} \mathbb{E} \left[ \norm{\p^* - \p^t}_1 \right], \nonumber \\
    &\stackrel{(e)}{\leq} \frac{2 \sqrt{m} \Bar{w}}{\underline{p}} \mathbb{E} \left[ \norm{\p^* - \p^t}_2 \right], \label{eq:helperGFunction}
\end{align}
where (a) follows since $j' \in \argmin_{j \in [m]} \{ \frac{p_{j}^{*}}{u_{t j}} \}$ and $\x_{t}(\p)$ corresponds to the optimal solution to the individual optimization Problem~\eqref{eq:Fisher1}-\eqref{eq:Fishercon3}, (b) follows by rearranging the right hand side of the equation in (a). Next, (c) follows from the fact that $\log(1+x) \leq x$ for $x > -1$ and that the difference between two indicators can be at most one. Inequality (d) follows by the upper bound on the budgets of users and the lower bound on the price vector. The final inequality (e) follows from the norm equivalence property which holds for the one and two norms. 

Finally, using Equations~\eqref{eq:helperGFunction} and~\eqref{eq:genericRegretBound}, we obtain the following generic upper bound on the regret of any online algorithm $\ppi$:
\begin{align}
    \mathbb{E}[U_n^* - U_n(\ppi)] &\leq \frac{2 \sqrt{m} \Bar{w}}{\underline{p}} \sum_{t = 1}^n \mathbb{E} \left[ \norm{\p^* - \p^t}_2 \right] + \mathbb{E} \left[ \Bar{p} \left| \sum_{j = 1}^m \left(\sum_{t=1}^n x_{tj} - c_j  \right) \right| \right], \label{eq:genericRegretBoundFINAL}
\end{align}
which proves our claim.

\subsection{Proof of Lemma~\ref{lem:capVioAlgo4}}

To prove this result, we first prove an upper bound on the expected constraint violation in terms of the stopping time $\tau$ of the algorithm. Then, we establish a lower bound on the expected value of the stopping time to establish the constant constraint violation bound.

\paragraph{Upper Bound on constraint violation in terms of stopping time:}
We begin by establishing that the constraint violation of Algorithm~\ref{alg:AlgoProbKnownDiscrete} is upper bounded by $O(\mathbb{E}[n-\tau])$, where the stopping time $\tau = \min \{t \leq n: \d_t \notin [\d - \Delta, \d + \Delta] \} \cup \{n\}$. To this end, first note by the definition of $\tau$ and that $\Delta < \d$ that no constraints are violated up until user $\tau$. Furthermore, since the consumption $x_{tj} \leq \frac{\Bar{w}}{\underline{p}}$ for all $t > \tau$, it follows that the constraint violation
\begin{align} \label{eq:capVioFirstRelation}
    \mathbb{E} \left[ \norm{\sum_{t = 1}^n x_{tj} - c_j}_2 \right] \leq \mathbb{E} \left[ \norm{\sum_{t = \tau+1}^n x_{tj}}_2 \right] \leq \mathbb{E}\left[(n-\tau) \sqrt{m} \frac{\Bar{w}}{\underline{p}}\right] = O(\mathbb{E}\left[n-\tau\right]).
\end{align}

\paragraph{Bound on Expected Stopping time $\tau$:}

From the above analysis, we observe that bounding the expected constraint violation amounts to obtaining a bound on the expected stopping time $\tau$. To this end, we first introduce some notation. In particular, as in~\cite{chen2021linear}, we define the following auxiliary process:
\begin{align*}
    \tilde{\d}_{t}= \begin{cases}\d_{t}, & t<\tau \\ \d_{\tau}, & t \geq \tau\end{cases}.
\end{align*}
Then, we can obtain a generic bound on the expected stopping time by observing that
\begin{align*}
    \mathbb{E}[\tau] &= \sum_{t = 1}^n t \Pp(\tau = t) = \sum_{t = 1}^n \Pp(\tau \geq t) = \sum_{t = 1}^n (1- \Pp(\tau < t)) \geq \sum_{t = 1}^n (1 - \Pp(\tau \leq t)), \\
    &\stackrel{(a)}{=} \sum_{t = 1}^n \left[1 - \Pp(\d_s \notin [\d - \Delta, \d + \Delta] \text{ for some } s \leq t) \right], \\
    &\stackrel{(b)}{\geq} n - \sum_{t = 1}^n \Pp \left(\Tilde{\d}_{s} \notin [\d - \Delta, \d + \Delta] \text { for some } s \leq t\right),
\end{align*}
where (a) follows by the definition of $\tau$, (b) follows since the auxiliary process $\Tilde{\d}_{s}$ is identical to $\d_s$ for all $s$ less than $\tau$. The above analysis implies that
\begin{align} \label{eq:nminustauUB}
    \mathbb{E}[n-\tau] \leq \sum_{t = 1}^n \Pp \left(\Tilde{\d}_{s} \notin [\d - \Delta, \d + \Delta] \text { for some } s \leq t\right).
\end{align}
Thus, to obtain an upper bound for $\mathbb{E}[n-\tau]$, we now proceed to finding an upper bound for the term $\Pp \left(\Tilde{\d}_{s} \notin [\d - \Delta, \d + \Delta ] \text { for some } s \leq t\right)$ for each user $t \in [n]$.

\paragraph{Upper bound on $\Pp \left(\Tilde{\d}_{s} \notin [\d - \Delta , \d + \Delta] \text { for some } s \leq t\right)$:}

To obtain an upper bound on this term, we leverage Hoeffding's inequality:

\begin{lemma} (Hoeffding's Inequality~\citep{van2002hoeffding}) \label{lem:hoeffding's}
Suppose there is a sequence of random variables. $\{ X_t \}_{t = 1}^n$ adapted to a filtration $\H_{t-1}$, and $\mathbb{E}[X_t|\H_{t-1}] = 0$ for all $t \in [n]$, where $\H_0 = \emptyset$. Suppose further that $L_t$ and $U_t$ are $\H_{t-1}$ measurable random variables such that $L_t \leq X_t \leq U_t$ almost surely for all $t \in [n]$. Then, letting $S_t = \sum_{s = 1}^t X_t$ and $V_t = \sum_{s = 1}^t (U_s - L_s)^2$, the following inequality holds for any constants $b, c>0$: $\mathbb{P}\left(\left|S_{t}\right| \geq b, V_{t} \leq c^{2} \text { for some } t \in\{1, \ldots, T\}\right) \leq 2 e^{-\frac{2 b^{2}}{c^{2}}}$.
\end{lemma}

To leverage Lemma~\ref{lem:hoeffding's}, we begin by introducing some notation. First define $Y_{tj}:=\tilde{d}_{j, t+1}-\tilde{d}_{j, t}$ and $X_{tj}:=Y_{tj}-\mathbb{E}\left[Y_{tj} \mid \mathcal{H}_{t-1}\right]$ for $t \geq 1$, where $\mathcal{H}_{t-1} = ((w_1, \u_1), \ldots, (w_{t-1}, \u_{t-1}))$ is the history of observed budget and utility parameters. 

Next, observe for $t \geq \tau $ that $\tilde{d}_{j, t+1} = \tilde{d}_{j, t}$ and when $1\leq t < \tau$ we have that:
\begin{align*}
    \tilde{d}_{j, t+1} = d_{j, t+1} = \frac{c_{j, t+1}}{n-t} = \frac{c_{jt} - x_{tj}}{n-t} = d_{jt} - \frac{1}{n-t}(x_{tj} - d_{jt}) = \Tilde{d}_{jt} - \frac{1}{n-t}(x_{tj} - \Tilde{d}_{jt})
\end{align*}
Next, noting that $\Tilde{d}_{jt}$ is $\mathcal{H}_{t-1}$ measurable, we have that:
\begin{align*}
    |X_{tj}| &= \left| \frac{1}{n-t}(x_{tj} - \Tilde{d}_{jt}) - \mathbb{E} \left[ \frac{1}{n-t}(x_{tj} - \Tilde{d}_{jt}) | \mathcal{H}_{t-1} \right] \right| \\
    &= \frac{1}{n-t} | x_{tj} - \mathbb{E} \left[ x_{tj} | \mathcal{H}_{t-1} \right] | \leq \frac{\bar{w}}{\underline{p}(n-t)}
\end{align*}
for each $t \leq n-1$ due to the boundedness of the allocations $x_{tj}$. Then, defining $L_t = -\frac{\bar{w}}{\underline{p}(n-t)}$ and $U_t = \frac{\bar{w}}{\underline{p}(n-t)}$, we obtain that
\begin{align*}
    V_t = \sum_{s = 1}^t (U_s - L_s)^2 = \sum_{s = 1}^t \frac{4 \bar{w}^2}{\underline{p}^2(n-s)^2} \leq \frac{4 \bar{w}^2}{\underline{p}^2 (n-t-1)},
\end{align*}
which holds for all $t \leq n-2$.

Then, from a direct application of Hoeffding's inequality (Lemma~\ref{lem:hoeffding's}) for some constant $\Delta'>0$ we have that
\begin{align} \label{eq:hoeffdingHelper}
    \mathbb{P}\left(\left|\sum_{i=1}^{s} X_{ij}\right| \geq \Delta^{\prime} \text { for some } s \leq t\right) \leq 2 e^{-\frac{ \underline{p}^2 \Delta^{\prime 2}(n-t-1)}{2 \bar{w}^{2}}}.
\end{align}
Next, we observe that
\begin{align}
    \left|X_{tj}-Y_{tj}\right| &=\left|\mathbb{E}\left[Y_{tj} \mid \mathcal{H}_{t-1}\right]\right| =\left|\mathbb{E}\left[\tilde{d}_{j, t+1}-\tilde{d}_{j, t} \mid \mathcal{H}_{t-1}\right]\right| \stackrel{(a)}{=}\left|\frac{1}{n-t} \mathbb{E}\left[\left(x_{tj}-\tilde{d}_{j, t}\right) I(t<\tau) \mid \mathcal{H}_{t-1}\right]\right| = 0, \label{eq:zero-x=y}
\end{align}
where (a) follows since the probability distribution is exactly known in Algorithm~\ref{alg:AlgoProbKnownDiscrete} and at the optimal solution of the certainty equivalent problem it holds that $\mathbb{E}[\x_t] = \d_t$ for all $t \leq \tau$. Consequently, it holds that the term $\mathbb{E}\left[\left(x_{tj}-\tilde{d}_{j, t}\right) I(t<\tau) \mid \mathcal{H}_{t-1}\right] = 0$ for all users $t<\tau$.

Then, to obtain a bound on $\mathbb{P}(\Tilde{\d}_s \notin [\d - \Delta, \d + \Delta ] \text{ for some } s \leq t)$, we first note the following key relation for the set $\{ \Tilde{\d}_s \notin [\d - \Delta , \d + \Delta ] \text{ for some } s \leq t \}$:
\begin{align*}
    \left\{\left|\tilde{d}_{j, s}-d_{j}\right|>\Delta_j \text { for some } s \leq t\right\}&\stackrel{(a)}{=}\left\{\left|\sum_{i=1}^{s-1} Y_{ij}\right|>\Delta_j \text { for some } s \leq t\right\}, \\
    &= \left\{\left|\sum_{i=1}^{s} Y_{ij}\right|>\Delta_j \text { for some } s \leq t-1\right\}, \\
    &\stackrel{(b)}{=} \left\{\left|\sum_{i=1}^{s} X_{ij}\right|>\Delta_j \text { for some } s \leq t-1\right\},
\end{align*}
where (a) follows from the definition of $Y_i$, and (b) follows since $\sum_{i=1}^{s} X_{ij} = \sum_{i=1}^{s} Y_{ij}$, as proved in Equation~\eqref{eq:zero-x=y}. Then setting $\Delta' = \min_{j \in [m]}\Delta_j = \underline{\Delta}$ in Equation~\eqref{eq:hoeffdingHelper}, and applying a union bound over all the goods $j \in [m]$, we obtain that
\begin{align} \label{eq:ProbUpperBdHelper}
    \mathbb{P}(\Tilde{\d}_s \notin [\d - \Delta, \d + \Delta ] \text{ for some } s \leq t) \leq 2m e^{-\frac{ \underline{p}^2 \underline{\Delta}^{ 2}(n-t-1)}{2 \bar{w}^{2}}},
\end{align}
which holds for all $t \leq n-2$.

\paragraph{Constant Bound on Expected Constraint Violation:}

We have already observed from our earlier analysis that the expected constraint violation is upper bounded by $O(\mathbb{E}[n-\tau])$, where
\begin{align*}
    \mathbb{E}[n-\tau] \leq \sum_{t = 1}^n \Pp \left(\Tilde{\d}_{s} \notin [\d - \Delta , \d + \Delta ] \text { for some } s \leq t\right)
\end{align*}
follows from Equation~\eqref{eq:nminustauUB}. Thus, we now use the obtained upper bound on $\mathbb{P}(\Tilde{\d}_s \notin [\d - \Delta , \d + \Delta ] \text{ for some } s \leq t)$ (Equation~\eqref{eq:ProbUpperBdHelper}) for any $t \leq n-2$ to show that $\mathbb{E}[n-\tau]$ is bounded above by a constant. To see this, observe that
\begin{align}
    \mathbb{E}[n-\tau] &\leq  2+ \sum_{t = 1}^{n-2} \mathbb{P}(\Tilde{\d}_s \notin [\d - \Delta , \d + \Delta ] \text{ for some } s \leq t), \nonumber \\
    &\leq 2 + \sum_{t = 1}^{n-2} 2m e^{-\frac{ \underline{p}^2 \underline{\Delta}^{2}(n-t-1)}{2 \bar{w}^{2}}}, \nonumber \\
    &= 2+ 2m \sum_{s = 1}^{n-2} e^{-\frac{ \underline{p}^2 \underline{\Delta}^{2} s}{2 \bar{w}^{2}}}, \nonumber \\
    &= 2 + 2m e^{-\frac{\underline{p}^2 \underline{\Delta}^2}{2 \Bar{w}^2}} \sum_{s = 0}^{n-3} e^{-\frac{ \underline{p}^2 \underline{\Delta}^{2} s}{2 \bar{w}^{2}}}, \nonumber \\
    &= 2 + 2m e^{-\frac{\underline{p}^2 \underline{\Delta}^2}{2 \Bar{w}^2}} \frac{1- e^{-\frac{ \underline{p}^2 \underline{\Delta}^2(n-3)}{2 \Bar{w}^2}}}{1- e^{-\frac{\underline{p}^2 \underline{\Delta}^2}{2 \Bar{w}^2}}}, \nonumber \\
    &\leq 2+ 2m \frac{1}{1- e^{-\frac{ \underline{p}^2 \underline{\Delta}^2}{2 \Bar{w}^2}}}, \nonumber \\
    &= O(m) \label{eq:nMinusTauRelation}
\end{align}
The above analysis for the upper bound on the term $\mathbb{E}[n-\tau]$ along with Equation~\eqref{eq:capVioFirstRelation} establishes the constant upper bound on the expected constraint violation for Algorithm~\ref{alg:AlgoProbKnownDiscrete}, as 
\begin{align*}
    \mathbb{E} \left[ \norm{\sum_{t = 1}^n x_{tj} - c_j}_2 \right] \leq O(\mathbb{E}\left[n-\tau\right]) \leq O(m).
\end{align*}
This completes the proof of our claim that the constraint violation of Algorithm~\ref{alg:AlgoProbKnownDiscrete} is bounded by a constant independent of the number of users $n$.

\subsection{Proof of Lemma~\ref{lem:lipshitzness}}

We begin by presenting the dual function $D(\p, \d)$, which is a function of the price vector $\p$ and parametrized by the per-user resource vector $\d$. In particular, the dual function is represented as
\begin{align} \label{eq:dual-function-restricted}
    D(\p, \d) = \sum_{j = 1}^m p_j d_j - \sum_{k= 1}^K q_k \Tilde{w}_k \log \left( \min_{j \in [m]} \frac{p_j}{\Tilde{u}_{kj}} \right)
\end{align}
Note that we have dropped the constant terms independent of the price vector $\p$ from the dual objective (see Section~\ref{sec:dualProblem}). We now re-parametrize the dual function with a variable $\alpha_k = \min_{j \in [m]} \frac{p_j}{\Tilde{u}_{kj}}$ to get the following:
\begin{align*}
    D(\alphaa, \d) = \sum_{j = 1}^m d_j \max_{k} \{ \alpha_k \Tilde{u}_{kj} \} - \sum_{k= 1}^K q_k \Tilde{w}_k \log \left( \alpha_k \right),
\end{align*}
where note that $\alpha_k \leq \frac{p_j}{\Tilde{u}_{kj}}$ for all $j$ and $k$.

Next, let $\d, \d'$ be two different resource consumption vectors and let $\alphaa^*(\d)$ (and $\alphaa^*(\d')$) be the optimal solution to the corresponding dual problems with resource vectors $\d$ (and $\d'$), respectively. Then, noting that the dual function $D(\alphaa, \d)$ is strongly convex in $\alphaa$ (for a bounded set of values of $\alphaa$), it holds that:
\begin{align} \label{eq:strong-convexity-alpha}
    D(\alphaa^*(\d'), \d) - D(\alphaa^*(\d), \d) \geq \eta \norm{\alphaa^*(\d') - \alphaa^*(\d)}^2,
\end{align}
where note that $\eta>0$ is a positive constant as the prices $\p$ remain bounded for all resource consumption vectors $\Tilde{\d} \in [\d - \Delta, \d + \Delta]$.

Next, note that
\begin{align*}
    D(\alphaa^*(\d'), \d) - D(\alphaa^*(\d), \d) &= [D(\alphaa^*(\d'), \d) - D(\alphaa^*(\d'), \d')] - [D(\alphaa^*(\d), \d) - D(\alphaa^*(\d), \d')] \\
    &+ D(\alphaa^*(\d'), \d') - D(\alphaa^*(\d), \d'), \\
    &\stackrel{(a)}{\leq} [D(\alphaa^*(\d'), \d) - D(\alphaa^*(\d'), \d')] - [D(\alphaa^*(\d), \d) - D(\alphaa^*(\d), \d')], \\
    &= \sum_{j = 1}^m d_j \max_{k} \{ \alpha_k^*(\d') \Tilde{u}_{kj} \} - \sum_{k= 1}^K q_k \Tilde{w}_k \log \left( \alpha_k^*(\d') \right) \\
    &- \sum_{j = 1}^m d_j' \max_{k} \{ \alpha_k^*(\d') \Tilde{u}_{kj} \} + \sum_{k= 1}^K q_k \Tilde{w}_k \log \left( \alpha_k^*(\d') \right) \\
    &- \sum_{j = 1}^m d_j \max_{k} \{ \alpha_k^*(\d) \Tilde{u}_{kj} \} + \sum_{k= 1}^K q_k \Tilde{w}_k \log \left( \alpha_k^*(\d) \right) \\
    &+ \sum_{j = 1}^m d_j' \max_{k} \{ \alpha_k^*(\d) \Tilde{u}_{kj} \} - \sum_{k= 1}^K q_k \Tilde{w}_k \log \left( \alpha_k^*(\d) \right), \\
    &= \sum_{j = 1}^m \max_{k} \{ \alpha_k^*(\d') \Tilde{u}_{kj} \} (d_j - d_j') + \sum_{j = 1}^m \max_{k} \{ \alpha_k^*(\d) \Tilde{u}_{kj} \} (d_j' - d_j), \\
    &= \sum_{j = 1}^m (d_j - d_j') (\max_{k} \{ \alpha_k^*(\d') \Tilde{u}_{kj} \} - \max_{k} \{ \alpha_k^*(\d) \Tilde{u}_{kj} \}), \\
    &\stackrel{(b)}{\leq} \sum_{j = 1}^m (d_j - d_j') (\alpha_{\Tilde{k}}^*(\d') \Tilde{u}_{\Tilde{k}j} - \alpha_{\Tilde{k}}^*(\d) \Tilde{u}_{\Tilde{k}j} ), \\
    &\leq \Bar{u} \norm{\alphaa^*(\d) - \alpha^*(\d')}_2 \norm{\d - \d'}_1, \\
    &\stackrel{(c)}{\leq} \sqrt{m} \Bar{u} \norm{\alphaa^*(\d) - \alpha^*(\d')}_2 \norm{\d - \d'}_2, \\
\end{align*}
where (a) follows as $\alphaa^*(\d')$ is a minimizer of the dual function $D(\cdot, \d')$, (b) follows as we define $\Tilde{k}$ to be the group such that $\Tilde{k} \in \argmax \{ \alpha_k^*(\d) \Tilde{u}_{kj} \}$ and (c) follows by the norm equivalence relation between the one and two norm.

From the above inequality and Equation~\eqref{eq:strong-convexity-alpha}, it follows that
\begin{align*}
    \eta \norm{\alphaa^*(\d') - \alphaa^*(\d)}_2^2 \leq D(\alphaa^*(\d'), \d) - D(\alphaa^*(\d), \d) \leq \sqrt{m} \Bar{u} \norm{\alphaa^*(\d) - \alpha^*(\d')}_2 \norm{\d - \d'}_2.
\end{align*}
Thus, it holds that
\begin{align} \label{eq:alpha-lipschitz}
    \norm{\alphaa^*(\d') - \alphaa^*(\d)}_2 \leq \frac{\sqrt{m} \Bar{u}}{\eta} \norm{\d - \d'}_2.
\end{align}
Finally, to establish our desired Lipschitzness result, we show that the prices are Lipschitz in $\alpha$. To show this, we denote $\p^*(\d)$ (and $\p^*(\d')$) be the optimal price vector corresponding to two different average resource consumption vectors $\d$ (and $\d'$), respectively. Then, we get that:
\begin{align}
    \norm{\p^*(\d) - \p^*(\d')}_2 &= \norm{\max_{k} \{ \alpha_k^*(\d) \u_{k} \} - \max_{k} \{ \alpha_k^*(\d') \u_{k} \} }_2, \nonumber \\
    &= \sqrt{ \sum_{j = 1}^m \left(  \max_{k} \{ \alpha_k^*(\d) u_{kj} \} - \max_{k} \{ \alpha_k^*(\d') u_{kj} \} \right)^2 }, \nonumber \\
    &\stackrel{(a)}{\leq} \sqrt{ \sum_{j = 1}^m \left(\alpha_{\Tilde{k}_j}^*(\d) u_{\Tilde{k}_j j} - \alpha_{\Tilde{k}_j}^*(\d') u_{\Tilde{k}_j j} \right)^2 }, \nonumber \\
    &\leq \sqrt{m } \Bar{u} \norm{\alphaa^*(\d) - \alphaa^*(\d')}_2, \label{eq:helper-strong-convexity}
\end{align}
where (a) follows for some $\Tilde{k}_j$. To see this, let $\Tilde{k}_j^1 = \argmax_{k} \{ \alpha_k^*(\d) u_{kj} \}$ and let $\Tilde{k}_j^2 = \max_{k} \{ \alpha_k^*(\d') u_{kj} \}$. Then, it follows that: $\max_{k} \{ \alpha_k^*(\d) u_{kj} \} - \max_{k} \{ \alpha_k^*(\d') u_{kj} \} \leq \max \{ \alpha_{\Tilde{k}_j^1}^*(\d) u_{\Tilde{k}_j^1 j} - \alpha_{\Tilde{k}_j^2}^*(\d') u_{\Tilde{k}_j^2 j}, \alpha_{\Tilde{k}_j^2}^*(\d') u_{\Tilde{k}_j^2 j} - \alpha_{\Tilde{k}_j^1}^*(\d) u_{\Tilde{k}_j^1 j} \}$ for all goods $j$.



Finally, combining the above relation with Equation~\eqref{eq:alpha-lipschitz}, we obtain the desired Lipschitzness relation between prices and the average resource consumption vectors:
\begin{align*}
    \norm{\p^*(\d) - \p^*(\d')}_2 \leq \sqrt{m } \Bar{u} \norm{\alphaa^*(\d) - \alphaa^*(\d')}_2 \leq \frac{m \Bar{u}^2}{\eta} \norm{\d - \d'}_2,
\end{align*}
which establishes our claim. We also note that the above Lipschitzness relation implies that when $\norm{\d - \d'} = O(\frac{1}{n-t})$, then even $\norm{\p^*(\d') - \p^*(\d)} = O(\frac{1}{n-t})$.

\subsection{Proof of Lemma~\ref{lem:firstTermUB}}

We use Lemma~\ref{lem:lipshitzness} to analyse the first term in the generic regret bound in Equation~\eqref{eq:genericRegretBoundFINAL} for Algorithm~\ref{alg:AlgoProbKnownDiscrete} and establish that $\frac{2 \sqrt{m} \Bar{w}}{\underline{p}} \sum_{t = 1}^n \mathbb{E} \left[ \norm{\p^* - \p^t}_2 \right] \leq O(\log(n))$.

To this end, we first show that $\mathbb{E}[\norm{\p^t - \p^*}_2] \leq O(\frac{1}{n-t+1})$ for all $t = \{1, \ldots, n-\tau\}$ for Algorithm~\ref{alg:AlgoProbKnownDiscrete}. To see this, we proceed by induction. For the base case, take $t = 1$, in which case Algorithm~\ref{alg:AlgoProbKnownDiscrete} initializes the price $\p^1 = \p^*$, as the adaptive expected equilibrium pricing algorithm sets the static expected equilibrium prices at $t = 1$ as $\d_1 = \frac{\c}{n}$. As a result, it clearly holds that $\mathbb{E}[\norm{\p^1 - \p^*}_2] = 0 \leq O(\frac{1}{n})$. For the inductive step, we now assume that $\mathbb{E}[\norm{\p^t - \p^*}_2] \leq O(\frac{1}{n-t+1})$ for all $t \leq k$. Then, we have for $t = k+1$ that
\begin{align}
    \mathbb{E} \left[\norm{\p^{k+1} - \p^*}_2 \right] &\leq \mathbb{E} \left[\norm{\p^{k+1} - \p^k}_2 \right] + \mathbb{E} \left[\norm{\p^{k} - \p^*}_2 \right], \nonumber \\ 
    &\leq \mathbb{E} \left[\norm{\p^{k+1} - \p^k}_2 \right] + O \left(\frac{1}{n-k+1} \right), \nonumber \\
    &\leq \mathbb{E} \left[\norm{\p^{k+1} - \p^k}_2 \right] + O \left(\frac{1}{n-k} \right). \label{eq:inductionIneq}
\end{align}
To bound $\mathbb{E}[\norm{\p^{k+1} - \p^k}_2]$, we note that $\d_{k+1} = \d_k + \frac{\d_k - \x_k(\p^k)}{n-k}$, i.e., $\norm{\d_{k+1} - \d_k} = O(\frac{1}{n-k})$. Then, using Lemma~\ref{lem:lipshitzness}, it follows that $\mathbb{E}[\norm{\p^{k+1} - \p^k}_2] = O(\frac{1}{n-k})$. This inequality, together with Equation~\eqref{eq:inductionIneq}, implies that
\begin{align}
    \mathbb{E} \left[\norm{\p^{k+1} - \p^*}_2 \right] &\leq O \left(\frac{1}{n-k} \right),
\end{align}
which establishes our inductive step and thus establishes our claim that $\mathbb{E}[\norm{\p^t - \p^*}_2] \leq O(\frac{1}{n-t+1})$ for all $t = \{1, \ldots, n-\tau\}$ for Algorithm~\ref{alg:AlgoProbKnownDiscrete}. Furthermore, observe that since $\p^t = \p^*$ for $t > \tau$, it holds that $\mathbb{E}[\norm{\p^t - \p^*}_2] \leq O(\frac{1}{n-t+1})$ for all $t = \{1, \ldots, n\}$. Using this result, we obtain the following upper bound on the first term of Equation~\eqref{eq:genericRegretBoundFINAL}
\begin{align}
    \frac{2 \sqrt{m} \Bar{w}}{\underline{p}} \sum_{t = 1}^n \mathbb{E} \left[ \norm{\p^* - \p^t}_2 \right] \leq \frac{2 \sqrt{m} \Bar{w}}{\underline{p}} \sum_{t = 1}^n \mathbb{E} \left[ O \left(\frac{1}{n-t+1} \right) \right] 
    \leq \frac{2 \sqrt{m} \Bar{w}}{\underline{p}} \sum_{t = 1}^n \mathbb{E} \left[ O \left(\frac{1}{t} \right) \right] \leq O(\log(n)), \label{eq:SecondHelperRelation}
\end{align}
which proves our claim.

\subsection{Proof of Corollary~\ref{cor:regretAlgo4}}

We now use the generic bound on the regret derived in Equation~\eqref{eq:genericRegretBoundFINAL} to obtain a $O(\log(n))$ bound on the regret of Algorithm~\ref{alg:AlgoProbKnownDiscrete}. In particular, we upper bound both the terms on the right hand side of Equation~\eqref{eq:genericRegretBoundFINAL} using the analysis performed in Lemmas~\ref{lem:capVioAlgo4} and~\ref{lem:firstTermUB} to establish that 
\begin{align*}
    \mathbb{E}[U_n^* - U_n(\ppi)] \leq \frac{2 \sqrt{m} \Bar{w}}{\underline{p}} \sum_{t = 1}^n \mathbb{E} \left[ \norm{\p^* - \p^t}_2 \right] + \mathbb{E} \left[ \Bar{p} \left| \sum_{j = 1}^m \left(\sum_{t=1}^n x_{tj} - c_j  \right) \right| \right] \leq O(\log(n)).
\end{align*}
To establish the above claim, we first observe by Lemma~\ref{lem:firstTermUB} that the first term of right hand side of the generic regret bound, i.e., Equation~\eqref{eq:genericRegretBoundFINAL}, is upper bounded by $O(\log(n))$. Next, noting that the second term on the right hand side of Equation~\eqref{eq:genericRegretBoundFINAL} is analogous to the constraint violation of Algorithm~\ref{alg:AlgoProbKnownDiscrete}, we observe that
\begin{align}
    \mathbb{E} \left[ \Bar{p} \left| \sum_{j = 1}^m \left(\sum_{t=1}^n x_{tj} - c_j  \right) \right| \right] &\stackrel{(a)}{\leq} \mathbb{E} \left[ \Bar{p} \left| \sum_{j = 1}^m \left(\sum_{t=\tau+1}^n x_{tj}  \right) \right| \right] \stackrel{(b)}{\leq} \Bar{p} \mathbb{E} \left[m (n-\tau) \frac{\Bar{w}}{\underline{p}} \right] \stackrel{(c)}{\leq} \frac{m \Bar{w} \Bar{p}}{\underline{p}} = O(m),  \label{eq:FirstHelperRelation}
\end{align}
where (a) follows since no constraints are violated up until the stopping time $\tau$, (b) follows as $x_{tj} \leq \frac{\Bar{w}}{\underline{p}}$, and (c) follows from Equation~\eqref{eq:nMinusTauRelation}. As a result, we have established that the second term in the generic regret bound is bounded above by a constant (and thus is also bounded above by $O(\log(n))$) for Algorithm~\ref{alg:AlgoProbKnownDiscrete}, which thus proves our claim.

\section{Derivation of Dual of Social Optimization Problem} \label{apdx:dual-pf}

In this section, we derive the dual of the social optimization Problem~\eqref{eq:FisherSocOpt}-\eqref{eq:FisherSocOpt3}. To this end, we first consider the following equivalent primal problem
\begin{maxi!}|s|[2]                   
    {\mathbf{x}_t \in \mathbb{R}^m, u_t}                               
    {U(\mathbf{x}_1, ..., \mathbf{x}_n) = \sum_{t = 1}^{n} w_t \log \left(u_t \right) , \label{eq:FisherSocOptLag}}   
    {\label{eq:FisherExample1}}             
    {}                                
    \addConstraint{\sum_{t = 1}^{n} x_{tj}}{ \leq c_j, \quad \forall j \in [m], \label{eq:FisherSocOpt1LAG}}    
    \addConstraint{x_{tj}}{\geq 0, \quad \forall t \in [n], j \in [m], \label{eq:FisherSocOpt3LAG}}  
    \addConstraint{u_t}{= \sum_{j = 1}^m u_{tj} x_{tj}, \quad \forall t \in [n], \label{eq:FisherSocOpt4LAG}}  
\end{maxi!}
where we replaced the linear utility $\sum_{j = 1}^m u_{tj} x_{tj}$ in the objective with the variable $u_t$ and added the constraint $u_t = \sum_{j = 1}^m u_{tj} x_{tj}$. Observe that the optimal solution of this problem is equal to that of the social optimization Problem~\eqref{eq:FisherSocOpt}-\eqref{eq:FisherSocOpt3}. We now formulate the Lagrangian of this problem and derive the first order conditions of this Lagrangian to obtain the dual Problem~\eqref{eq:dual}.

To formulate the Lagrangian of Problem~\eqref{eq:FisherSocOptLag}-\eqref{eq:FisherSocOpt4LAG}, we introduce the dual variables $p_j$ for each good $j \in [m]$ for the capacity Constraints~\eqref{eq:FisherSocOpt1LAG}, $\lambda_{tj}$ for each user $t \in [n]$ and good $j \in [m]$ for the non-negativity Constraints~\eqref{eq:FisherSocOpt3LAG}, and $\beta_t$ for each user $t \in [n]$ for the linear utility Constraints~\eqref{eq:FisherSocOpt4LAG}. For conciseness, we denote $\p \in \mathbb{R}^m$ as the vector of dual variables of the capacity Constraints~\eqref{eq:FisherSocOpt1LAG}, $\Lambda \in \mathbb{R}^{n \times m}$ as the matrix of dual variables of the non-negativity Constraints~\eqref{eq:FisherSocOpt3LAG}, and $\bbeta$ as the vector of dual variables of the linear utility Constraints~\eqref{eq:FisherSocOpt4LAG}. 
Then, we have the following Lagrangian:
\begin{align*}
    \mathcal{L}(\{ \x_t, u_t \}_{t = 1}^n, \p, \Lambda, \bbeta) &= \sum_{t = 1}^n w_t \log(u_t)  -  \sum_{j = 1}^m p_j  \bigg( \sum_{t = 1}^n x_{tj} - c_j \bigg) - \sum_{t = 1}^n \sum_{j = 1}^m \lambda_{ij} x_{ij} \\&- \sum_{t = 1}^n \beta_t (u_t - \sum_{j = 1}^m u_{tj} x_{tj})
\end{align*}
Next, we observe from the first order derivative condition of the Lagrangian that
\begin{align*}
    &\frac{\partial \mathcal{L}}{\partial u_t} = \frac{w_t}{u_t} - \beta_t = 0, \quad \forall t \in [n], \text{ and }\\
    &\frac{\partial \mathcal{L}}{\partial x_{tj}} = -p_j - \lambda_{tj} + \beta_t u_{tj} = 0, \quad \forall t \in [n], j \in [m].
\end{align*}
Note that we can rearrange the first equation to obtain that $u_t = \frac{w_t}{\beta_t}$ for all $t \in [n]$. Furthermore, by the sign constraint that $\lambda_{tj} \leq 0$ for all $t \in [n]$, $j \in [m]$ it follows from the second equation that $\beta_t u_{tj} \leq p_j$ for all $t \in [n]$, $j \in [m]$. Using the above equations, we can write the following dual problem:

\begin{equation}
\begin{aligned}
\min_{\p, \bbeta} \quad & \sum_{t  = 1}^n w_t \log(w_t) - \sum_{t = 1}^n w_t \log(\beta_t) + \sum_{j = 1}^m p_j c_j - \sum_{t = 1}^n w_t  \\
& \beta_t u_{tj} \leq p_j, \quad \forall t \in [n], j \in [m]
\end{aligned}
\end{equation}
Note that at the optimal solution to the above problem $\beta_t = \min_{j \in [m]} \{ \frac{p_j}{u_{tj}} \}$. Using this observation, we can rewrite the above problem as
\begin{equation}
\begin{aligned}
\min_{\mathbf{p}} \quad & \sum_{t  = 1}^n w_t \log(w_t) - \sum_{t = 1}^n w_t \log \bigg(\min_{j \in [m]} \frac{p_j}{u_{tj}} \bigg) + \sum_{j = 1}^m p_j c_j - \sum_{t = 1}^n w_t,
\end{aligned}
\end{equation}
which is the dual Problem~\eqref{eq:dual}.

\section{Detailed Proof Sketch of Theorem~\ref{thm:PrivacyPreserving}} \label{sec:pfThm3}

The proof of Theorem~\ref{thm:PrivacyPreserving} relies on two intermediate arguments. First, we show that if the price vector at every step of the algorithm is bounded above and below by some positive constant, then the $O(\sqrt{n})$ upper bounds on both the regret and expected constraint violation hold.

\begin{lemma} [Regret and Constraint Violation of Algorithm~\ref{alg:PrivacyPreserving} under Positivity and Boundedness of Prices] \label{lem:PrivacyPreserving}
Suppose users' budget and utility parameters are drawn i.i.d. from a distribution $\D$ satisfying Assumption~\ref{asmpn:mainRestriction}. Furthermore, let $\ppi$ denote the online pricing policy described by Algorithm~\ref{alg:PrivacyPreserving}, $\x_1, \ldots, \x_n$ be the corresponding allocations for the $n$ users, and suppose that the price vector $\p^t$ corresponding to Algorithm~\ref{alg:PrivacyPreserving} is such that $\0< \underline{\p} \leq \p^t \leq \Bar{\p}$ for all users $t \in [n]$. Then, for a step size $\gamma = \gamma_t = \frac{\Bar{D}}{\sqrt{n}}$ for some constant $\Bar{D}>0$ for all users $t \in [n]$, the regret $R_n(\ppi) \leq O(\sqrt{n})$ and the constraint violation $V_n(\ppi) \leq O(\sqrt{n})$.
\end{lemma}

\hproof
To establish this result, we proceed in three steps. First, we prove an $O(\sqrt{n})$ upper bound on the constraint violation. To do so, we sum the price update equation in Algorithm~\ref{alg:PrivacyPreserving} across all users to establish that the excess demand for any good $j$ is upper bounded by $\frac{p_j^{n+1}}{\gamma}$, i.e., $\sum_{t = 1}^n x_{tj} - c_j \leq \frac{p_j^{n+1}}{\gamma}$. Using this relation and the fact that the prices are upper bounded by $\Bar{p}$ and the step size $\gamma = O(\frac{1}{\sqrt{n}})$, we obtain the $O(\sqrt{n})$ upper bound on the constraint violation. Next, we derive a generic upper bound on the regret (different from that in the proof of Theorem~\ref{thm:RegretCapVioAlgo4}) of any online algorithm $\ppi$ using duality (see Section~\ref{sec:dualProblem}), and show that $\mathbb{E}[U_n^* - U_n(\ppi)] \leq \mathbb{E} \left[ \sum_{t = 1}^n \sum_{j = 1}^m p_j^{t} d_j -  w_t \right]$. Finally, we apply the price update rule in Algorithm~\ref{alg:PrivacyPreserving} with a step size $\gamma = O(\frac{1}{\sqrt{n}})$ to establish an $O(\sqrt{n})$ upper bound on the term $\mathbb{E} \left[ \sum_{t = 1}^n \sum_{j = 1}^m p_j^{t} d_j -  w_t \right]$, i.e., the right hand side of the generic regret bound, which establishes our claim. 


We refer to Appendix~\ref{apdx:ubpf} for a complete proof of Lemma~\ref{lem:PrivacyPreserving} and note that its proof does not rely on Assumption~\ref{asmpn:regularity}.


Our second intermediary result states that if the distribution $\D$ satisfies Assumption~\ref{asmpn:regularity}, then the price vector $\p^t$ in Algorithm~\ref{alg:PrivacyPreserving} remains strictly positive and bounded for all users $t \in [n]$.

\begin{lemma} [Strictly Positive and Bounded Prices for Algorithm~\ref{alg:PrivacyPreserving}] \label{lem:PositivePrices}
Suppose that the budget and utility parameters of users are drawn i.i.d. from a distribution $\D$ satisfying Assumptions~\ref{asmpn:mainRestriction} and~\ref{asmpn:regularity}. Then, the price vector $\p^t$ corresponding to Algorithm~\ref{alg:PrivacyPreserving} will remain strictly positive and bounded for all users $t \in [n]$ when $\gamma = \gamma_t = \frac{\Bar{D}}{\sqrt{n}}$ for some constant $\Bar{D}>0$ for all users $t \in [n]$.
\end{lemma}

\hproof
To prove this claim, we proceed in two steps. First, we show that if the price vector $\p^t$ at each iteration of Algorithm~\ref{alg:PrivacyPreserving} is bounded below by some vector $\underline{\p}$, then the price vector also remains bounded above by $\Bar{\p}$, where $\Bar{p}>0$ is a constant, as we show in Lemma~\ref{lem:PosImplBounded} in Appendix~\ref{apdx:positivePricesPf}. In other words, the positivity of prices during the operation of Algorithm~\ref{alg:PrivacyPreserving} implies the boundedness of the prices. Next, we show that the prices of the goods will always remain positive under Assumption~\ref{asmpn:regularity}. To this end, we first consider the setting of one and two goods in the market, and then extend our analysis for the two good setting to the more general setting of $m$ goods. We present here the main ideas to prove this result for the two good case. In particular, it directly follows from Assumption~\ref{asmpn:regularity} that if the price of one good is small while that of another good is large, as specified by a certain price threshold $p^{\text{thresh}}$, then the price of the good that is small cannot become lower than $\frac{p^{\text{thresh}} \underline{u}}{2 \Bar{u}}$, as users will always purchase the good with the lower price given their strictly positive utilities. Next, in the case that the price of both goods is smaller than the specified threshold during the operation of Algorithm~\ref{alg:PrivacyPreserving}, we define a potential $V_t = (\p^t)^T \d$, and show that this potential is non-decreasing, i.e., $V_{t+1} \geq V_t$, if the prices of both goods are less than $p^{\text{thresh}}$ for user $t$. We then use this result along with Assumption~\ref{asmpn:regularity} to show that the price of both goods cannot go below a constant $\underline{p}$ during the operation of Algorithm~\ref{alg:PrivacyPreserving}. 


Note that Lemmas~\ref{lem:PrivacyPreserving} and~\ref{lem:PositivePrices} jointly imply Theorem~\ref{thm:PrivacyPreserving}. For a complete proof of Lemma~\ref{lem:PositivePrices}, we refer to Appendix~\ref{apdx:positivePricesPf}. 

Finally, we reiterate that the key to establishing Lemma~\ref{lem:PositivePrices} lies in constructing a potential function that is non-decreasing between subsequent users when the prices of all goods are below a particular threshold during the operation of Algorithm~\ref{alg:PrivacyPreserving}. Since this result is fundamental to the proof of Lemma~\ref{lem:PositivePrices} and elucidates a close connection between the price update rule in Algorithm~\ref{alg:PrivacyPreserving} and Fisher markets, we believe it is of independent interest. In particular, we formalize the non-decreasing potential function property of Algorithm~\ref{alg:PrivacyPreserving} through the following lemma.

\begin{lemma} [Non-Decreasing Potential] \label{lem:potentialFunction}
Let $p^{\text{thresh}} = \frac{\underline{w} \min_{j \in [m]} \{ d_j \}}{\sum_{j = 1}^m d_j^2}$ and define the potential $V_t = (\p^t)^{T} \d$, where $\p^t$ is the price vector corresponding to user $t$ in Algorithm~\ref{alg:PrivacyPreserving}. Then, the potential for Algorithm~\ref{alg:PrivacyPreserving} is non-decreasing for user $t+1$, i.e., $V_{t+1} \geq V_t$.
\end{lemma}

We refer to Appendix~\ref{apdx:positivityPf} for a proof of Lemma~\ref{lem:potentialFunction}.

\section{Proof of Lemma~\ref{lem:PrivacyPreserving}} \label{apdx:ubpf}

To establish this result, we proceed in three steps. First, we first prove an $O(\sqrt{n})$ upper bound on the constraint violation for the price update rule in Algorithm~\ref{alg:PrivacyPreserving}. Then, to establish an upper bound on the regret, we establish a generic bound on the regret (different from that in Lemma~\ref{lem:genericRegret2} in the proof of Theorem~\ref{thm:RegretCapVioAlgo4}) of any online algorithm as long as the prices $\p_t$ are strictly positive and bounded for all users $t \in [n]$. Finally, we apply the price update rule in Algorithm~\ref{alg:PrivacyPreserving} to establish an $O(\sqrt{n})$ upper bound on the regret for $\gamma = \gamma_t = \frac{\Bar{D}}{\sqrt{n}}$ for all users $t \in [n]$ for some constant $\Bar{D}>0$.

\paragraph{Expected Constraint Violation Bound:} 

To establish an $O(\sqrt{n})$ upper bound on the constraint violation, we utilize the price update rule in Algorithm~\ref{alg:PrivacyPreserving} where $\gamma_t = \frac{\Bar{D}}{\sqrt{n}}$ for some constant $\Bar{D}>0$. In particular, the price update step
\begin{align*}
    p_j^{t+1} = p_j^t - \frac{\Bar{D}}{\sqrt{n}} \left( d_j - x_{tj} \right)
\end{align*}
in Algorithm~\ref{alg:PrivacyPreserving} can be rearranged to obtain
\begin{align*}
    x_{tj} - d_j = \frac{\sqrt{n}}{\Bar{D}} \left(p_j^{t+1} - p_j^t \right).
\end{align*}
Summing this equation over all arriving users $t \in [n]$, it follows that
\begin{align*}
    \sum_{t =1}^n x_{tj} - c_j \leq \frac{\sqrt{n}}{\Bar{D}} \sum_{t = 1}^n \left(p_j^{t+1} - p_j^t \right) = \frac{\sqrt{n}}{\Bar{D}} \left(p_j^{n+1} - p_j^1 \right) \leq \frac{\sqrt{n}}{\Bar{D}} p_j^{n+1} \leq \frac{\Bar{p}}{\Bar{D}} \sqrt{n},
\end{align*}
where the last inequality follows since $p_j^{n+1} \leq \Bar{p}$ by the boundedness assumption on the price vector. Using this relation, the norm of the constraint violation can be bounded as
\begin{align*}
    \norm{\left(\sum_{t =1}^n \mathbf{x}_{t} - \c \right)_+}_2 \leq \norm{\sum_{t =1}^n \mathbf{x}_{t} - \c}_2 = \sqrt{\sum_{j = 1}^m \left(\sum_{t =1}^n x_{tj} - c_j \right)^2} \leq \sqrt{\sum_{j = 1}^m \left( \frac{\Bar{p}}{\Bar{D}} \right)^2 n } = \sqrt{m \left( \frac{\Bar{p}}{\Bar{D}} \right)^2 n } 
    \leq O(\sqrt{n}).
\end{align*}
Taking an expectation of the above quantity, we obtain a $O(\sqrt{n})$ upper bound on the expected constraint violation, where $\mathbb{E}[V_n(\x_1, \ldots, \x_n)] \leq \frac{\Bar{p}}{\Bar{D}} \sqrt{mn} = O(\sqrt{n})$.

\paragraph{Generic Bound on the Regret:}

We now turn to establishing a generic bound on the regret of any online algorithm for which the price vector $\p^t$ is strictly positive and bounded for each user $t \in [n]$. To perform our analysis, let $\p^*$ be the optimal price vector for the following stochastic program
\begin{equation} \label{eq:stochastic222}
\begin{aligned}
\min_{\mathbf{p}} \quad & D(\p) = \sum_{j = 1}^m p_j d_j + \mathbb{E}\left[\left(w \log \left(w\right)-w \log \left(\min _{j \in [m]} \frac{p_{j}}{u_{j}}\right)-w\right)\right].
\end{aligned}
\end{equation}

Then, by duality we have that the primal objective value $U_n^*$ is no more than the dual objective value with $\p = \p^*$, which gives the following upper bound on the optimal objective
\begin{align*}
    U_n^* &\leq \sum_{j = 1}^m p_j^{*} c_j + \sum_{t = 1}^n \left( w_t \log(w_t) - w_t \log \bigg(\min_{j \in [m]} \frac{p_j^{*}}{u_{tj}} \bigg)  -  w_t \right).
\end{align*}
Then, taking an expectation on both sides of the above inequality, it follows that
\begin{align*} 
    \mathbb{E}[U_n^*] &\leq  \mathbb{E} \left[ \sum_{j = 1}^m p_j^{*} c_j + \sum_{t = 1}^n \left( w_t \log(w_t) - w_t \log \bigg(\min_{j \in [m]} \frac{p_j^{*}}{u_{tj}} \bigg)  -  w_t \right) \right], \\
    &= n D(\p^*),
\end{align*}
by the definition of $D(\p)$ in Problem~\eqref{eq:stochastic222}. Finally, noting that $\p^*$ is the optimal solution to the stochastic Program~\eqref{eq:stochastic222}, it follows that
\begin{align*}
    \mathbb{E} \left[ U_n^* \right] &\leq n D(\p^*) \stackrel{(a)}{\leq} \sum_{t = 1}^n \mathbb{E} \left[ D(\p^t) \right] \stackrel{(b)}{=} \sum_{t = 1}^n \mathbb{E} \left[ \sum_{j = 1}^m p_j^t d_j + w_t \log(w_t) - w_t \log \bigg(\min_{j \in [m]} \frac{p_j^{t}}{u_{tj}} \bigg) -  w_t \right], \\ \label{eq:upperBdOptimalHelper}
    &\stackrel{(c)}{=}  \mathbb{E} \left[ \sum_{t = 1}^n \left( \sum_{j = 1}^m p_j^t d_j + w_t \log(w_t) - w_t \log \bigg(\min_{j \in [m]} \frac{p_j^{t}}{u_{tj}} \bigg) -  w_t \right) \right],
\end{align*}
where (a) follows by the optimality of $\p^*$ for the stochastic Program~\eqref{eq:stochastic222}, (b) follows by the definition of $D(\p^t)$, and (c) follows from the linearity of expectations.

Next, let $j_t$ be a good in the optimal consumption set $S_t^*$ for user $t$ given the price vector $\p^t$. Then, the true accumulated social welfare objective under an algorithm $\ppi$ can be expressed as
\begin{align*}
    U_n(\ppi) &= \sum_{t = 1}^n w_t \log \left( \sum_{j =1}^m u_{tj} x_{tj} \right), \\
    &= \sum_{t = 1}^n w_t \log \left( \sum_{j =1}^m u_{tj} \mathbbm{1}_{j = j_t} \frac{w_t}{p_j^t} \right),
\end{align*}
which follows since the utility when consuming any feasible bundle of goods in their optimal consumption set equals their utility when purchasing $\frac{w_t}{p_{j_t}^t}$ units of good $j_t \in S_t^*(\p^t)$. Finally combining the upper bound on the expected optimal objective and the above obtained relation on the accumulated objective under an algorithm $\ppi$, we obtain the following upper bound on the expected regret
\begin{align}
    \mathbb{E}[U_n^* - U_n(\ppi)] &\leq \mathbb{E} \left[ \sum_{t = 1}^n \left( \sum_{j = 1}^m p_j^t d_j + w_t \log(w_t) - w_t \log \bigg(\min_{j \in [m]} \frac{p_j^{t}}{u_{tj}} \bigg) -  w_t \right) \right] \\&- \mathbb{E} \left[ \sum_{t = 1}^n w_t \log \left( u_{tj_t} \frac{w_t}{p_{j_t}^t} \right) \right], \\
    &= \mathbb{E} \left[ \sum_{t = 1}^n \sum_{j = 1}^m p_j^{t} d_j -  w_t \right], \label{eq:genericBound2}
\end{align}
where the final equality follows as $j_t \in S_t^*(\p^t)$.

\paragraph{Square Root Regret Bound:} We now use the generic regret bound derived in Equation~\eqref{eq:genericBound2} for any online algorithm with bounded prices that are always strictly positive for each $t \in [n]$ to obtain an $O(\sqrt{n})$ upper bound on the regret of Algorithm~\ref{alg:PrivacyPreserving}. In particular, we use the price update equation in Algorithm~\ref{alg:PrivacyPreserving} to derive the $O(\sqrt{n})$ regret bound. We begin by observing from the price update equation that
\begin{align*}
    \norm{\p^{t+1}}_2^2 = \norm{\p^{t} - \frac{\Bar{D}}{\sqrt{n}} \left(\d - \x_t \right) }_2^2.
\end{align*}
Expanding the right hand side of the above equation, we obtain that
\begin{align*}
    \norm{\p^{t+1}}^2 \leq \norm{\p^{t}}^2 - \frac{2 \Bar{D}}{\sqrt{n}} \left( \sum_{j = 1}^m p_j^t d_j - \sum_{j =1}^m p_j^t x_{tj} \right) + \frac{\Bar{D}^2}{n} \norm{\d - \x_t}^2.
\end{align*}
We can then rearrange the above equation to obtain
\begin{align*}
    \sum_{j = 1}^m p_j^t d_j - \sum_{j =1}^m p_j^t x_{tj} \leq \frac{\sqrt{n}}{2 \Bar{D}} \left( \norm{\p^{t}}^2 - \norm{\p^{t+1}}^2  \right) + \frac{\Bar{D}}{2\sqrt{n}} \norm{\d- \x_t}^2
\end{align*}
Finally, summing both sides of the above equation over $t \in [n]$, we get
\begin{align}
    \sum_{t =1}^n \sum_{j = 1}^m p_j^t d_j - \sum_{t = 1}^n \sum_{j = 1}^m p_j^t x_{tj} &\leq \frac{\sqrt{n}}{2 \Bar{D}} \sum_{t = 1}^n \left( \norm{\p^{t}}^2 - \norm{\p^{t+1}}^2 \right) + \sum_{t = 1}^n \frac{\Bar{D}}{2\sqrt{n}} \norm{\d - \x_t}_2^2, \\
    &\stackrel{(a)}{\leq} \frac{\sqrt{n}}{2 \Bar{D}} \norm{p^1}^2 + \frac{\Bar{D}}{2\sqrt{n}} \sum_{t = 1}^n m \left(\max_{j \in [m]} d_j + \frac{\Bar{w}}{\underline{p}} \right)^2, \\
    &\leq \sqrt{n} \left( \frac{\norm{p^1}^2}{2 \Bar{D}} + \frac{\Bar{D} m}{2}  \left(\max_{j \in [m]} d_j + \frac{\Bar{w}}{\underline{p}} \right)^2 \right), \label{eq:upperBdHelper} \\
    &\leq O(\sqrt{n}), \label{eq:finalubHelper}
\end{align}
where the (a) follows by the boundedness of the consumption vector for each agent, since the prices are strictly positive and bounded below by $\underline{p}>0$. Finally, noting that all agents completely spend their budget at the optimal solution of the individual optimization problem, i.e., $\sum_{j \in [M]} p_j^t x_{tj} = w_t$, we obtain from the generic regret bound in Equation~\eqref{eq:genericBound2} that
\begin{align*}
    \mathbb{E}[U_n^* - U_n(\ppi)] &\leq \mathbb{E} \left[ \sum_{t = 1}^n \sum_{j = 1}^m p_j^{t} d_j -  w_t \right] = \sum_{t =1}^n \mathbb{E} \left[ \sum_{j = 1}^m p_j^t d_j - \sum_{t = 1}^n \sum_{j = 1}^m p_j^t x_{tj} \right], \\
    &\stackrel{(a)}{\leq} \sqrt{n} \left( \frac{\norm{p^1}^2}{2 \Bar{D}} + \frac{\Bar{D} m}{2}  \left(\max_{j \in [m]} d_j + \frac{\Bar{w}}{\underline{p}} \right)^2 \right),  \\
    &= O(\sqrt{n}),
\end{align*}
where (a) follows from Equation~\eqref{eq:upperBdHelper}. Thus, we have proven the $O(\sqrt{n})$ upper bound on the expected regret of Algorithm~\ref{alg:PrivacyPreserving} under the assumed conditions on the price vectors $\p^t$ for all users $t \in [n]$.

\section{Proof of Lemma~\ref{lem:PositivePrices}} \label{apdx:positivePricesPf}

To establish this result, we proceed in two steps. In particular, we first show that the strict positivity of prices during the operation of Algorithm~\ref{alg:PrivacyPreserving} implies that the prices are bounded for all $t \in [n]$ in Appendix~\ref{subsec:pfPosImplBounded}. Then, we show that the prices of the goods will always remain positive under Assumption~\ref{asmpn:regularity} in Appendix~\ref{apdx:positivityPf}.

\subsection{Positivity of Prices Implies Boundedness} \label{subsec:pfPosImplBounded}

We show through the following lemma that if the price vector $\p^t$ is bounded below by some vector $\underline{\p}$ at each iteration of Algorithm~\ref{alg:PrivacyPreserving}, then the price vector also remains bounded above by $\Bar{\p}$, where each component $\Bar{p}>0$ of the vector $\Bar{\p}$ is a constant.

\begin{lemma} [Positivity Implies Price Boundedness in Algorithm~\ref{alg:PrivacyPreserving}] \label{lem:PosImplBounded}
Suppose that the budget and utility parameters of users are drawn i.i.d. from a distribution $\D$ satisfying Assumption~\ref{asmpn:mainRestriction}, and the price vector $\p^t \geq \underline{\p} > \0$ for all users $t \in [n]$. Then, the price vector $\p^t$ corresponding to Algorithm~\ref{alg:PrivacyPreserving} is bounded at each time an agent $t \in [n]$ arrives, i.e., $\p^t \leq \Bar{\p} $ for all $t \in [n]$ for some vector $\Bar{\p}\geq \underline{\p}$,  when the step-size $\gamma = \gamma_t = \frac{\Bar{D}}{\sqrt{n}}$ for some $0<\Bar{D}\leq1$.
\end{lemma}

\proof{Proof.}
We establish that the prices of all goods are always bounded above at each step of Algorithm~\ref{alg:PrivacyPreserving} if the prices of the goods are bounded below by $\underline{\p}$ at each step. To show that the prices are bounded above, we consider the settings when (i) $\norm{\p^t}_2 \geq \frac{m \left( \Bar{d} + \frac{\Bar{w}}{\underline{p}} \right)^2 + 2 \Bar{w}}{2 \underline{d}}$, and (ii) $\norm{\p^t}_2 \leq \frac{m \left( \Bar{d} + \frac{\Bar{w}}{\underline{p}} \right)^2 + 2 \Bar{w}}{2 \underline{d}}$, where $\Bar{d} = \max_{j \in [m]} d_j$ and $\underline{d} = \min_{j \in [m]} d_j > 0$. In case (i), we observe that
\begin{align*}
    \norm{\p^{t+1}}_2^2 &= \norm{\p^t - \gamma (\d - \x_t)}_2^2 = \norm{\p^{t}}_2^2 - 2 \gamma (\p^t)^{\top} (\d - \x^t) + \gamma^2 \norm{\d - \x^t}_2^2, \\
    &\stackrel{(a)}{\leq} \norm{\p^{t}}_2^2 + 2 \gamma \Bar{w} + \gamma^2 m \left( \Bar{d} + \frac{\Bar{w}}{\underline{p}} \right)^2 - 2 \gamma (\p^t)^{\top} \d, \\
    &\leq \norm{\p^{t}}_2^2 + 2 \gamma \Bar{w} + \gamma^2 m \left( \Bar{d} + \frac{\Bar{w}}{\underline{p}} \right)^2 - 2 \gamma \underline{d} \norm{\p^t}_1, \\
    &\stackrel{(b)}{\leq} \norm{\p^{t}}_2^2 + 2 \gamma \Bar{w} + \gamma^2 m \left( \Bar{d} + \frac{\Bar{w}}{\underline{p}} \right)^2 - 2 \gamma \underline{d} \norm{\p^t}_2, \\
    &\stackrel{(c)}{\leq} \norm{\p^{t}}_2^2,
\end{align*}
where (a) follows from the fact that $(\p^t)^{\top} \x^t = w_t \leq \Bar{w}$ and $\x^t \leq \frac{\Bar{w}}{\underline{p}} \1$, where $\1$ is an $m$-dimensional vector of all ones, (b) follows from the norm equivalence relation between the one and the two norms, and (c) follows for any step-size $\gamma \leq 1$ and the fact that $\norm{\p^t}_2 \geq \frac{m \left( \Bar{d} + \frac{\Bar{w}}{\underline{p}} \right)^2 + 2 \Bar{w}}{2 \underline{d}}$. 

Next, in case (ii) it holds that
\begin{align*}
    \norm{\p^{t+1}}_2 &= \norm{\p^t - \gamma (\d - \x_t)}_2 \stackrel{(a)}{\leq} \norm{\p^t}_2 + \gamma \norm{\d - \x_t}_2 \stackrel{(b)}{\leq} \norm{\p^t}_2 + \gamma \norm{\d - \x_t}_1, \\
    &\stackrel{(c)}{\leq} \frac{m \left( \Bar{d} + \frac{\Bar{w}}{\underline{p}} \right)^2 + 2 \Bar{w}}{2 \underline{d}} + m \left(\Bar{d} + \frac{\Bar{w}}{\underline{p}} \right),
\end{align*}
where (a) follows by the triangle inequality, (b) follows from the norm equivalence relation between the one and to norms, and (c) holds since $\norm{\p^t}_2 \leq \frac{m \left( \Bar{d} + \frac{\Bar{w}}{\underline{p}} \right)^2 + 2 \Bar{w}}{2 \underline{d}}$ and $\x^t \leq \frac{\Bar{w}}{\underline{p}} \1$.

From the above inequalities, we observe that $\norm{\p^t}_2 \leq \frac{m \left( \Bar{d} + \frac{\Bar{w}}{\underline{p}} \right)^2 + 2 \Bar{w}}{2 \underline{d}} + m (\Bar{d} + \frac{\Bar{w}}{\underline{p}})$ for all $t$. This relation implies that the price vector of Algorithm~\ref{alg:PrivacyPreserving} is always bounded above when the price vector of Algorithm~\ref{alg:PrivacyPreserving} is bounded below by $\underline{\p}$ at each step, which completes the proof of Lemma~\ref{lem:PosImplBounded}.

\subsection{Positivity of Prices} \label{apdx:positivityPf}

We now show that under Assumption~\ref{asmpn:regularity} the prices of the goods remain strictly positive during the operation of Algorithm~\ref{alg:PrivacyPreserving}. To this end, we first prove this claim for the setting of one good. Then, we leverage Assumption~\ref{asmpn:regularity} to construct an argument for the setting of two goods and finally, extend this argument to the general setting of $m$ goods.

\subsubsection{One Good}

Suppose that there is exactly one good in the market with a capacity $c_1$. Then, we claim that under Algorithm~\ref{alg:PrivacyPreserving}, the price of each good for each user $t \in [n]$ remains bounded between $\underline{p}$ and $\Bar{p}$, where $\underline{p}>0$ and $\Bar{p}$ are constants. In particular, we prove this claim for $\underline{p} = \frac{\underline{w}}{2d_1}$, and $\Bar{p} = \frac{\Bar{w}}{d_1} + d_1 + \frac{\Bar{w}d_1}{\underline{w}}$ that $\underline{p} \leq p_1^t \leq \Bar{p}$ for all $t \in [n]$, when $\gamma = \frac{\Bar{D}}{\sqrt{n}}$, where $\Bar{D} = \frac{\frac{\underline{w}}{2d_1}}{d_1 + \frac{\Bar{w}d_1}{\underline{w}}}$, and the initial price vector $p_1^1 \in [\frac{\underline{w}}{2d_1}, \frac{\Bar{w}}{d_1} + d_1 + \frac{\Bar{w}d_1}{\underline{w}}]$.


To this end, first observe by the price update rule that
\begin{align*}
    p_1^{t+1} = p_1^t - \gamma \left( d_1 - x_{t1} \right) = p_1^t - \gamma \left( d_1 - \frac{w_t}{p_{1}^t} \right)
\end{align*}
since it is optimal for each user to purchase $\frac{w_t}{p_{1}^t}$ units of good one as there is only one good in the market. Next, to establish the the bounds on the price, we show that if $p_1^t$ is large (small), then it must hold that $p_1^{t+1} \leq p_1^t$ ($p_1^{t+1} \geq p_1^t$). In particular, observe that if $p_1^{t} \geq \frac{\Bar{w}}{d_1}$, then
\begin{align*}
    p_1^{t+1} = p_1^t - \gamma (d_1 - \frac{w_t}{p_1^t}) \leq p_1^t - \gamma d_1 + \gamma \frac{w_t d_1}{\Bar{w}} \leq p_1^t.
\end{align*}
and if $p_1^{t+1} \leq \frac{\underline{w}}{d_1}$, then $p_1^{t+1} \geq p_1^t$. On the other hand, when $\frac{\underline{w}}{d_1} \leq p_1^t \leq \frac{\Bar{w}}{d_1}$, then it holds that
\begin{align*}
    p_1^{t+1} = p_1^t - \gamma (d_1 - \frac{w_t}{p_1^t}) \leq \frac{\Bar{w}}{d_1} + d_1 + \frac{\Bar{w}d_1}{\underline{w}},
\end{align*}
since $\gamma \leq 1$, which holds true for large $n$. This establishes the upper bound on the price. For the lower bound on the price, observe that when $\frac{\underline{w}}{d_1} \leq p_1^t \leq \frac{\Bar{w}}{d_1}$, then it holds that
\begin{align*}
    p_1^{t+1} = p_1^t - \gamma (d_1 - \frac{w_t}{p_1^t}) \geq \frac{\underline{w}}{d_1} - \gamma (d_1 + \frac{\Bar{w}d_1}{\underline{w}}) \geq \frac{\underline{w}}{2d_1},
\end{align*}
since $\gamma = \frac{\Bar{D}}{\sqrt{n}}$, where $\Bar{D} = \frac{\frac{\underline{w}}{2d_1}}{d_1 + \frac{\Bar{w}d_1}{\underline{w}}}$. This establishes the lower bound on the price, which proves our claim that for the setting of one good the price for each user $t \in [n]$ under Algorithm~\ref{alg:PrivacyPreserving} is always bounded away from zero and is bounded above by some constant.

\subsubsection{Two Goods}

We now consider the setting where there are two goods in the market with capacities $c_1$ and $c_2$, respectively. Furthermore, we consider the setting when the support of the utilities and budgets of users is strictly positive. In particular, let $\rho = \frac{\bar{u}}{\underline{u}}$ be the maximum ratio of the utilities in the support of the distribution $\D$. In this case, we claim that under Algorithm~\ref{alg:PrivacyPreserving}, the price of both goods for each user $t \in [n]$ is always bounded away from zero. To establish this claim, we analyse several cases for when the prices of the goods is above or below certain thresholds. In each of these cases, we show that the prices of both goods for each user are always bounded away from zero by some constant $\Tilde{\underline{p}} = \frac{1}{(4 \rho)^2} \frac{\underline{w} \min \{ d_1, d_2 \}}{d_1^2 + d_2^2}$.

To prove our claim, we first let the initial price vector be $p_j^1 \in \left[ \frac{\underline{w}d_j}{\sum_{j = 1}^2 d_j^2}, \frac{\Bar{w}d_j}{\sum_{j = 1}^2 d_j^2} \right]$. Next, observe that at each time when a user $t$ arrives, the prices of the goods in the market must fall within one of the following cases:
\begin{enumerate}
    \item Case 1 (Both Prices are Large): $p_1^t \geq \frac{\Bar{w} d_1}{d_1^2 + d_2^2}$ and $p_2^t \geq \frac{\Bar{w} d_2}{d_1^2 + d_2^2}$;
    \item Case 2 (Both Prices are Small): $p_1^t < \frac{\underline{w} d_1}{d_1^2 + d_2^2}$ and $p_2^t < \frac{\underline{w} d_2}{d_1^2 + d_2^2}$;
    \item Case 3 (Intermediate Prices): One of the prices $p_j^t < \frac{\underline{w} d_j}{d_1^2 + d_2^2}$ while the other price $p_{j'}^t \geq \frac{\underline{w} d_{j'}}{d_1^2 + d_2^2}$ 
\end{enumerate}
Here case 1 corresponds to the setting when the prices of both goods is large, while case 2 represents the setting when the prices of both goods is small. On the other hand, case 3 captures all the other intermediate cases, where either both prices are bounded above and below or one of the prices is bounded above and the other is bounded below. To show that the prices are bounded below by the above defined $\Tilde{\underline{p}}$, we first note that in Case (i) that the prices of both goods are large and since the amount by which prices can drop is at most $O(\frac{1}{\sqrt{T}})$ at each step that the price at step $t+1$ will clearly be bounded below by $\Tilde{\underline{p}}$. Next, observe in case 3 that since the price of one of the goods is above the specified threshold. Without loss of generality, suppose that this is good one. Then, by the boundedness of utilities, we know that the price of good two must be at least $\frac{1}{2 \rho} \frac{\underline{w} d_1}{d_1^2 + d_2^2}$, as users would only purchase good two if their price fell below $\frac{1}{\rho} \frac{\underline{w} d_1}{d_1^2 + d_2^2}$. Thus, if we are in case 3, then it must also be that the prices of both goods are bounded below by $\Tilde{\underline{p}}$. Finally, we proceed to analysing case 2.

\paragraph{Analysis of Case 2:}

In case 2, we establish that the prices of the goods at each time a user arrives is bounded below in the setting when the good prices are low. In particular, we now show that both the good prices at step $t+1$ can be no lower than $\Tilde{\underline{p}}$. To this end, we proceed in the following steps. First, we define a potential $V_{t} = (\p^t)^{\top} \d$, and show that this potential is non-decreasing, i.e., $V_{t+1} \geq V_t$ if the price vector $\p^t$ satisfies the condition of case two. This claim establishes that the potential $V_t$ is monotonically non-decreasing until the price vector $\p^t$ exits case two. Then, we use this result and the fact that the utilities and budgets of users are bounded below to establish that the prices of the two goods are always strictly positive.

\paragraph{Proof of $V_{t+1} \geq V_t$ in Case 2:} Let $V_t = p_1^t d_1 + p_2^t d_2$, where $p_1^t < \frac{ \underline{w} d_1}{d_1^2 + d_2^2}$ and $p_2^t < \frac{\underline{w} d_2}{d_1^2 + d_2^2}$. Then, we show that $V_{t+1} \geq V_t$. To see this, suppose, without loss of generality, that buyer $t$ consumes $\frac{a w_t}{p_1^t}$ units of good one and $\frac{(1-a) w_t}{p_2^t}$ units of good two, where $a \in [0, 1]$. Then the price of good one for user $t+1$ is given by
\begin{align*}
    p_1^{t+1} = p_1^t - \gamma \left( d_1 - \frac{a w_t}{p_1^t} \right) \geq p_1^t - \gamma \left( d_1 - \frac{a w_t (d_1^2 + d_2^2)}{\underline{w} d_1} \right) \geq p_1^t - \gamma \left( d_1 - \frac{a(d_1^2+d_2^2)}{d_1} \right).
\end{align*}
Similarly, the price of good two for user $t+1$ is given by
\begin{align*}
    p_2^{t+1} = p_2^t - \gamma \left( d_2 - \frac{(1-a) w_t}{p_2^t} \right) \geq p_2^t - \gamma \left( d_2 - \frac{(1-a)(d_1^2+d_2^2)}{d_2} \right).
\end{align*}
Using the above inequalities for the prices of the two goods for user $t+1$, we obtain that $V_{t+1} \geq V_t$ since
\begin{align*}
    V_{t+1} &= p_1^{t+1} d_1 + p_2^{t+1} d_2, \\ 
    &\geq p_1^t d_1 + p_2^t d_2 - \gamma d_1 \left( d_1 - \frac{a(d_1^2+d_2^2)}{d_1} \right) - \gamma d_2 \left( d_2 - \frac{(1-a)(d_1^2+d_2^2)}{d_2} \right), \\
    &= V_t - \gamma \left( d_1^2 - a(d_1^2 + d_2^2) + d_2^2 - (1-a)(d_1^2 + d_2^2) \right), \\
    &= V_t,
\end{align*}
which proves our claim that the potential is non-decreasing when the price vector $\p^t$ lies in case two.

\paragraph{$V_t$ forms a monotonic sequence in case 2:} 
We have observed that the potential is non-decreasing for each user $t$ when $\p^t$ is in case two. Now, let $\tau_1$ be the index of the first user when the price vector belongs to case two, and  it holds that $\tau_2>\tau_1$ is the user index for which the price vector exits case 2 or at which the algorithm ends, i.e., $\tau_2 = \min \{ \{t>\tau_1:p_1^t \geq \frac{ \underline{w}d_1}{d_1^2+d_2^2} \text{ or } p_2^t \geq \frac{\underline{w}d_2}{d_1^2+d_2^2} \}, n+1 \}$. Then, from the above analysis that $V_{t+1} \geq V_t$ if the price $\p^t$ is in case two, it holds that
\begin{align*}
    V_{\tau_1} \leq V_{\tau_1 + 1} \leq \ldots \leq V_{\tau_2}.
\end{align*}

\paragraph{Prices of both goods are strictly positive in Case 2:}

Since $\tau_1$ is the first user index for which the prices of the goods belongs to case two, it must hold that the price of at least one of the goods exceeds the respective threshold for user $\tau_1-1$. Without loss of generality, suppose that $p_1^{\tau_1-1} \geq \frac{\underline{w} d_1}{d_1^2 + d_2^2}$. 

Next, observe that the price of good one for user $\tau_1$ must be such that 
\begin{align*}
    p_1^{ \tau_1} = p_1^{\tau_1-1} - \gamma (d_1 - x_{\tau_1-1,1}) \geq \frac{\underline{w} d_1}{d_1^2 + d_2^2} - \gamma d_1.
\end{align*}
Since we can take $\gamma \leq \frac{\frac{\underline{w} d_1}{d_1^2 + d_2^2}}{2d_1}$, it follows that $p_1^{ \tau_1} \geq \frac{\underline{w} d_1}{2(d_1^2 + d_2^2)}$.

Furthermore, since the utilities are bounded below, it follows that $p_2^{\tau_1} \geq \frac{1}{2 \rho} p_1^{ \tau_1} \geq \frac{1}{4 \rho} \frac{\underline{w} d_1}{2(d_1^2 + d_2^2)}$, as good two will be the only one consumed when its price is lower than $\frac{1}{\rho}$ times the price of good one. We now show that at all points between $\tau_1$ and $\tau_2-1$ that the prices of both goods is bounded below by $\Tilde{\underline{p}}$. To see this, first note that the price of good two must always be at least $\frac{1}{4 \rho} \frac{\underline{w} d_1}{2(d_1^2 + d_2^2)} \geq \Tilde{\underline{p}}$ by the monotonicity property of the potential function (as at least one of the good prices must increase but the price of good two cannot fall below $\frac{1}{4 \rho} \frac{\underline{w} d_1}{2(d_1^2 + d_2^2)}$ by the boundedness of utilities below). Analogously, since the price of good two cannot fall below $\frac{1}{4 \rho} \frac{\underline{w} d_1}{2(d_1^2 + d_2^2)}$ between $\tau_1$ and $\tau_2$, it also follows that the price of good one cannot fall below $\frac{1}{8 \rho^2} \frac{\underline{w} \min \{d_1, d_2 \}}{2(d_1^2 + d_2^2)} \geq \Tilde{\underline{p}}$ by the the boundedness of utilities below. Thus, we have established that both $p_1^t, p_2^t \geq \Tilde{\underline{p}}$ for all users $t \in \{ \tau_1, \ldots \tau_2 \}$. We note that we can repeat the above line of reasoning for all periods when the price vector belongs to case 2 and thus have shown that the prices are bounded below by a constant, which establishes our claim.

\subsubsection{Extending Above Argument to Multiple Goods}

To extend our analysis from the setting of two goods to that of $m$ goods, we first let $\p_j^1 \in \left[ \frac{\underline{w} d_j}{\sum_{j \in [m]} d_j^2}, \frac{\Bar{w} d_j}{\sum_{j \in [m]} d_j^2} \right]$ for all goods $j$. Then, we also consider multiple cases as in the two-good setting, and observe that to establish a lower bound on the prices, by the boundedness of utilities from below it suffices to consider the case when all goods have a price strictly below $\frac{\underline{w} d_j}{\sum_{j \in [m]} d_j^2}$.

To show that the prices are also bounded by below, we follow a similar line of reasoning as for the two good case. To this end, first observe that as in the two good case, there must be a good $j$ (without loss of generality suppose this is good one) that has a price just below $\frac{\underline{w} d_1}{\sum_{j \in [m]} d_j^2}$, e.g., $\frac{\underline{w} d_1}{2 \sum_{j \in [m]} d_j^2}$, when the price vector satisfies the condition that all goods have a price strictly below $\frac{\underline{w} d_1}{\sum_{j \in [m]} d_j^2}$ for the first time. This implies by by the boundedness of utilities from below that the price of the other goods are at least $p_2 \geq \frac{\underline{w} d_1}{4 \rho \sum_{j \in [m]} d_j^2}$, $p_3 \geq \frac{\underline{w} d_1}{(4\rho)^2\sum_{j \in [m]} d_j^2}$, $\ldots$, $p_m \geq \frac{\underline{w} d_j}{(4 \rho)^{m-1}\sum_{j \in [m]} d_j^2}$.

Next, it can again be shown in the multiple good case that the potential $V_{t+1} = (\p^{t+1})^T \d \geq (\p^{t})^T \d = V_t$ in the case when all goods $j$ have a price strictly below $\frac{\underline{w} d_j}{\sum_{j \in [m]} d_j^2}$ at each time a user $t$ arrives. Using the fact that $V_{t+1} \geq V_t$, it follows that the price of one of the goods must always be above their respective lower bounds. However, the price of good $m$ must be above its threshold since otherwise we would violate the boundedness of utilities from below. Since the price of good $m$ is bounded from below by a positive constant it follows that the prices of all the other goods must be at least $\frac{\underline{w} d_j}{ (4 \rho)^{2m}  \sum_{j \in [m]} d_j^2}$ by the boundedness of utilities from below. This completes the claim that the prices of the goods are always bounded below, which proves our claim.

\section{Remarks on the Positivity of Prices in Algorithm~\ref{alg:PrivacyPreserving}} \label{apdx:ubPrices}

In this section, we show that the price vector $\p^t$ is strictly positive during the operation of Algorithm~\ref{alg:PrivacyPreserving} for all $t \in [n]$ with high probability for distributions $\D$ such as in the counterexample used to prove Theorem~\ref{thm:lbStatic}.

In particular, we consider the class of distributions $\D$ that satisfy the following natural assumption, which states that the expected consumption of a good by any user is strictly greater than their market share of that good if the price of the good is small. 


\begin{assumption} \label{asmpn:newAsmpn}
There exists $\Tilde{\underline{p}}$ such that if $p_j < \Tilde{\underline{p}}$ for any good $j$, then the distribution $\D$ is such that the expected consumption of that good is at least $\frac{d_j}{1-\delta}$ for some $\delta>0$.
\end{assumption}

We note that Assumption~\ref{asmpn:newAsmpn} imposes a mild restriction on the set of allowable distributions from which the utility parameters of users are drawn. In particular, the assumption on the distribution $\D$ implies that for each good there are a certain fraction of the arriving users with a sufficiently high utility for that good. As a result if the price of a good drops too low then a certain fraction of users will purchase large quantities of that good that is far greater than their market share for that good. For instance, the distribution $\D$ constructed in the counterexample in the proof of Theorem~\ref{thm:lbStatic} satisfies Assumption~\ref{asmpn:newAsmpn}, as the expected consumption of each good is strictly greater than each user's market share $d_j$ of that good if its price drops strictly below $0.5$. As a result, Assumption~\ref{asmpn:newAsmpn} intuitively implies that the price of any good cannot drop ``too far'' below some specified price $\tilde{\underline{p}}$ during the operation of Algorithm~\ref{alg:PrivacyPreserving}. 

We now apply the Chernoff bound and use Assumption~\ref{asmpn:newAsmpn} to claim that the price vector $\p^t$ for all users $t \in [n]$ is lower bounded by $\underline{\p}$ with high probability for some constant $\underline{p}>0$. 
To this end, as in Assumption~\ref{asmpn:newAsmpn}, let $\Tilde{\underline{p}}$ be a constant and let $0<\underline{p} \leq (1-\epsilon)(\Tilde{\underline{p}} - \gamma d_j)$ for small $\epsilon>0$. We now provide a bound on the probability that the price $p_j^t$ of some good $j$ for some user $t$ drops below $\underline{p}$ during the operation of Algorithm~\ref{alg:PrivacyPreserving}. Here, we assume that the initial price $\p^1$ in Algorithm~\ref{alg:PrivacyPreserving} is sufficiently higher than $\underline{\p}$. We now suppose that $t$ is the first time step at which the price of some good $j$ falls below $\Tilde{\underline{p}}$, and that the price of that good stays below $\Tilde{\underline{p}}$ for another $k$ steps. Then, we can upper bound the probability that the price $p_j^t$ of some good $j$ for some user $t$ drops below $\underline{p}$ as follows
\begin{align*}
    \mathbb{P}[p_j^t \leq \underline{p} \text{ for some } j, t ] &\stackrel{(a)}{\leq} \mathbb{P}[p_j^t \leq (1-\epsilon) (\Tilde{\underline{p}} - \gamma d_j) \text{ for some good } j \text{ for some user } t ], \\
    &\stackrel{(b)}{\leq} \mathbb{P} \left[\Tilde{\underline{p}} - \gamma d_j - \gamma k d_j + \gamma \sum_{t' = t}^{t+k-1} x_{t'j} \leq (1-\epsilon) (\Tilde{\underline{p}} - \gamma d_j) \text{ for some } j , t \right], \\
    &\stackrel{(c)}{=} \mathbb{P} \left[ \gamma \sum_{t' = t}^{t+k-1} x_{t'j} \leq -\epsilon (\Tilde{\underline{p}} - \gamma d_j) + \gamma k d_j \text{ for some } j , t \right], \\
    &\stackrel{(d)}{\leq} \mathbb{P} \left[ \sum_{t' = t}^{t+k-1} x_{t'j} \leq k d_j \text{ for some } j , t \right],
\end{align*}
where (a) follows as $\underline{p} \leq (1-\epsilon) (\Tilde{\underline{p}} - \gamma d_j)$ for small $\epsilon > 0$, (b) follows by the price update rule in Algorithm~\ref{alg:PrivacyPreserving} and the fact that $t$ is the first step at which the price of some good $j$ falls below $(\Tilde{\underline{p}} - \gamma d_j)$, (c) follows by rearranging the terms in the inequality, and (d) follows as $-\epsilon (\Tilde{\underline{p}} - \gamma d_j)< 0$. To upper bound the right hand side term $\mathbb{P}[ \sum_{t' = t-1}^{t+k-1} x_{t'j} \leq (k+1) d_j \text{ for some } j , t]$, we begin by noting that the user consumption $x_{t'j}$ is not i.i.d. since user's consumption bundles depend on the price, which is inherently dependent on the budget and utility parameters of earlier users by the price update equation of Algorithm~\ref{alg:PrivacyPreserving}. However, defining $\mathbb{E}_{t-1}[\cdot] = \mathbb{E}[\cdot | (w_1, \u_1), \ldots, (w_1, \u_{t-1})]$ as the conditional expectation of the allocations of Algorithm~\ref{alg:PrivacyPreserving} depending on the realizations of the users' parameters, we can rewrite $\mathbb{P}[ \sum_{t' = t-1}^{t+k-1} x_{t'j} \leq (k+1) d_j \text{ for some } j , t]$ as follows:
\begin{align*}
    \mathbb{P} \left[ \sum_{t' = t}^{t+k-1} \! \! x_{t'j} \leq k d_j \text{ for some } j , t \right] &= \mathbb{P} \left[ \sum_{t' = t}^{t+k-1} \! \! x_{t'j} \! - \! \sum_{t' = t}^{t+k-1} \mathbb{E}_{t'-1}[x_{t'j}] \leq k d_j \! - \! \sum_{t' = t}^{t+k-1} \! \! \mathbb{E}_{t'-1}[x_{t'j}] \text{ for some } j , t \right], \\
    &\leq \mathbb{P} \left[ \sum_{t' = t}^{t+k-1} x_{t'j} - \sum_{t' = t}^{t+k-1} \mathbb{E}_{t'-1}[x_{t'j}] \leq k d_j - k \frac{d_j}{1-\delta} \text{ for some } j , t \right], \\
    &= \mathbb{P} \left[ \sum_{t' = t}^{t+k-1} x_{t'j} - \sum_{t' = t}^{t+k-1} \mathbb{E}_{t'-1}[x_{t'j}] \leq -\frac{k d_j \delta}{1-\delta} \text{ for some } j , t \right],
\end{align*}
where the inequality follows by Assumption~\ref{asmpn:newAsmpn} that $\mathbb{E}_{t'-1}[x_{t'j}] \geq \frac{d_j}{1-\delta}$ for all $t\leq t' \leq t+k-1$ as $p_j^{t'} \leq \Tilde{\underline{p}}$ for this range of values of $t'$. Furthermore, defining $\kappa_{t'j} = x_{t'j} - \mathbb{E}_{t'-1}[x_{t'j}]$ and $S_{n} = \sum_{t' = t}^{t+n-1} \kappa_{t'j}$, we note that $S_{n}$ is a martingale with respect to the filtration $\sigma((w_1, \u_1), \ldots, (w_t, \u_{t}), \ldots, (w_{t+n-1}, \u_{t+n-1}))$. To see this, observe that
\begin{align*}
    \mathbb{E}[S_{n+1}|S_1, \ldots, S_n] = S_n + \mathbb{E}[\kappa_{t+n-1,j}] = S_n + \mathbb{E}[x_{t+n-1,j} - \mathbb{E}_{t+n-2}[x_{t+n-1,j}]] = S_n.
\end{align*}
noting that $k$ is the number of steps for which the price of a good $j$ remains below $\Tilde{\underline{p}}$ and that the step size is $O(\frac{1}{\sqrt{n}})$, it follows that for a constant reduction in the price of good $j$, i.e., for $p_j^{t'} \leq (1-\epsilon)\Tilde{\underline{p}}$, it must hold that $k = O(\sqrt{n})$. Thus, for any $k = o(\sqrt{n})$, it must hold that $x_{t'j}$ is bounded for all $t\leq t' \leq t+k-1$ (as the price remains strictly positive for $k = o(\sqrt{n})$) and thus the corresponding martingale has bounded differences, i.e., $|S_n - S_{n-1}| \leq L$ for some constant $L$. Then, by Azuma's inequality for martingales with bounded differences~\citep{lalley2013concentration}, it follows that
\begin{align*}
    \mathbb{P} \left[ \sum_{t' = t}^{t+k-1} x_{t'j} - \sum_{t' = t}^{t+k-1} \mathbb{E}_{t'-1}[x_{t'j}] \leq -\frac{k d_j \delta}{1-\delta} \text{ for some } j , t \right] \leq e^{\frac{-k^2 d_j^2 \delta^2}{2kL^2 (1-\delta)^2}} = e^{\frac{-k d_j^2 \delta^2}{2 L^2 (1-\delta)^2}}
\end{align*}
for $k = o(\sqrt{n})$. Finally, combining the above derived sequence of inequalities, we obtain that
\begin{align*}
    \mathbb{P}[p_j^t \leq \underline{p} \text{ for some good } j \text{ for some user } t ] &\leq \mathbb{P} \left[ \sum_{t' = t}^{t+k-1} x_{t'j} - \sum_{t' = t}^{t+k-1} \mathbb{E}_{t'-1}[x_{t'j}] \leq -\frac{k d_j \delta}{1-\delta} \text{ for some } j , t \right], \\  
    &\leq e^{\frac{-k d_j^2 \delta^2}{2 L^2 (1-\delta)^2}},
\end{align*}
for $k = o(\sqrt{n})$ which implies that the probability that the price $p_j^t$ for some good $j$ for some user $t$ drops below $\underline{p}$ exponentially decays in $k$. Since this inequality holds for all $k = o(\sqrt{n})$, in particular, we have that the above inequality holds for $k = n^{\frac{1}{3}}$. Thus, it holds that the right hand side term goes to zero as $n \rightarrow \infty$ and so, for large $n$, it follows that the price of each good will always remain bounded below by $\underline{p}$ with high probability.

In particular, suppose that $\epsilon$ is the desired probability that we want to ensure that $\mathbb{P}[ \sum_{t' = t}^{t+k-1} x_{t'j} \leq k d_j \text{ for some } j , t] \leq \epsilon$, then we require that $e^{\frac{-k d_j^2 \delta^2}{2 L^2 (1-\delta)^2}} \leq \epsilon$, which implies that $k \geq \log(\frac{1}{\epsilon}) \frac{2 L^2 (1-\delta)^2}{d_j^2 \delta^2}$ ensures the high probability bound. In particular, this holds if $n \geq \left( \log(\frac{1}{\epsilon}) \frac{2 L^2 (1-\delta)^2}{d_j^2 \delta^2} \right)^3$.

\section{Proof of Theorem~\ref{thm:PrivacyPreservingImproved}} \label{apdx:pf-PrivacyPreservingImproved}

As with the proof of Theorem~\ref{thm:PrivacyPreserving}, the proof of Theorem~\ref{thm:PrivacyPreservingImproved} relies on two intermediate results. In particular, we first show in Section~\ref{apdx:regret-bd-variable-step-size-bdd-prices} that if the price vector is bounded above and below by some positive constant at each iteration of Algorithm~\ref{alg:PrivacyPreserving} with a two-stage adjustment of the step size of the price updates, then the $O(n^{2/5})$ upper bounds on both the regret and constraint violation hold. Then, in Section~\ref{apdx:positivity-bdd-prices-variable-step-size}, we modify our earlier developed potential function argument in the proof of Theorem~\ref{thm:PrivacyPreserving} to show that if the distribution $\D$ satisfies Assumption~\ref{asmpn:regularity}, then the price vector $\p^t$ remains strictly positive and bounded throughout the operation of Algorithm~\ref{alg:PrivacyPreserving} with a two-stage adjustment of the step size of the price updates. Note that the above two claims establish Theorem~\ref{thm:PrivacyPreservingImproved}.

\subsection{Regret and Constraint Violation Bound under Positivity and Boundedness of Prices} \label{apdx:regret-bd-variable-step-size-bdd-prices}

To prove this claim, we begin by introducing some notion. In particular, we suppose that the price vector $\p^t$ through the operation of Algorithm~\ref{alg:PrivacyPreserving} with a two-stage adjustment of the price updates is such that $0< \underline{p} \leq p_j^t \leq \Bar{p}$ for all goods $j$ and users $t \in [n]$. Furthermore, we define the $\gamma_e = O(\frac{1}{n^{2/5}})$ as the step size of the price updates in the first stage of Algorithm~\ref{alg:PrivacyPreserving}, $\gamma_p = O(\frac{1}{n^{3/5}})$ as the step size of the price updates in the second stage of Algorithm~\ref{alg:PrivacyPreserving}, and $T_e = n^{4/5}$ as the period at which the step size of Algorithm~\ref{alg:PrivacyPreserving} changes. Moreover, for brevity, we let $T_p = n - T_e$ be the number of periods corresponding to the second stage of Algorithm~\ref{alg:PrivacyPreserving} for which the step size of the price updates is $\gamma_p$. Finally, we let $\x_1, \ldots, \x_n$ be the allocations for the $n$ users under the pricing policy $\ppi$ and let $\p^*$ be the price corresponding to the solution of the stochastic Program~\eqref{eq:stochastic222}. 

Then, we establish our desired regret and constraint violation bounds in three steps. First, we obtain bounds on the regret and constraint violation of Algorithm~\ref{alg:PrivacyPreserving} in terms of the differences in the prices $\norm{\p^{T_e+1} - \p^*}_2$ and $\norm{\p^{n+1} - \p^*}_2$. Then, we present upper bounds on the norm of the differences in the prices $\norm{\p^{T_e+1} - \p^*}_2$ and $\norm{\p^{n+1} - \p^*}_2$. Finally, we utilize the obtained bounds on the norms of the difference in prices to obtain our desired $O(n^{2/5})$ upper bound on the regret and constraint violation.

\paragraph{Regret Upper Bound:} We first derive an upper bound on the regret of Algorithm~\ref{alg:PrivacyPreserving} with a two-stage adjustment of the step size of the price updates. To do so, we first recall from the generic regret bound derived in Equation~\eqref{eq:genericBound2} for any online algorithm that the expected regret:
\begin{align*}
    \mathbb{E}[R_n(\ppi)] &\leq \mathbb{E} \left[ \sum_{t = 1}^n \sum_{j = 1}^m p_j^{t} d_j -  w_t \right] =  \sum_{t = 1}^{T_e} \mathbb{E} \left[ \sum_{j = 1}^m p_j^{t} d_j -  w_t \right] + \sum_{t = T_e+1}^{n} \mathbb{E} \left[ \sum_{j = 1}^m p_j^{t} d_j -  w_t \right] = r_e+r_p,
\end{align*}
where we define $r_e = \sum_{t = 1}^{T_e} \mathbb{E} \left[ \sum_{j = 1}^m p_j^{t} d_j -  w_t \right]$ and $r_p = \sum_{t = T_e+1}^{n} \mathbb{E} \left[ \sum_{j = 1}^m p_j^{t} d_j -  w_t \right]$.

Next, to upper bound the terms $r_e$ and $r_p$, we recall that for any step size $\gamma_t$ of the price updates that:
\begin{align*}
    \norm{\p^{t+1}}_2^2 = \norm{\p^{t} - \gamma_t \left(\d - \x_t \right) }_2^2 = \norm{\p^{t}}_2^2- 2 \gamma_t \left(\d - \x_t \right)^T \p^t + \gamma_t^2 \norm{\d - \x_t}_2^2.
\end{align*}
Rearranging the above equation, we obtain that:
\begin{align*}
    \left(\d - \x_t \right)^T \p^t &= \frac{\norm{\p^{t}}_2^2 - \norm{\p^{t+1}}_2^2}{2\gamma_t} + \frac{\gamma_t}{2} \norm{\d - \x_t}_2^2.
\end{align*}

Next, summing the above equation for $t \in \{1, \ldots, T_e \}$, i.e., for the first-stage of Algorithm~\ref{alg:PrivacyPreserving} for which the step size $\gamma_t = \gamma_e$ for all $t \in [T_e]$, we obtain that:
\begin{align}
    r_e &= \sum_{t = 1}^{T_e} \mathbb{E} \left[ \sum_{j = 1}^m p_j^{t} d_j -  w_t \right] = \sum_{t = 1}^{T_e} \mathbb{E} \left[ (\d-\x_t)^T \p^t \right] \leq \frac{\gamma_e}{2} \sum_{t = 1}^{T_e} \mathbb{E} \left[ \norm{\d - \x_t}_2^2 \right] + \mathbb{E} \left[ \frac{\norm{\p^1}_2^2 - \norm{\p^{T_e+1}}_2^2}{2\gamma_e} \right], \nonumber \\
    &\stackrel{(a)}{\leq} \frac{\gamma_e}{2} T_e m \left( \max_{j \in [m]} d_j + \frac{\Bar{w}}{\underline{p}} \right)^2 + \frac{\norm{\p^1}_2^2}{2\gamma_e}, \label{eq:r_e-ub}
\end{align}
where (a) follows by the boundedness of the consumption vector of each agent since the prices are strictly positive and bounded below by $\underline{p}>0$.

Similarly, we can derive the following upper bound for $r_p$:
\begin{align}
    r_p &= \sum_{t = T_e+1}^{n} \mathbb{E} \left[ \sum_{j = 1}^m p_j^{t} d_j -  w_t \right] \stackrel{(a)}{\leq} \frac{\gamma_p}{2} T_p m \left( \max_{j \in [m]} d_j + \frac{\Bar{w}}{\underline{p}} \right)^2 + \frac{\mathbb{E} \left[ \norm{\p^{T_e+1}}_2^2 - \norm{\p^{n+1}}_2^2 \right]}{2\gamma_p}, \nonumber \\
    &= \frac{\gamma_p}{2} T_p m \left( \max_{j \in [m]} d_j + \frac{\Bar{w}}{\underline{p}} \right)^2 + \frac{\mathbb{E} \left[ (\p^{T_e+1} + \p^{T+1}) \cdot (\p^{T_e+1} - \p^{n+1}) \right]}{2\gamma_p}, \nonumber \\
    &\stackrel{(b)}{\leq} \frac{\gamma_p}{2} T_p m \left( \max_{j \in [m]} d_j + \frac{\Bar{w}}{\underline{p}} \right)^2 + \frac{\mathbb{E} \left[ \norm{\p^{T_e+1} + \p^{T+1}}_2  \norm{\p^{T_e+1} - \p^{n+1}}_2 \right]}{2\gamma_p}, \nonumber \\
    &\stackrel{(c)}{\leq} \frac{\gamma_p}{2} T_p m \left( \max_{j \in [m]} d_j + \frac{\Bar{w}}{\underline{p}} \right)^2 + \frac{\Bar{p} \sqrt{m} \mathbb{E} \left[\norm{\p^{T_e+1} - \p^{n+1}}_2 \right]}{\gamma_p}, \nonumber \\
    &\stackrel{(d)}{\leq} \frac{\gamma_p}{2} T_p m \left( \max_{j \in [m]} d_j + \frac{\Bar{w}}{\underline{p}} \right)^2 + \frac{\Bar{p} \sqrt{m} \mathbb{E} \left[\norm{\p^{T_e+1} - \p^{*}}_2 + \norm{\p^{n+1} - \p^{*}}_2 \right]}{\gamma_p}, \label{eq:r_p-ub}
\end{align}
where (a) follows from an analogous line of reasoning to the earlier analysis for $r_e$, (b) follows from the Cauchy-Schwarz inequality, (c) follows from the fact that the prices are bounded above by $\Bar{p}$, and (d) follows from the triangle inequality where $\p^*$ is the price corresponding to the stochastic Program~\eqref{eq:stochastic222}.

Summing Equations~\eqref{eq:r_e-ub} and~\eqref{eq:r_p-ub}, we have the following upper bound on the regret:
\begin{align}
    \mathbb{E}[R_n(\ppi)] &= r_e + r_p, \nonumber \\
    &\leq \frac{\gamma_e}{2} T_e m \left( \max_{j \in [m]} d_j + \frac{\Bar{w}}{\underline{p}} \right)^2 + \frac{\norm{\p^1}_2^2}{2\gamma_e} \nonumber \\&+ \frac{\gamma_p}{2} T_p m \left( \max_{j \in [m]} d_j + \frac{\Bar{w}}{\underline{p}} \right)^2 + \frac{\Bar{p} \sqrt{m} \mathbb{E} \left[\norm{\p^{T_e+1} - \p^{*}}_2 + \norm{\p^{n+1} - \p^{*}}_2 \right]}{\gamma_p}. \label{eq:rp_re_ub}
\end{align}

\paragraph{Constraint Violation Upper Bound:} Next, we provide an upper bound on the constraint violation of algorithm $\ppi$. In particular, we have that:
\begin{align}
    \mathbb{E} \left[ \norm{ \left( \sum_{t = 1}^n \x_t - \c \right)_+ }_2 \right] &\leq \mathbb{E} \left[ \norm{ \sum_{t = 1}^n (\x_t - \d) }_2 \right] = \mathbb{E} \left[ \norm{ \sum_{t = 1}^{T_e} (\x_t - \d) + \sum_{t = T_e+1}^{T} (\x_t - \d) }_2 \right], \nonumber \\
    &\stackrel{(a)}{\leq} \mathbb{E} \left[ \norm{ \sum_{t = 1}^{T_e} \frac{\p^{t+1}-\p^t}{\gamma_e} + \sum_{t = T_e+1}^{n} \frac{\p^{t+1}-\p^t}{\gamma_p} }_2 \right], \nonumber \\
    &= \mathbb{E} \left[ \norm{ \frac{\p^{T_e+1}-\p^1}{\gamma_e} + \frac{\p^{n+1}-\p^{T_e+1}}{\gamma_p} }_2 \right], \nonumber \\
    &\stackrel{(b)}{\leq} \mathbb{E} \left[ \norm{ \frac{\p^{T_e+1} - \p^1}{\gamma_e}}_2 + \norm{\frac{\p^{n+1}-\p^{T_e+1}}{\gamma_p} }_2 \right], \nonumber \\
    &\leq \frac{1}{\gamma_e} \mathbb{E} \left[ \norm{ \p^{T_e+1} + \p^1 }_2 \right] + \frac{1}{\gamma_p} \mathbb{E} \left[ \norm{ \p^{n+1}-\p^{T_e+1} }_2 \right], \nonumber \\
    &\stackrel{(c)}{\leq} \frac{2 \Bar{p} \sqrt{m}}{\gamma_e} + \frac{1}{\gamma_p} \mathbb{E} \left[ \norm{ \p^{T_e+1}-\p^{*} }_2 + \norm{ \p^{n+1}-\p^{*} }_2 \right], \label{eq:ub-con-viol-twophase}
\end{align}
where (a) follows from the price update rule of Algorithm~\ref{alg:PrivacyPreserving} and (b) and (c) follow by the triangle inequality.

\paragraph{Bound on Regret and Constraint Violation:} Combining the above obtained relations in Equations~\eqref{eq:rp_re_ub} and~\eqref{eq:ub-con-viol-twophase}, we obtain the following bound on the sum of the regret and constraint violation of Algorithm~\ref{alg:PrivacyPreserving} with a two-stage adjustment in the step size:
\begin{align} \label{reg_con-vio-bound-twoPhase}
    \mathbb{E}[R_n(\ppi) + V_n(\ppi)] \leq &\frac{2 \Bar{p} \sqrt{m}}{\gamma_e} + \frac{1}{\gamma_p} \mathbb{E} \left[ \norm{ \p^{T_e+1}-\p^{*} }_2 + \norm{ \p^{n+1}-\p^{*} }_2 \right] \\ &+\frac{\gamma_e}{2} T_e m \left( \max_{j \in [m]} d_j + \frac{\Bar{w}}{\underline{p}} \right)^2 + \frac{\norm{\p^1}_2^2}{2\gamma_e} \nonumber \\&+ \frac{\gamma_p}{2} T_p m \left( \max_{j \in [m]} d_j + \frac{\Bar{w}}{\underline{p}} \right)^2 + \frac{\Bar{p} \sqrt{m} \mathbb{E} \left[\norm{\p^{T_e+1} - \p^{*}}_2 + \norm{\p^{n+1} - \p^{*}}_2 \right]}{\gamma_p}.
\end{align}
In the above upper bound, we combine the terms containing $\gamma_e$ and those containing $\gamma_p$ to obtain two relations:
\begin{align*}
    &W_e = \frac{2 \Bar{p} \sqrt{m}}{\gamma_e} + \frac{\gamma_e}{2} T_e m \left( \max_{j \in [m]} d_j + \frac{\Bar{w}}{\underline{p}} \right)^2 + \frac{\norm{\p^1}_2^2}{2\gamma_e} = \frac{\gamma_e}{2} T_e m \left( \max_{j \in [m]} d_j + \frac{\Bar{w}}{\underline{p}} \right)^2 + \frac{4 \Bar{p} \sqrt{m} + \norm{\p^1}_2^2}{2\gamma_e}. 
\end{align*}
\begin{align*}
    W_p &= 
    \frac{\gamma_p}{2} T_p m \left( \max_{j \in [m]} d_j + \frac{\Bar{w}}{\underline{p}} \right)^2 + \frac{\Bar{p} \sqrt{m} + 1}{\gamma_p} \mathbb{E} \left[\norm{\p^{T_e+1} - \p^{*}}_2 + \norm{\p^{n+1} - \p^{*}}_2 \right].
\end{align*}
From the above relation for $W_e$, note that setting $\gamma_e = \frac{1}{\sqrt{T_e}}$, we obtain that $W_e = O(\sqrt{T_e}) = O(n^{2/5})$. 

\paragraph{Bound on Difference in Prices:} To upper bound $W_p$, we now present bounds on the terms $\mathbb{E} \left[\norm{\p^{T_e+1} - \p^{*}}_2 \right]$ and $ \mathbb{E} \left[ \norm{\p^{n+1} - \p^{*}}_2 \right]$. To this end, consider the following upper bound for the square of the first term:
\begin{align*}
    \mathbb{E} \left[\norm{\p^{T_e+1} - \p^{*}}_2^2 \right] &= \mathbb{E} \left[ \norm{\p^{T_e} - \gamma_e (\d - \x_{T_e}) - \p^{*}}_2^2 \right], \\ 
    &= \mathbb{E}\left[\norm{\p^{T_e} - \p^{*}}_2^2 + \gamma_e^2 \norm{(\d - \x_{T_e})}_2^2 - 2 \gamma_e (\d - \x_{T_e})^T (\p^{T_e} - \p^*) \right], \\
    &\stackrel{(a)}{\leq} \mathbb{E}\left[\norm{\p^{T_e} - \p^{*}}_2^2\right] + \gamma_e^2 m \left( \max_{j \in [m]} d_j + \frac{\Bar{w}}{\underline{p}} \right)^2 - 2 \gamma_e \mathbb{E}\left[(\d - \x_{T_e})^T (\p^{T_e} - \p^*) \right], \\
    &\stackrel{(b)}{\leq} \mathbb{E} \left[\norm{\p^{T_e} - \p^{*}}_2^2 \right] + \gamma_e^2 m \left( \max_{j \in [m]} d_j + \frac{\Bar{w}}{\underline{p}} \right)^2 - 2 \gamma_e \mathbb{E} \left[(D(\p^{T_e}) - D(\p^*)) \right],
\end{align*}
where (a) follows by the boundedness of the consumption vector as the prices are strictly positive and lower bounded by $\underline{p}>0$ through the course of the Algorithm~\ref{alg:PrivacyPreserving} and (b) follows by conditioning on the history and taking the expectation and by the convexity of the dual function $D(\p)$ (see Equation~\eqref{eq:dual-function-restricted} and recall that $\mathbb{E}[\d - \x_{T_e}]$ is a sub-gradient of the dual function given that the price $\p^{T_e}$ for user $T_e$ as noted in Section~\ref{sec:algoMain}).

Next, to upper bound the right hand side of the above term, as in the proof of Theorem~\ref{thm:RegretCapVioAlgo4}, we reparametrize the dual objective by introducing a variable $\alpha_k = \min_{j} \frac{p_j}{\Tilde{u}_{kj}}$ for all utility vectors $\Tilde{\u}_k$ drawn from the discrete distribution $\D$ with finite support. Then, from our earlier analysis in Equations~\eqref{eq:strong-convexity-alpha} and~\eqref{eq:helper-strong-convexity}, we have that:
\begin{align*}
    \mathbb{E}\left[\norm{\p^{T_e+1} - \p^{*}}_2^2 \right] &\stackrel{(c)}{\leq} \mathbb{E}\left[\norm{\p^{T_e} - \p^{*}}_2^2 \right] + \gamma_e^2 m \left( \max_{j \in [m]} d_j + \frac{\Bar{w}}{\underline{p}} \right)^2 - 2 \gamma_e \mathbb{E}\left[ (D(\alphaa^{T_e}) - D(\alphaa^*)) \right], \\ 
    &\stackrel{(d)}{\leq} \mathbb{E}\left[ \norm{\p^{T_e} - \p^{*}}_2^2 \right] + \gamma_e^2 m \left( \max_{j \in [m]} d_j + \frac{\Bar{w}}{\underline{p}} \right)^2 - 2 \gamma_e \eta \mathbb{E}\left[ \norm{ \alphaa^{T_e} - \alphaa^* }_2^2 \right], \\
    &\stackrel{(e)}{\leq } \mathbb{E}\left[ \norm{\p^{T_e} - \p^{*}}_2^2 \right] + \gamma_e^2 m \left( \max_{j \in [m]} d_j + \frac{\Bar{w}}{\underline{p}} \right)^2 - \frac{2 \gamma_e \eta}{\sqrt{m} \Bar{u}}  \mathbb{E}\left[ \norm{ \p^{T_e} - \p^* }_2^2 \right], \\
    &= \left( 1 - \frac{2 \gamma_e \eta}{\sqrt{m} \Bar{u}} \right) \mathbb{E}\left[ \norm{ \p^{T_e} - \p^* }_2^2 \right] + \gamma_e^2 m \left( \max_{j \in [m]} d_j + \frac{\Bar{w}}{\underline{p}} \right)^2, \\
    &\leq \left( 1 - \frac{2 \gamma_e \eta}{\sqrt{m} \Bar{u}} \right)^{T_e} \norm{ \p^{1} - \p^* }_2^2 + \sum_{l = 0}^{T_e-1} \gamma_e^2 m \left( \max_{j \in [m]} d_j + \frac{\Bar{w}}{\underline{p}} \right)^2 \left( 1 - \frac{2 \gamma_e \eta}{\sqrt{m} \Bar{u}} \right)^l,
\end{align*}
where (c) follows from the variable transformation $\alpha_k = \min_j \frac{p_j}{\Tilde{u}_{kj}}$ as in the proof of Lemma~\ref{lem:lipshitzness}, (d) follows from Equation~\eqref{eq:strong-convexity-alpha} and (e) follows from Equation~\eqref{eq:helper-strong-convexity}, derived in our analysis in the proof of Lemma~\ref{lem:lipshitzness} for discrete distributions with finite support. 

Next, defining $\zeta_1 = \frac{2\eta}{\sqrt{m} \Bar{u}}$ and $\zeta_2 = m \left( \max_{j \in [m]} d_j + \frac{\Bar{w}}{\underline{p}} \right)^2$, we have from the above relations that:
\begin{align}
    \mathbb{E}\left[\norm{\p^{T_e+1} - \p^{*}}_2^2 \right] &\leq \left( 1 - \gamma_e \zeta_1 \right)^{T_e} \norm{ \p^{1} - \p^* }_2^2 + \gamma_e^2 \zeta_2 \sum_{l = 0}^{T_e-1}  \left( 1 - \gamma_e \zeta_1 \right)^l, \nonumber \\
    &\stackrel{(a)}{=} \left( 1 - \gamma_e \zeta_1 \right)^{T_e} \norm{ \p^{1} - \p^* }_2^2 + \gamma_e^2 \zeta_2 \frac{1 - (1 - \gamma_e \zeta_1)^{T_e}}{\gamma_e \zeta_1}, \nonumber \\
    &\stackrel{(b)}{\leq} \left( 1 - \gamma_e \zeta_1 \right)^{T_e} \norm{ \p^{1} - \p^* }_2^2 + \frac{\gamma_e \zeta_2 }{\zeta_1}, \nonumber \\
    &\stackrel{(c)}{\leq} \frac{\norm{ \p^{1} - \p^* }_2^2 }{1+\gamma_e \zeta_1 T_e} + \frac{\gamma_e \zeta_2 }{\zeta_1}, \nonumber \\
    &\leq \frac{\norm{ \p^{1} - \p^* }_2^2 }{\gamma_e \zeta_1 T_e} + \frac{\gamma_e \zeta_2 }{\zeta_1}, \label{eq:p_te_p_star_diff}
\end{align}
where (a) follows from the formula of the sum of a geometric series, (b) follows as $1-\gamma_e \zeta_1 \geq 0$, as we can select $\gamma_e = \frac{D_e}{n^{2/5}}$ such that the constant $D_e \leq \frac{1}{\zeta_1}$, and (c) follows as $(1-x)^r \leq \frac{1}{1+rx}$ for $x \in (0, 1)$.

Following the above analysis, we can analogously derive that:
\begin{align} \label{eq:p_T_p_star_diff}
    \mathbb{E}\left[\norm{\p^{n+1} - \p^{*}}_2^2 \right] \leq \frac{\mathbb{E}[\norm{ \p^{T_e+1} - \p^* }_2^2] }{\gamma_p \zeta_1 T_p} + \frac{\gamma_p \zeta_2 }{\zeta_1}.
\end{align}

\paragraph{Bound on $W_p$:} We now use Equations~\eqref{eq:p_te_p_star_diff} and~\eqref{eq:p_T_p_star_diff} to compute the bound for $V_p$. In particular, we get that:
\begin{align*}
    W_p &= \frac{\gamma_p}{2} T_p \zeta_2 + \frac{\Bar{p} \sqrt{m}+1}{\gamma_p} \mathbb{E} \left[\norm{\p^{T_e+1} - \p^{*}}_2 + \norm{\p^{n+1} - \p^{*}}_2 \right], \\
    &\stackrel{(a)}{\leq} \frac{\gamma_p}{2} T_p \zeta_2 + \frac{\Bar{p} \sqrt{m} + 1}{\gamma_p} \left[ \sqrt{\mathbb{E} \left[ \norm{\p^{T_e+1} - \p^{*}}_2^2 \right]} + \sqrt{\mathbb{E} \left[ \norm{\p^{n+1} - \p^{*}}_2^2 \right]} \right] \\
    &\stackrel{(b)}{\leq} \frac{\gamma_p}{2} T_p \zeta_2 + \frac{\Bar{p} \sqrt{m}+1}{\gamma_p} \left( \sqrt{ \frac{\norm{ \p^{1} - \p^* }_2^2 }{\gamma_e \zeta_1 T_e} + \frac{\gamma_e \zeta_2 }{\zeta_1}} + \sqrt{\frac{\mathbb{E}[\norm{ \p^{T_e+1} - \p^* }_2^2] }{\gamma_p \zeta_1 T_p} + \frac{\gamma_p \zeta_2 }{\zeta_1} } \right), \\
    &\stackrel{(c)}{\leq} \frac{\gamma_p}{2} T_p \zeta_2 + \frac{\Bar{p} \sqrt{m}+1}{\gamma_p} \left( \sqrt{ \frac{\norm{ \p^{1} - \p^* }_2^2 }{\gamma_e \zeta_1 T_e}} + \sqrt{\frac{\gamma_e \zeta_2 }{\zeta_1}} + \sqrt{\frac{\mathbb{E}[\norm{ \p^{T_e+1} - \p^* }_2^2] }{\gamma_p \zeta_1 T_p}} + \sqrt{\frac{\gamma_p \zeta_2 }{\zeta_1} } \right), \\
    &\stackrel{(d)}{=} O(\gamma_p T_p + \gamma_p^{-1}(\gamma_e^{-1/2} T_e^{-1/2} + \gamma_p^{-1/2} T_p^{-1/2} + \gamma_e^{1/2} + \gamma_p^{1/2}), \\
    &= O(\gamma_p T_p + \gamma_p^{-1}\gamma_e^{-1/2} T_e^{-1/2} + \gamma_p^{-3/2} T_p^{-1/2} + \gamma_p^{-1} \gamma_e^{1/2} + \gamma_p^{-1/2}), \\
    &\stackrel{(e)}{=} O(n^{2/5}),
\end{align*}
where (a) follows by Jensen's inequality, (b) follows by our derived relations in Equations~\eqref{eq:p_te_p_star_diff} and~\eqref{eq:p_T_p_star_diff}, (c) follows as $\sqrt{a+b} \leq \sqrt{a}+\sqrt{b}$, (d) follows by dropping all the constants, and (e) follows by plugging in the expressions for $\gamma_p, \gamma_e, T_p$, and $T_e$ and simplifying.

\paragraph{$O(n^{2/5})$ Bound:} Thus, we have shown in our above analysis that both $W_e$ and $W_p$ are upper bounded by $O(n^{2/5})$. Consequently, from Equation~\eqref{reg_con-vio-bound-twoPhase} we obtain the following bound on the regret and constraint violation of Algorithm~\ref{alg:PrivacyPreserving} with a two-stage adjustment in the step size of the price updates:
\begin{align*}
    \mathbb{E}[R_n(\ppi) + V_n(\ppi)] \leq W_e + W_p \leq O(n^{2/5}),
\end{align*}
which implies that both the regret and constraint violation of Algorithm~\ref{alg:PrivacyPreserving} with a two-stage adjustment in the step size of the price updates are bounded by $O(n^{2/5})$, establishing our desired result.

\subsection{Positivity and Boundedness of Prices} \label{apdx:positivity-bdd-prices-variable-step-size}

We note that the proof of the positivity and boundedness of prices throughout the operation of Algorithm~\ref{alg:PrivacyPreserving} with a two-stage adjustment in the step size of the price updates follows almost entirely analogously to the proof of Lemma~\ref{lem:PositivePrices} for the case of Algorithm~\ref{alg:PrivacyPreserving} with a fixed step size. To see this, consider the two stages corresponding to the Algorithm~\ref{alg:PrivacyPreserving} when the step size is fixed to $\gamma_e$ and $\gamma_p$, respectively. In this case, note that the step size is fixed until the arrival of the first $T_e$ users. Consequently, following an entirely analogous line of reasoning to the proof of Lemma~\ref{lem:PositivePrices}, the price vectors $\p^1, \p^2, \ldots, \p^{T_e}, \p^{T_e+1}$ remain strictly positive and bounded as the step size is fixed to $\gamma_e$ across the first $T_e$ periods. Next, taking the price vector $\p^{T_e+1}$ as the initial price vector for the second stage, which is strictly positive and bounded as noted above, it also follows from an entirely analogous line of reasoning to the proof of Lemma~\ref{lem:PositivePrices} that the price vectors $\p^{T_e+2}, \p^{T_e+3}, \ldots, \p^{n}$, i.e., the prices in the second stage of Algorithm~\ref{alg:PrivacyPreserving}, remain strictly positive and bounded as the step size is fixed to $\gamma_p$ for the remaining periods. Hence, from an almost entirely analogous line of reasoning to the proof of Lemma~\ref{lem:PositivePrices}, it follows that the prices are strictly positive and bounded throughout the operation of Algorithm~\ref{alg:PrivacyPreserving} with a two-stage adjustment of the step size of the price updates.





\section{Proofs of Results in Section~\ref{sec:feasible-algo-design}} \label{apdx:nocapvioPfs}

\subsection{Proof of Theorem~\ref{thm:general-regret-stoppingtime}} \label{apdx:pf-thm-no-capvio-stoppintime}

To prove this result, we begin by defining $U_{\tau^{\ppi}+1:n}$ as the budget weighted logarithmic utility objective ranging from the periods $\tau^{\ppi}+1$ to $n$. Then, the expected regret for algorithm $\ppi^f$ can be expressed as follows:
\begin{align} 
    R_n(\ppi^f) &= \mathbb{E}[U_{n}^* - U_{n}(\ppi^f)], \nonumber \\
    &= \mathbb{E}[U_{\tau^{\ppi}}^* - U_{\tau^{\ppi}}(\ppi^f)] + \mathbb{E}[U_{\tau^{\ppi}+1:n}^* - U_{\tau^{\ppi}+1:n}(\ppi^f)], \nonumber \\
    &= R_{\tau^{\ppi}}(\ppi) + \mathbb{E}[U_{\tau^{\ppi}+1:n}^* - U_{\tau^{\ppi}+1:n}(\ppi^f)], \label{eq:fullregretformula}
\end{align}
where the final equality follows as the algorithms $\ppi$ and $\ppi^f$ are identical up until period $\tau^{\ppi}$. Then, to establish the desired upper bound on the regret, we now bound second term on the right hand side of the above equation. 

\paragraph{Regret from period $\tau^{\ppi}+1$ to $n$:} 

Recall that users from periods $\tau^{\ppi}+1$ through $n$ are provided $\frac{\epsilon}{n}$ units of each resource. Thus, it holds that
\begin{align}
    \mathbb{E}[U_{\tau^{\ppi}+1:n}^* - U_{\tau^{\ppi}+1:n}(\ppi^f)] &= \mathbb{E} \left[\sum_{t = \tau^{\ppi}+1}^n w_t \log(\sum_{j = 1}^m u_{tj} x_{tj}^*) \right] - \mathbb{E} \left[\sum_{t = \tau^{\ppi}+1}^n w_t \log(\sum_{j = 1}^m u_{tj} x_{tj}) \right], \nonumber \\
    &\stackrel{(a)}{\leq} \mathbb{E}[(n -\tau^{\ppi})] \Bar{w} \log\left(\sum_{j = 1}^m \Bar{u} \frac{\Bar{w}}{\underline{p}}\right) - \mathbb{E}[(n -\tau^{\ppi})] \left[ \underline{w} \log\left( \underline{u} \frac{\epsilon}{n} \right) \right], \nonumber \\
    &= \mathbb{E}[(n - \tau^{\ppi})] \left[ \Bar{w} \log\left(m \Bar{u} \frac{\Bar{w}}{\underline{p}}\right) - \underline{w} \log\left( \underline{u} \epsilon \right) + \underline{w} \log(n) \right], \nonumber  \\
    &= O(\mathbb{E}[(n - \tau^{\ppi})] \log(n)), \label{eq:tailRegret}
\end{align}
where (a) follows as $\Bar{w}$, $\underline{w}$, $\Bar{u}$, $\underline{u}$ are strictly positive upper and lower bound on the budgets and utilities, respectively, and the fact that all users have at least one resource for which $u_{tj}\geq \underline{u}>0$. Furthermore, $\underline{p}>0$ corresponds to a lower bound on the optimal price at any period, which holds under Assumption~\ref{asmpn:mainRestriction}.

\paragraph{Final Regret Bound for Algorithms~\ref{alg:AlgoProbKnownDiscrete} and~\ref{alg:PrivacyPreserving}:} Combining the above obtained relationship for the second component of the regret on the right hand side of Equation~\eqref{eq:fullregretformula}, we obtain from Equation~\eqref{eq:tailRegret} the following upper bound on the regret:
\begin{align*}
    R_n(\ppi^f) &= R_{\tau^{\ppi}}(\ppi) + \mathbb{E}[U_{\tau^{\ppi}+1:n}^* - U_{\tau^{\ppi}+1:n}(\ppi^f)], \\
    &\leq R_{\tau^{\ppi}}(\ppi) + O(\mathbb{E}[(n - \tau^{\ppi})] \log(n)),
\end{align*}
which establishes our result.

\subsection{Proof of Corollary~\ref{cor:noCapVioAlgo2}} \label{apdx:pf-cor-no-capVioAlgo2}



In this proof, for brevity, we use Algorithm~\ref{alg:PrivacyPreserving} to refer to the revealed preference algorithm with a fixed step size of $O(\frac{1}{\sqrt{n}})$.

To prove our desired regret bound, we first note from Theorem~\ref{thm:general-regret-stoppingtime} that we have the following upper bound on the regret of the feasible variant $\ppi^f$ of Algorithm~\ref{alg:PrivacyPreserving}:
\begin{align} \label{eq:generic-bound-pf-cor}
    R_n(\ppi^f) &\leq R_{\tau^{\ppi}}(\ppi) + O(\mathbb{E}[(n - \tau^{\ppi})] \log(n)),
\end{align}
where $\ppi$ corresponds to Algorithm~\ref{alg:PrivacyPreserving}. To bound the terms on the right hand side of the above equation, first note by the analysis in the proof of Theorem~\ref{thm:PrivacyPreserving} that $R_{\tau^{\ppi}}(\ppi) \leq O(\sqrt{\tau^{\ppi}})$. Next, we obtain an upper bound on the second term on the right hand side of the above equation by obtaining a bound on $\mathbb{E}[(n - \tau^{\ppi})]$.

\paragraph{Upper Bound on $\mathbb{E}[(n - \tau^{\ppi})]$:}
To bound $\mathbb{E}[(n - \tau^{\ppi})]$, first note by the definition of $\tau^{\ppi}$ that there is some good $j \in [m]$ for which the following inequality holds:
\begin{align*}
    \sum_{t = 1}^{\tau^{\ppi}} x_{tj} + \frac{\Bar{w}}{\underline{p}} \geq c_j = n d_j,
\end{align*}
where $\underline{p}>0$ corresponds to a lower bound on the price in the update step of Algorithm~\ref{alg:PrivacyPreserving} (see Lemma~\ref{lem:PositivePrices}). Subtracting $\tau^{\ppi} d_j$ to both sides of the above inequality and rearranging, we obtain the following upper bound on $n - \tau^{\ppi}$:
\begin{align}
    \tau^{\ppi} d_j - \sum_{t = 1}^{\tau^{\ppi}} x_{tj} \leq \tau^{\ppi} d_j - d_j n + \frac{\Bar{w}}{\underline{p}}, \nonumber \\
    \implies (n-\tau^{\ppi}) \leq \frac{\frac{\Bar{w}}{\underline{p}} -(\tau^{\ppi} d_j - \sum_{t = 1}^{\tau^{\ppi}} x_{tj}) }{d_j} \label{eq:tausBound}
\end{align}

Next, to bound the right hand side of the above equation, we note by the price update rule in Algorithm~\ref{alg:PrivacyPreserving} that:
\begin{align*}
    \sum_{t = 1}^{\tau^{\ppi}} -(d_j - x_{tj}) &= \frac{1}{\gamma} \sum_{t = 1}^{\tau^{\ppi}} (p_j^{t+1} - p_j^t) = \frac{1}{\gamma} (p_j^{\tau^{\ppi}+1} - p_j^1) \leq \frac{1}{\gamma}(\Bar{p} - p_j^1),
\end{align*}
where recall that $\Bar{p}$ represents an upper bound on the price at any stage of Algorithm~\ref{alg:PrivacyPreserving} as established in the proof of Theorem~\ref{thm:PrivacyPreserving}. Then, combining the above inequality with Equation~\eqref{eq:tausBound}, we obtain that:
\begin{align} \label{eq:second-term-logn-guarantee}
    n - \tau^{\ppi} \leq \frac{\frac{\Bar{w}}{\underline{p}} + \frac{1}{\gamma}(\Bar{p} - p_j^1)}{d_j} = O(\sqrt{n}).
\end{align}

\paragraph{Final Regret Bound:} Combining the above obtained relationship for $\mathbb{E}[(n - \tau^{\ppi})]$, we obtain the following upper bound on the regret of $\ppi^f$:
\begin{align*}
    R_n(\ppi^f) &\leq R_{\tau^{\ppi}}(\ppi) + O(\mathbb{E}[(n - \tau^{\ppi})] \log(n)) \stackrel{(a)}{\leq} O(\sqrt{\tau^{\ppi}}) + O(\sqrt{n} \log(n)), \\
    &\stackrel{(b)}{\leq} O(\sqrt{n}) +  O(\sqrt{n} \log(n)) = O(\sqrt{n} \log(n)),
\end{align*}
where (a) follows from Equation~\eqref{eq:second-term-logn-guarantee} and the fact that $R_{\tau^{\ppi}}(\ppi) \leq O(\sqrt{\tau^{\ppi}})$ from the analysis in Theorem~\ref{thm:PrivacyPreserving}. Furthermore, (b) follows from the fact that $\tau^{\ppi} \leq n$. The above inequality establishes our desired result that $R_n(\ppi^f) \leq O(\sqrt{n} \log(n))$.

\subsection{Proof of Corollary~\ref{cor:noCapVioAlgo2-variable}} \label{apdx:pf-cor-no-capVioAlgo2-variable}

In this proof, for brevity, we use Algorithm~\ref{alg:PrivacyPreserving} to refer to the revealed preference algorithm with a two-stage adjustment of the step size.

We first note From Theorem~\ref{thm:general-regret-stoppingtime} that we have the following upper bound on the regret of the feasible variant $\ppi^f$ of Algorithm~\ref{alg:PrivacyPreserving}:
\begin{align} \label{eq:generic-bound-pf-cor}
    R_n(\ppi^f) &\leq R_{\tau^{\ppi}}(\ppi) + O(\mathbb{E}[(n - \tau^{\ppi})] \log(n)),
\end{align}
where $\ppi$ corresponds to Algorithm~\ref{alg:PrivacyPreserving}. To bound the terms on the right hand side of the above equation, first note by the analysis in the proof of Theorem~\ref{thm:PrivacyPreservingImproved} that $R_{\tau^{\ppi}}(\ppi) \leq O((\tau^{\ppi})^{2/5})$. Next, we obtain an upper bound on the second term on the right hand side of the above equation by obtaining a bound on $\mathbb{E}[(n - \tau^{\ppi})]$.

\paragraph{Upper Bound on $\mathbb{E}[(n - \tau^{\ppi})]$:}
To bound $\mathbb{E}[(n - \tau^{\ppi})]$, first note by the definition of $\tau^{\ppi}$ that there is some good $j \in [m]$ for which the following inequality holds:
\begin{align} \label{eq:tau-pi-inequality-basic-variable}
    \sum_{t = 1}^{\tau^{\ppi}} x_{tj} + \frac{\Bar{w}}{\underline{p}} \geq c_j = n d_j,
\end{align}
where $\underline{p}>0$ corresponds to a lower bound on the price in the update step of Algorithm~\ref{alg:PrivacyPreserving} (see Lemma~\ref{lem:PositivePrices}). The above inequality implies that:
\begin{align*}
    \frac{\tau^{\ppi} \Bar{w}}{\underline{p}} + \frac{\Bar{w}}{\underline{p}} \geq n d_j \implies \tau^{\ppi} \geq n \frac{d_j \underline{p}}{\Bar{w}} - 1.
\end{align*}
Then, under the condition that $n \geq \max \{ 1, \left( \frac{2 \Bar{w}}{\underline{p} \min_j d_j} \right)^5 \}$, using the above inequality, it is straightforward to check that $\tau^{\ppi} - T_e = \tau^{\ppi} - n^{4/5} \geq  n \frac{d_j \underline{p}}{\Bar{w}} - 1 - n^{4/5} \geq 0$, i.e., $\tau^{\ppi} - T_e \geq \Omega(n)$ for all $n \geq \max \{ 1, \left( \frac{2 \Bar{w}}{\underline{p} \min_j d_j} \right)^5 \}$, where recall from the proof of Theorem~\ref{thm:PrivacyPreservingImproved} that $T_e = n^{4/5}$ represents the period at which the step size of Algorithm~\ref{alg:PrivacyPreserving} is adjusted.


Next, subtracting $\tau^{\ppi} d_j$ on both sides of the inequality in Equation~\eqref{eq:tau-pi-inequality-basic-variable} and rearranging, we obtain the following upper bound on $n - \tau^{\ppi}$:
\begin{align}
    \tau^{\ppi} d_j - \sum_{t = 1}^{\tau^{\ppi}} x_{tj} \leq \tau^{\ppi} d_j - d_j T + \frac{\Bar{w}}{\underline{p}}, \nonumber \\
    \implies (n-\tau^{\ppi}) \leq \frac{\frac{\Bar{w}}{\underline{p}} -(\tau^{\ppi} d_j - \sum_{t = 1}^{\tau^{\ppi}} x_{tj}) }{d_j} \label{eq:tausBound-variable}
\end{align}

Next, to bound the right hand side of the above equation, we note by the price update rule in Algorithm~\ref{alg:PrivacyPreserving} that:
\begin{align}
    \mathbb{E} \left[\sum_{t = 1}^{\tau^{\ppi}} -(d_j - x_{tj}) \right] &= \mathbb{E} \left[ \frac{1}{\gamma_e} \sum_{t = 1}^{T_e} (p_j^{t+1} - p_j^t) + \frac{1}{\gamma_p} \sum_{t = T_e+1}^{\tau^{\ppi}} (p_j^{t+1} - p_j^t) \right], \nonumber \\
    &= \frac{1}{\gamma_e} \mathbb{E} \left[ (p_j^{T_e+1} - p_j^1) \right] + \frac{1}{\gamma_p} \mathbb{E} \left[ (p_j^{\tau^{\ppi}+1} - p_j^{T_e+1}) \right], \nonumber \\
    &\stackrel{(a)}{\leq} \frac{\Bar{p}}{\gamma_e} + \frac{1}{\gamma_p} \mathbb{E} \left[ |p_j^{\tau^{\ppi}+1} - p_j^{T_e+1}| \right], \nonumber \\
    &\leq \frac{\Bar{p}}{\gamma_e} + \frac{1}{\gamma_p} \mathbb{E} \left[ \norm{\p^{\tau^{\ppi}+1} - \p^{T_e+1}}_2 \right], \nonumber \\
    &\stackrel{(b)}{\leq} \frac{\Bar{p}}{\gamma_e} + \frac{1}{\gamma_p} \left[ \mathbb{E} \left[\norm{\p^{\tau^{\ppi}+1} - \p^{*}}_2 \right] + \mathbb{E} \left[\norm{\p^{T_e+1} - \p^{*}}_2 \right] \right] , \label{eq:helper-two-phase-1-variable}
\end{align}
where (a) follows as $\Bar{p}$ represents an upper bound on the price at any stage of Algorithm~\ref{alg:PrivacyPreserving} and (b) follows by the triangle inequality. 

From our earlier analysis (see Equations~\eqref{eq:p_te_p_star_diff} and~\eqref{eq:p_T_p_star_diff}) in the proof of Theorem~\ref{thm:PrivacyPreservingImproved}, we have that the terms $\mathbb{E} \left[ \norm{\p^{\tau^{\ppi}+1} - \p^{*}}_2 \right]$ and $\mathbb{E} \left[ \norm{\p^{T_e+1} - \p^{*}}_2 \right]$ on the right hand side of the above inequality can be bounded as follows:
\begin{align} \label{eq:p_T_ppi_star_diff2-variable}
    \mathbb{E} \left[\norm{\p^{T_e+1} - \p^{*}}_2^2 \right] \leq \frac{\norm{ \p^{1} - \p^* }_2^2 }{\gamma_e \zeta_1 T_e} + \frac{\gamma_e \zeta_2 }{\zeta_1}
\end{align}
\begin{align} \label{eq:p_T_ppi_star_diff3-variable}
    \mathbb{E}\left[\norm{\p^{\tau^{\ppi}} - \p^{*}}_2^2 \right] \leq \frac{\mathbb{E}\left[\norm{ \p^{T_e+1} - \p^* }_2^2 \right] }{\gamma_p \zeta_1 (\tau^{\ppi} - T_e)} + \frac{\gamma_p \zeta_2 }{\zeta_1},
\end{align}
where recall that $\zeta_1, \zeta_2$ are constants, $\gamma_e = O(\frac{1}{n^{2/5}})$ is the step size of the price updates in the first stage of Algorithm~\ref{alg:PrivacyPreserving}, $\gamma_p = O(\frac{1}{n^{3/5}})$ is the step size of the price updates in the second stage of Algorithm~\ref{alg:PrivacyPreserving}, and $T_e$ is the period at which the step size of Algorithm~\ref{alg:PrivacyPreserving} changes.

Finally, plugging in Equations~\eqref{eq:p_T_ppi_star_diff2-variable} and~\eqref{eq:p_T_ppi_star_diff3-variable} into the right hand side of Equation~\eqref{eq:helper-two-phase-1-variable}, we get:
\begin{align*}
    \mathbb{E} \left[\sum_{t = 1}^{\tau^{\ppi}} -(d_j - x_{tj}) \right] &\leq \frac{\Bar{p}}{\gamma_e} + \frac{1}{\gamma_p} \left[ \mathbb{E} \left[\norm{\p^{\tau^{\ppi}+1} - \p^{*}}_2 \right] + \mathbb{E} \left[\norm{\p^{T_e+1} - \p^{*}}_2 \right] \right], \\
    &\stackrel{(a)}{\leq} \frac{\Bar{p}}{\gamma_e} + \frac{1}{\gamma_p} \left[ \sqrt{\frac{\norm{ \p^{1} - \p^* }_2^2 }{\gamma_e \zeta_1 T_e} + \frac{\gamma_e \zeta_2 }{\zeta_1}} + \sqrt{\frac{\mathbb{E} \left[\norm{ \p^{T_e+1} - \p^* }_2^2 \right] }{\gamma_p \zeta_1 (\tau^{\ppi} - T_e)} + \frac{\gamma_p \zeta_2 }{\zeta_1}} \right], \\
    &\stackrel{(b)}{\leq} \frac{\Bar{p}}{\gamma_e} + \frac{1}{\gamma_p} \left[ \sqrt{\frac{\norm{ \p^{1} - \p^* }_2^2 }{\gamma_e \zeta_1 T_e}} + \sqrt{ \frac{\gamma_e \zeta_2 }{\zeta_1}} + \sqrt{\frac{ \mathbb{E} \left[ \norm{ \p^{T_e+1} - \p^* }_2^2 \right] }{\gamma_p \zeta_1 (\tau^{\ppi} - T_e)}} + \sqrt{\frac{\gamma_p \zeta_2 }{\zeta_1}} \right], \\
    &\stackrel{(c)}{=} O(\gamma_e^{-1} + \gamma_p^{-1}(\gamma_e^{-1/2} T_e^{-1/2} + \gamma_p^{-1/2} (\tau^{\ppi} - T_e)^{-1/2} + \gamma_e^{1/2} + \gamma_p^{1/2}), \\
    &\stackrel{(d)}{=} O(\gamma_e^{-1} + \gamma_p^{-1}\gamma_e^{-1/2} T_e^{-1/2} + \gamma_p^{-3/2} n^{-1/2} + \gamma_p^{-1} \gamma_e^{1/2} + \gamma_p^{-1/2}), \\
    &\stackrel{(e)}{=} O(n^{2/5}),
\end{align*}
where (a) follows by Cauchy-Schwarz inequality and plugging in Equations~\eqref{eq:p_T_ppi_star_diff2-variable} and~\eqref{eq:p_T_ppi_star_diff3-variable} into the right hand side of Equation~\eqref{eq:helper-two-phase-1-variable}, (b) follows as $\sqrt{a+b} \leq \sqrt{a} + \sqrt{b}$, and (c) follows by dropping the constants and only retaining the terms that depend on the number of users $n$. Moreover, (d) follows as $\tau^{\ppi} - T_e = O(n)$, as $\tau^{\ppi} - T_e \geq \Omega(n)$ as noted earlier when $n \geq \max \{ 1, \left( \frac{2 \Bar{w}}{\underline{p} \min_j d_j} \right)^5 \}$
and $T_e = O(n^{4/5})$. Finally, (e) follows by plugging in the expressions for $\gamma_p$, $\gamma_e$, and $T_e$ (see proof of Theorem~\ref{thm:PrivacyPreservingImproved}) and simplifying.

Then, combining the above inequality with Equation~\eqref{eq:tausBound-variable}, we obtain that:
\begin{align} \label{eq:second-term-logn-guarantee-variable}
    \mathbb{E} \left[n - \tau^{\ppi} \right] \leq \frac{\frac{\Bar{w}}{\underline{p}} -\left(\mathbb{E} \left[\tau^{\ppi} d_j - \sum_{t = 1}^{\tau^{\ppi}} x_{tj} \right] \right) }{d_j} = O(n^{2/5}).
\end{align}

\paragraph{Final Regret Bound:} Combining the above obtained relationship for $\mathbb{E}[(n - \tau^{\ppi})]$, we obtain the following upper bound on the regret of $\ppi^f$:
\begin{align*}
    R_n(\ppi^f) &\leq R_{\tau^{\ppi}}(\ppi) + O(\mathbb{E}[(n - \tau^{\ppi})] \log(n)) \stackrel{(a)}{\leq} O((\tau^{\ppi})^{2/5}) + O(n^{2/5} \log(n)), \\
    &\stackrel{(b)}{\leq} O(n^{2/5}) +  O(n^{2/5} \log(n)) = O(n^{2/5} \log(n)),
\end{align*}
where (a) follows from Equation~\eqref{eq:second-term-logn-guarantee-variable} and the fact that $R_{\tau^{\ppi}}(\ppi) \leq O((\tau^{\ppi})^{2/5})$ from the analysis in Theorem~\ref{thm:PrivacyPreservingImproved}. Furthermore, (b) follows from the fact that $\tau^{\ppi} \leq n$. The above inequality establishes our desired result that $R_n(\ppi^f) \leq O(n^{2/5} \log(n))$.

\subsection{Proof of Corollary~\ref{cor:noCapVioAlgo1}} \label{apdx:pf-cocapVioAlgo1}

From Theorem~\ref{thm:general-regret-stoppingtime}, we have the following upper bound on the regret of the feasible variant of Algorithm~\ref{alg:AlgoProbKnownDiscrete}:
\begin{align} \label{eq:generic-bound-pf-cor2}
    R_n(\ppi^f) &\leq R_{\tau^{\ppi}}(\ppi) + O(\mathbb{E}[(n - \tau^{\ppi})] \log(n)),
\end{align}
where $\ppi$ corresponds to Algorithm~\ref{alg:AlgoProbKnownDiscrete}. To bound the terms on the right hand side of the equation, we first recall that $O(\mathbb{E}[(n - \tau^{\ppi})]$ is constant for Algorithm~\ref{alg:AlgoProbKnownDiscrete} (see Equation~\eqref{eq:nMinusTauRelation} in the proof of Lemma~\ref{lem:capVioAlgo4}). Next, following our analysis in the proof of Theorem~\ref{thm:RegretCapVioAlgo4}, note that $R_{\tau^{\ppi}}(\ppi) \leq O(\log(\tau^{\ppi}))$. Thus, we get the following upper bound on the regret of $\ppi^f$:
\begin{align*}
    R_n(\ppi^f) &\leq R_{\tau^{\ppi}}(\ppi) + O(\mathbb{E}[(n - \tau^{\ppi})] \log(n)) \stackrel{(a)}{\leq} O(\log(\tau^{\ppi})) + O(\log(n)) \stackrel{(b)}{\leq} O(\log(n)),
\end{align*}
where (a) follows as $\mathbb{E}[(n - \tau^{\ppi})]$ is a constant for Algorithm~\ref{alg:AlgoProbKnownDiscrete} as noted above, and (b) follows as $\tau^{\ppi} \leq n$. The above inequality establishes our desired result that $R_n(\ppi^f) \leq O(\log(n))$.

\section{Additional Details on Benchmarks} \label{apdx:numericalBenchmarks}



In this section, we provide more details on two of the benchmarks (i.e., the \emph{Stochastic Program} and \emph{Dynamic Learning SAA} benchmarks) to which we compare our revealed preference algorithms in our experiments. Both these benchmarks are akin to several classical algorithms developed in the online resource allocation literature~\citep{li2021online,online-agrawal} and assume access to additional information on users' utility and budget parameters. 
In particular, the first benchmark assumes knowledge of the distribution $\D$ from which the budget and utility parameters are drawn, as is the case for an algorithm that sets expected equilibrium prices. The second benchmark assumes that users' utility and budget parameters are revealed to the central planner when they enter the market and can be used to set prices for subsequent users. We mention that these algorithms are solely for benchmark purposes, and thus we do not discuss the practicality of the corresponding informational assumptions of these benchmarks. We also reiterate that, as opposed to these benchmarks, the price updates in Algorithm~\ref{alg:PrivacyPreserving} only rely on users' revealed preferences rather than relying on additional information on their budget and utility parameters.

\paragraph{Stochastic Program:}

We begin with the benchmark wherein the distribution $\D$ from which the budget and utility parameters are generated i.i.d. is known. In this case, the SAA Problem~\eqref{eq:SAA2} is related to the following stochastic program
\begin{equation} \label{eq:stochastic2}
\begin{aligned}
\min_{\mathbf{p}} \quad & D(\p) = \sum_{j = 1}^m p_j d_j + \mathbb{E}_{(w,\u) \sim \D }\left[\left(w \log \left(w\right)-w \log \left(\min _{j \in [m]} \frac{p_{j}}{u_{j}}\right)-w\right)\right],
\end{aligned}
\end{equation}
which can be solved to give an optimal price vector $\p^*$. Note that this price vector $\p^*$ corresponds to the static expected equilibrium price, as it takes an expectation over the distribution $\D$. The corresponding pricing policy $\ppi$ only depends on the distribution $\D$ is thus given by $\p^t = \p^* = \pi_t(\D)$ for all users $t \in [n]$. Given the price vector $\p^*$, all arriving users will purchase an affordable utility-maximizing bundle of goods by solving their individual optimization Problem~\eqref{eq:Fisher1}-\eqref{eq:Fishercon3}. Note here that the price vector $\p^*$ is computed before the online procedure, which is possible due to the prior knowledge of the distribution $\D$. For numerical implementation purposes, we consider a sample average approximation to compute the expectation in Problem~\eqref{eq:stochastic2}, as elucidated in \ifarxiv Section~\ref{sec:experimental-setup-details}\fi.

\paragraph{Dynamic Learning using SAA:}

In this benchmark, we consider the setting wherein users' budget and utility parameters are revealed to the central planner each time a user arrives. In this context, the prices are set based on the dual variables of the capacity constraints of the sampled Eisenberg-Gale program with the observed budget and utility parameters of agents that have previously arrived. That is, the pricing policy $\ppi$ depends on the history of users' budget and utility parameters, i.e., $\p^t = \pi_t((w_{t'}, \u_{t'})_{t' = 1}^{t-1})$. We note that to improve on the computational complexity, we update the dual prices at geometric intervals, as in earlier work~\citep{li2021online,online-agrawal}. Users arriving in each interval observe the corresponding price vector for that interval and solve their individual optimization problems to obtain their most favorable goods under the set prices. This process is presented formally in Algorithm~\ref{alg:ParamsRevealed}.

\begin{algorithm}
\SetAlgoLined
\SetKwInOut{Input}{Input}\SetKwInOut{Output}{Output}
\Input{Vector of Capacities $\c$}
Set $\delta \in (1, 2]$ and $L>0$ such that $\floor*{\delta^L} = n$ \;
Let $t_{k}=\left\lfloor\delta^{k}\right\rfloor, k=1,2, \ldots, L-1$ and $t_{L}=n+1$ \;
Initialize $\p^{t_1} > \0$ \;
Each user $t \in [t_1]$ purchases a bundle of goods $\x_t$ by solving Problem~\eqref{eq:Fisher1}-\eqref{eq:Fishercon3} given the price $\p^{t_1}$ \;
 \For{$k = 1, \ldots, L-1$}{
    \textbf{Phase I: Set Price for Geometric Interval} \\
    Set price $\p^{t_k}$ based on dual variables of the capacity constraints of the sampled social optimization problem:
    \begin{maxi*}|s|[2]                   
    {\mathbf{x}_t \in \mathbb{R}^m, \forall t \in [t_k]}                               
    {U(\mathbf{x}_1, ..., \mathbf{x}_{t_k}) = \sum_{t = 1}^{t_k} w_t \log \left( \sum_{j = 1}^m u_{tj} x_{tj} \right) , }   
    {}             
    {}                                
    \addConstraint{\sum_{t = 1}^{t_k} x_{tj}}{ \leq \frac{t_k}{n} c_j, \quad \forall j \in [m], }    
    \addConstraint{x_{tj}}{\geq 0, \quad \forall t \in [t_k], j \in [m], }  
    \end{maxi*}
    \textbf{Phase II: Each User in Interval Consumes Optimal Bundle} \\
    Each user $t \in \{ t_k+1, \ldots, t_{k+1} \}$ purchases an optimal bundle of goods $\x_t$ by solving Problem~\eqref{eq:Fisher1}-\eqref{eq:Fishercon3} given the price $\p^{t_k}$ \;
  }
\caption{Dynamic Learning SAA Algorithm}
\label{alg:ParamsRevealed}
\end{algorithm}

\section{Additional Numerical Experiments}

\subsection{Numerical Validation of Lipschitzness Relation} \label{apdx:priceStabilization}

In this section, we present the results of a numerical experiment to validate the Lipschitzness relation established in Lemma~\ref{lem:lipshitzness}. In particular, we consider the instance described in the proof of Theorem~\ref{thm:lbStatic} with $n = 10,000$ users, where all users have a fixed budget of one, and two goods, each with a capacity of $c_j = n = 10,000$. The utility parameters of users are drawn i.i.d. from a distribution $\D$, where users have an equal 0.5 probability of having the utility (1, 0) or (0, 1). 

Figure~\ref{fig:priceStability} depicts the change in the dual prices of the certainty equivalent problem between subsequent iterations of Algorithm~\ref{alg:AlgoProbKnownDiscrete} for this instance. To see that the Lipschitzness relation is satisfied, first note that the norm of the difference between the average remaining resource capacities between subsequent time steps is $O(\frac{1}{n-t})$, i.e., $\norm{\d_{t+1} - \d_t}_2 \leq O(\frac{1}{n-t})$ as $\d_{t+1} = \d_t + \frac{\d_t - \x_k(\p_t)}{n-t}$. Then, Figure~\ref{fig:priceStability} implies that the two norm of the change in the dual prices of the certainty equivalent problem, i.e., $\mathbb{E}[\norm{\p^{t+1} - \p^t}]$ is always upper bounded by $O(\frac{1}{n-t})$ for all $t \in [n-1]$, which thus implies that the obtained Lipschitzness relation in Lemma~\ref{lem:lipshitzness} is satisfied. We note that we present the results on a log plot for readability purposes.


\begin{figure}[tbh!]
    \centering
%
%
\definecolor{mycolor3}{rgb}{0.00000,0.44700,0.74100}%
\definecolor{mycolor2}{rgb}{1.0, 0.88, 0.21}
\definecolor{mycolor1}{rgb}{0.89, 0.0, 0.13}
\begin{tikzpicture}

\begin{axis}[%
width=2in,
height=2in,
legend style={
legend cell align=left, align=left,
  fill opacity=0.8,
  draw opacity=1,
  text opacity=1,
  at={(0.02,0.95)},
  anchor=north west,
  draw=white!80!black
},
scale only axis,
xmin=0,
xmax=11000,
xlabel style={font=\color{white!15!black}},
xlabel={Number of Users (t)},
ymin=-10,
ymax=2,
legend style={
legend cell align=left, align=left,
  fill opacity=0.8,
  draw opacity=1,
  text opacity=1,
  anchor=north west,
  at = {(0.07, 0.87)},
  draw=white!80!black
},
xtick=\empty,
    extra x ticks={0,2000, 4000, 6000, 8000, 10000},
    extra x tick labels={0, 2, 4, 6, 8, 10},
ylabel style={font=\color{white!15!black}, align=center},
ylabel={$\log(\mathbb{E}[\norm{\p^{t+1} - \p^t}])$},
axis background/.style={fill=white}
]
\addplot[color=mycolor3, line width=1.5pt] coordinates {
    (0.00, -6.21)
    (50.00, -6.21)
    (100.00, -6.20)
    (150.00, -6.20)
    (200.00, -6.19)
    (250.00, -6.19)
    (300.00, -6.18)
    (350.00, -6.18)
    (400.00, -6.17)
    (450.00, -6.17)
    (500.00, -6.16)
    (550.00, -6.16)
    (600.00, -6.15)
    (650.00, -6.15)
    (700.00, -6.14)
    (750.00, -6.14)
    (800.00, -6.13)
    (850.00, -6.13)
    (900.00, -6.12)
    (950.00, -6.11)
    (1000.00, -6.11)
    (1050.00, -6.10)
    (1100.00, -6.10)
    (1150.00, -6.09)
    (1200.00, -6.09)
    (1250.00, -6.08)
    (1300.00, -6.08)
    (1350.00, -6.07)
    (1400.00, -6.06)
    (1450.00, -6.06)
    (1500.00, -6.05)
    (1550.00, -6.05)
    (1600.00, -6.04)
    (1650.00, -6.03)
    (1700.00, -6.03)
    (1750.00, -6.02)
    (1800.00, -6.02)
    (1850.00, -6.01)
    (1900.00, -6.00)
    (1950.00, -6.00)
    (2000.00, -5.99)
    (2050.00, -5.99)
    (2100.00, -5.98)
    (2150.00, -5.97)
    (2200.00, -5.97)
    (2250.00, -5.96)
    (2300.00, -5.95)
    (2350.00, -5.95)
    (2400.00, -5.94)
    (2450.00, -5.93)
    (2500.00, -5.93)
    (2550.00, -5.92)
    (2600.00, -5.91)
    (2650.00, -5.91)
    (2700.00, -5.90)
    (2750.00, -5.89)
    (2800.00, -5.89)
    (2850.00, -5.88)
    (2900.00, -5.87)
    (2950.00, -5.86)
    (3000.00, -5.86)
    (3050.00, -5.85)
    (3100.00, -5.84)
    (3150.00, -5.84)
    (3200.00, -5.83)
    (3250.00, -5.82)
    (3300.00, -5.81)
    (3350.00, -5.81)
    (3400.00, -5.80)
    (3450.00, -5.79)
    (3500.00, -5.78)
    (3550.00, -5.78)
    (3600.00, -5.77)
    (3650.00, -5.76)
    (3700.00, -5.75)
    (3750.00, -5.74)
    (3800.00, -5.74)
    (3850.00, -5.73)
    (3900.00, -5.72)
    (3950.00, -5.71)
    (4000.00, -5.70)
    (4050.00, -5.70)
    (4100.00, -5.69)
    (4150.00, -5.68)
    (4200.00, -5.67)
    (4250.00, -5.66)
    (4300.00, -5.65)
    (4350.00, -5.64)
    (4400.00, -5.63)
    (4450.00, -5.63)
    (4500.00, -5.62)
    (4550.00, -5.61)
    (4600.00, -5.60)
    (4650.00, -5.59)
    (4700.00, -5.58)
    (4750.00, -5.57)
    (4800.00, -5.56)
    (4850.00, -5.55)
    (4900.00, -5.54)
    (4950.00, -5.53)
    (5000.00, -5.52)
    (5050.00, -5.51)
    (5100.00, -5.50)
    (5150.00, -5.49)
    (5200.00, -5.48)
    (5250.00, -5.47)
    (5300.00, -5.46)
    (5350.00, -5.45)
    (5400.00, -5.44)
    (5450.00, -5.43)
    (5500.00, -5.42)
    (5550.00, -5.40)
    (5600.00, -5.39)
    (5650.00, -5.38)
    (5700.00, -5.37)
    (5750.00, -5.36)
    (5800.00, -5.35)
    (5850.00, -5.33)
    (5900.00, -5.32)
    (5950.00, -5.31)
    (6000.00, -5.30)
    (6050.00, -5.29)
    (6100.00, -5.27)
    (6150.00, -5.26)
    (6200.00, -5.25)
    (6250.00, -5.23)
    (6300.00, -5.22)
    (6350.00, -5.21)
    (6400.00, -5.19)
    (6450.00, -5.18)
    (6500.00, -5.16)
    (6550.00, -5.15)
    (6600.00, -5.14)
    (6650.00, -5.12)
    (6700.00, -5.11)
    (6750.00, -5.09)
    (6800.00, -5.07)
    (6850.00, -5.06)
    (6900.00, -5.04)
    (6950.00, -5.03)
    (7000.00, -5.01)
    (7050.00, -4.99)
    (7100.00, -4.98)
    (7150.00, -4.96)
    (7200.00, -4.94)
    (7250.00, -4.92)
    (7300.00, -4.90)
    (7350.00, -4.89)
    (7400.00, -4.87)
    (7450.00, -4.85)
    (7500.00, -4.83)
    (7550.00, -4.81)
    (7600.00, -4.79)
    (7650.00, -4.77)
    (7700.00, -4.74)
    (7750.00, -4.72)
    (7800.00, -4.70)
    (7850.00, -4.68)
    (7900.00, -4.65)
    (7950.00, -4.63)
    (8000.00, -4.60)
    (8050.00, -4.58)
    (8100.00, -4.55)
    (8150.00, -4.53)
    (8200.00, -4.50)
    (8250.00, -4.47)
    (8300.00, -4.44)
    (8350.00, -4.41)
    (8400.00, -4.38)
    (8450.00, -4.35)
    (8500.00, -4.32)
    (8550.00, -4.28)
    (8600.00, -4.25)
    (8650.00, -4.21)
    (8700.00, -4.17)
    (8750.00, -4.13)
    (8800.00, -4.09)
    (8850.00, -4.05)
    (8900.00, -4.01)
    (8950.00, -3.96)
    (9000.00, -3.91)
    (9050.00, -3.86)
    (9100.00, -3.81)
    (9150.00, -3.75)
    (9200.00, -3.69)
    (9250.00, -3.62)
    (9300.00, -3.55)
    (9350.00, -3.48)
    (9400.00, -3.40)
    (9450.00, -3.31)
    (9500.00, -3.22)
    (9550.00, -3.11)
    (9600.00, -2.99)
    (9650.00, -2.86)
    (9700.00, -2.70)
    (9750.00, -2.52)
    (9800.00, -2.30)
    (9850.00, -2.01)
    (9900.00, -1.60)
    (9950.00, -0.90)
};
\addlegendentry{$O(\log(\frac{1}{n-t}))$}
\addplot[color=mycolor1, line width=1.5pt] coordinates {
    (0.00, -9.21)
    (50.00, -9.55)
    (100.00, -9.55)
    (150.00, -9.54)
    (200.00, -9.54)
    (250.00, -9.53)
    (300.00, -9.53)
    (350.00, -9.52)
    (400.00, -9.52)
    (450.00, -9.51)
    (500.00, -9.51)
    (550.00, -9.50)
    (600.00, -9.49)
    (650.00, -9.49)
    (700.00, -9.48)
    (750.00, -9.48)
    (800.00, -9.47)
    (850.00, -9.47)
    (900.00, -9.46)
    (950.00, -9.46)
    (1000.00, -9.45)
    (1050.00, -9.45)
    (1100.00, -9.44)
    (1150.00, -9.43)
    (1200.00, -9.43)
    (1250.00, -9.42)
    (1300.00, -9.42)
    (1350.00, -9.41)
    (1400.00, -9.41)
    (1450.00, -9.40)
    (1500.00, -9.39)
    (1550.00, -9.39)
    (1600.00, -9.38)
    (1650.00, -9.38)
    (1700.00, -9.37)
    (1750.00, -9.36)
    (1800.00, -9.36)
    (1850.00, -9.35)
    (1900.00, -9.35)
    (1950.00, -9.34)
    (2000.00, -9.33)
    (2050.00, -9.33)
    (2100.00, -9.32)
    (2150.00, -9.31)
    (2200.00, -9.31)
    (2250.00, -9.30)
    (2300.00, -9.30)
    (2350.00, -9.29)
    (2400.00, -9.28)
    (2450.00, -9.28)
    (2500.00, -9.27)
    (2550.00, -9.26)
    (2600.00, -9.26)
    (2650.00, -9.25)
    (2700.00, -9.24)
    (2750.00, -9.24)
    (2800.00, -9.23)
    (2850.00, -9.22)
    (2900.00, -9.21)
    (2950.00, -9.21)
    (3000.00, -9.20)
    (3050.00, -9.19)
    (3100.00, -9.19)
    (3150.00, -9.18)
    (3200.00, -9.17)
    (3250.00, -9.16)
    (3300.00, -9.16)
    (3350.00, -9.15)
    (3400.00, -9.14)
    (3450.00, -9.13)
    (3500.00, -9.13)
    (3550.00, -9.12)
    (3600.00, -9.11)
    (3650.00, -9.10)
    (3700.00, -9.09)
    (3750.00, -9.09)
    (3800.00, -9.08)
    (3850.00, -9.07)
    (3900.00, -9.06)
    (3950.00, -9.05)
    (4000.00, -9.05)
    (4050.00, -9.04)
    (4100.00, -9.03)
    (4150.00, -9.02)
    (4200.00, -9.01)
    (4250.00, -9.00)
    (4300.00, -8.99)
    (4350.00, -8.99)
    (4400.00, -8.98)
    (4450.00, -8.97)
    (4500.00, -8.96)
    (4550.00, -8.95)
    (4600.00, -8.94)
    (4650.00, -8.93)
    (4700.00, -8.92)
    (4750.00, -8.91)
    (4800.00, -8.90)
    (4850.00, -8.89)
    (4900.00, -8.88)
    (4950.00, -8.87)
    (5000.00, -8.86)
    (5050.00, -8.85)
    (5100.00, -8.84)
    (5150.00, -8.83)
    (5200.00, -8.82)
    (5250.00, -8.81)
    (5300.00, -8.80)
    (5350.00, -8.79)
    (5400.00, -8.78)
    (5450.00, -8.77)
    (5500.00, -8.76)
    (5550.00, -8.75)
    (5600.00, -8.74)
    (5650.00, -8.72)
    (5700.00, -8.71)
    (5750.00, -8.70)
    (5800.00, -8.69)
    (5850.00, -8.68)
    (5900.00, -8.67)
    (5950.00, -8.65)
    (6000.00, -8.64)
    (6050.00, -8.63)
    (6100.00, -8.62)
    (6150.00, -8.60)
    (6200.00, -8.59)
    (6250.00, -8.58)
    (6300.00, -8.56)
    (6350.00, -8.55)
    (6400.00, -8.53)
    (6450.00, -8.52)
    (6500.00, -8.51)
    (6550.00, -8.49)
    (6600.00, -8.48)
    (6650.00, -8.46)
    (6700.00, -8.45)
    (6750.00, -8.43)
    (6800.00, -8.42)
    (6850.00, -8.40)
    (6900.00, -8.39)
    (6950.00, -8.37)
    (7000.00, -8.35)
    (7050.00, -8.34)
    (7100.00, -8.32)
    (7150.00, -8.30)
    (7200.00, -8.28)
    (7250.00, -8.26)
    (7300.00, -8.25)
    (7350.00, -8.23)
    (7400.00, -8.21)
    (7450.00, -8.19)
    (7500.00, -8.17)
    (7550.00, -8.15)
    (7600.00, -8.13)
    (7650.00, -8.11)
    (7700.00, -8.09)
    (7750.00, -8.06)
    (7800.00, -8.04)
    (7850.00, -8.02)
    (7900.00, -8.00)
    (7950.00, -7.97)
    (8000.00, -7.95)
    (8050.00, -7.92)
    (8100.00, -7.90)
    (8150.00, -7.87)
    (8200.00, -7.84)
    (8250.00, -7.81)
    (8300.00, -7.78)
    (8350.00, -7.75)
    (8400.00, -7.72)
    (8450.00, -7.69)
    (8500.00, -7.66)
    (8550.00, -7.63)
    (8600.00, -7.59)
    (8650.00, -7.55)
    (8700.00, -7.52)
    (8750.00, -7.48)
    (8800.00, -7.43)
    (8850.00, -7.39)
    (8900.00, -7.35)
    (8950.00, -7.30)
    (9000.00, -7.25)
    (9050.00, -7.20)
    (9100.00, -7.15)
    (9150.00, -7.09)
    (9200.00, -7.03)
    (9250.00, -6.96)
    (9300.00, -6.90)
    (9350.00, -6.82)
    (9400.00, -6.74)
    (9450.00, -6.65)
    (9500.00, -6.56)
    (9550.00, -6.45)
    (9600.00, -6.33)
    (9650.00, -6.19)
    (9700.00, -6.04)
    (9750.00, -5.85)
    (9800.00, -5.62)
    (9850.00, -5.32)
    (9900.00, -4.88)
    (9950.00, -4.19)
};
\addlegendentry{Algorithm~\ref{alg:AlgoProbKnownDiscrete}}
\end{axis}

\begin{axis}[%
width=0in,
height=0in,
at={(0in,0in)},
scale only axis,
xmin=0,
xmax=1,
ymin=0,
ymax=1,
axis line style={draw=none},
ticks=none,
axis x line*=bottom,
axis y line*=left
]
\end{axis}

\end{tikzpicture}%
    \vspace{-25pt}
    \caption{{\small \sf  Validation of Lemma~\ref{lem:lipshitzness} for an instance with $n = 10,000$ users, where all users have a fixed budget of one, and two goods, each with a capacity of $c_j = n = 10,000$. 
    }} 
    \label{fig:priceStability}
\end{figure}

\subsection{Numerical Experiments Comparing Static and Adaptive Variants of Expected Equilibrium Pricing} \label{apdx:staticVdynamic}

In this section, we numerically evaluate the performance of the static expected equilibrium pricing algorithm and its dynamic counterpart (Algorithm~\ref{alg:AlgoProbKnownDiscrete}) on the counterexample in the proof of Theorem~\ref{thm:lbStatic}. In particular, we considered a setting of $n$ users, where all users have a fixed budget of one, and two goods, each with a capacity of $n$. The utility parameters of users are drawn i.i.d. from a distribution $\D$, where users have an equal $0.5$ probability of having the utility $(1, 0)$ or $(0, 1)$. For the experiments, we let the number of users $n$ range between $100$ to $10,000$ users. 


Figure~\ref{fig:stativVAdaptive} depicts both the constraint violation and the regret of the two algorithms. From the figure, it can be observed that the static expected equilibrium pricing approach achieves negative regret for a large constraint violation, while Algorithm~\ref{alg:AlgoProbKnownDiscrete} achieves a small positive regret for almost no constraint violation. Recall here from the proof of Theorem~\ref{thm:lbStatic} that the expected optimal social welfare objective $\mathbb{E}[U_n^*] \in [n \log(2) - 1, n \log(2)]$, and thus a regret of less than five for 10,000 users is negligible. As a result, Figure~\ref{fig:stativVAdaptive} clearly depicts the benefit of adaptivity in online Fisher markets.


We also note that the regret of the static expected equilibrium pricing algorithm is in the range $[-1, 0]$, as the accumulated online objective is $n \log(2)$, as each user obtains two units of the good for which they have positive utility under the static expected equilibrium prices of $(0.5, 0.5)$ for this instance. As a result, observe that the numerically observed regret in the range $[-1, 0]$ aligns with the tight bound for the expected optimal social welfare objective obtained in the proof of Theorem~\ref{thm:lbStatic}, i.e., $n \log(2) - 1 \leq \mathbb{E}[U_n^*] \leq n \log(2)$.

\begin{figure}[tbh!]
    \centering
    \begin{subfigure}[t]{0.45\columnwidth}
        \centering
%
%
\definecolor{mycolor3}{rgb}{0.00000,0.44700,0.74100}%
\definecolor{mycolor2}{rgb}{1.0, 0.88, 0.21}
\definecolor{mycolor1}{rgb}{0.89, 0.0, 0.13}
\begin{tikzpicture}

\begin{axis}[%
width=2in,
height=2in,
legend style={
legend cell align=left, align=left,
  fill opacity=0.8,
  draw opacity=1,
  text opacity=1,
  at={(0.02,0.95)},
  anchor=north west,
  draw=white!80!black
},
scale only axis,
xmin=0,
xmax=11000,
xlabel style={font=\color{white!15!black}},
xlabel={Number of Users (Thousands)},
ymin=-50,
ymax=50,
legend style={
legend cell align=left, align=left,
  fill opacity=0.8,
  draw opacity=1,
  text opacity=1,
  anchor=north west,
  at = {(0.07, 0.27)},
  draw=white!80!black
},
xtick=\empty,
    extra x ticks={0,2000, 4000, 6000, 8000, 10000},
    extra x tick labels={0, 2, 4, 6, 8, 10},
ylabel style={font=\color{white!15!black}, align=center},
ylabel={Regret},
axis background/.style={fill=white}
]
\addplot [color=mycolor3, line width=1.5pt]
  table{%
100 0.0
250 -2.842170943040401e-14
500 -5.684341886080802e-14
750 0.0
1000 -1.1368683772161603e-13
1250 -1.1368683772161603e-13
1500 0.0
2000  -2.2737367544323206e-13
2500 -2.2737367544323206e-13
3000 0.0
5000 -4.547473508864641e-13
10000 0.0
};
\addlegendentry{Static}
\addplot [color=mycolor1, line width=1.5pt]
  table{%
100 1.7865433268605813
250 1.466347288863119
500 1.7726527539442145
750 7.783496373274033
1000 3.8709331178401953
1250 16.020500523852206
1500 2.656467566017227
2000 3.249807257939665
2500 25.737985289769313
3000 2.570098000242524
5000 3.220299645972318
10000 3.8731048666168135
};
\addlegendentry{Adaptive}
\end{axis}

\begin{axis}[%
width=0in,
height=0in,
at={(0in,0in)},
scale only axis,
xmin=0,
xmax=1,
ymin=0,
ymax=1,
axis line style={draw=none},
ticks=none,
axis x line*=bottom,
axis y line*=left
]
\end{axis}

\end{tikzpicture}%
    \end{subfigure} \hspace{5pt}
    \begin{subfigure}[t]{0.45\columnwidth}
        \centering
        \vspace{-6pt}
%
%
\definecolor{mycolor3}{rgb}{0.00000,0.44700,0.74100}%
\definecolor{mycolor2}{rgb}{1.0, 0.88, 0.21}
\definecolor{mycolor1}{rgb}{0.89, 0.0, 0.13}
\begin{tikzpicture}

\begin{axis}[%
width=2in,
height=2in,
legend style={
legend cell align=left, align=left,
  fill opacity=0.8,
  draw opacity=1,
  text opacity=1,
  at={(0.02,0.95)},
  anchor=north west,
  draw=white!80!black
},
scale only axis,
xmin=0,
xmax=11000,
xlabel style={font=\color{white!15!black}},
xlabel={Number of Users (Thousands)},
ymin=-10,
ymax=150,
legend style={
legend cell align=left, align=left,
  fill opacity=0.8,
  draw opacity=1,
  text opacity=1,
  anchor=north west,
  at = {(0.07, 0.27)},
  draw=white!80!black
},
xtick=\empty,
    extra x ticks={0,2000, 4000, 6000, 8000, 10000},
    extra x tick labels={0, 2, 4, 6, 8, 10},
ylabel style={font=\color{white!15!black}, align=center},
ylabel={Two Norm of Excess Demand},
axis background/.style={fill=white}
]
\addplot [color=mycolor3, line width=1.5pt]
  table{%
100 10.0
250 6
500 8
750 6
1000 16
1250 44
1500 2.0
2000  24
2500 92
3000 14.0
5000 36
10000 138.0
};

\addplot [color=mycolor1, line width=1.5pt]
  table{%
100 0
250 4.813464674817425
500 0
750 0
1000 0
1250 0
1500 0
2000 0
2500 0
3000 0
5000 0
10000 0
};
\end{axis}

\begin{axis}[%
width=0in,
height=0in,
at={(0in,0in)},
scale only axis,
xmin=0,
xmax=1,
ymin=0,
ymax=1,
axis line style={draw=none},
ticks=none,
axis x line*=bottom,
axis y line*=left
]
\end{axis}

\end{tikzpicture}%
    \end{subfigure}
    \vspace{-25pt}
    \caption{{\small \sf  Comparison between the static expected equilibrium pricing algorithm and its dynamic counterpart (Algorithm~\ref{alg:AlgoProbKnownDiscrete}) on regret and constraint violation metrics. 
    }} 
    \label{fig:stativVAdaptive}
\end{figure} 

\subsection{Numerical Comparison between the Additive and Multiplicative Price Updates in Algorithm~\ref{alg:PrivacyPreserving}} \label{apdx:AddVMult}

We now compare Algorithm~\ref{alg:PrivacyPreserving} (with a fixed step size) that has an additive price update step to a corresponding algorithm with a multiplicative price update step (see Section~\ref{sec:algoMain}). To this end, we consider instance two described in \ifarxiv Section~\ref{sec:experimental-setup-details} \fi with a step size of $\gamma = \frac{1}{10 \sqrt{n}}$. 



Figure~\ref{fig:comparisonPlot3} depicts the regret and constraint violation for algorithms with the two price update steps given an initial price of $0.5$ for all goods. We can observe from Figure~\ref{fig:comparisonPlot3} that Algorithm~\ref{alg:PrivacyPreserving} with an additive price update rule has a higher regret but a lower constraint violation as compared to the corresponding algorithm with a multiplicative price update rule. This observation highlights the fundamental trade-off between the regret and constraint violation metrics. Yet, we note that since the multiplicative price update rule achieves a lower regret (despite achieving a higher constraint violation) compared to the additive price update rule in Algorithm~\ref{alg:PrivacyPreserving}, our results motivate a deeper study of the regret and constraint violation bounds under the multiplicative price update rule.

\begin{figure}[tbh!]
    \centering
    \begin{subfigure}[t]{0.45\columnwidth}
        \centering
%
%
\definecolor{mycolor3}{rgb}{0.00000,0.44700,0.74100}%
\definecolor{mycolor2}{rgb}{1.0, 0.88, 0.21}
\definecolor{mycolor1}{rgb}{0.89, 0.0, 0.13}
\definecolor{mycolor4}{rgb}{0.20, 0.63, 0.17} 
\definecolor{mycolor5}{rgb}{0.58, 0.44, 0.86} 
\definecolor{mycolor6}{rgb}{0.90, 0.60, 0.00} 
\definecolor{mycolor7}{rgb}{0.67, 0.85, 0.90} 

\begin{tikzpicture}

\begin{axis}[%
width=2in,
height=2in,
scale only axis,
xmin=0,
xmax=3500,
xlabel style={font=\color{white!15!black}},
xlabel={Number of Users},
ymin=-2200,
ymax=0,
ylabel style={font=\color{white!15!black}, align=center},
ylabel={Regret},
legend style={
legend cell align=left, align=left,
  fill opacity=0.8,
  draw opacity=1,
  text opacity=1,
  anchor=north west,
  font = \scriptsize,
  at = {(0.17, 0.97)},
  draw=white!80!black
},
axis background/.style={fill=white},
]
\addplot [color=mycolor4, line width=1.5pt]
  table[row sep=crcr]{%
100 -180.48895488316657\\
250 -396.64063054535654\\
500 -598.6328469906421\\
750 -785.1843986471977\\
1000 -903.0175488809764\\
1250 -1007.717194191151\\
1500 -1004.4871921709055\\
2000 -1325.154832013075\\
2500 -1406.4195065940366\\
3000 -1589.627413162205\\
};
\addlegendentry{Additive Price Update}

\addplot [color=mycolor3, line width=1.5pt]
  table[row sep=crcr]{%
100 -225.46452925859103\\
250 -500.88942284784935\\
500 -762.5158099697728\\
750 -987.8095596630628\\
1000 -1145.9364384733344\\
1250 -1275.8605118842243\\
1500 -1303.793763349444\\
2000 -1674.860217718553\\
2500 -1797.1713905919096\\
3000 -2019.6504384643777\\
};
\addlegendentry{Multiplicative Price Update}
\end{axis} 

\begin{axis}[%
width=0in,
height=0in,
at={(0in,0in)},
scale only axis,
xmin=0,
xmax=1,
ymin=0,
ymax=1,
axis line style={draw=none},
ticks=none,
axis x line*=bottom,
axis y line*=left
]
\end{axis}

\end{tikzpicture}%
    \end{subfigure} \hspace{5pt}
    \begin{subfigure}[t]{0.45\columnwidth}
        \centering
        \vspace{6pt}
%
%
\definecolor{mycolor3}{rgb}{0.00000,0.44700,0.74100}%
\definecolor{mycolor2}{rgb}{1.0, 0.88, 0.21}
\definecolor{mycolor1}{rgb}{0.89, 0.0, 0.13}
\definecolor{mycolor4}{rgb}{0.20, 0.63, 0.17} 
\definecolor{mycolor5}{rgb}{0.58, 0.44, 0.86} 
\definecolor{mycolor6}{rgb}{0.90, 0.60, 0.00} 
\definecolor{mycolor7}{rgb}{0.67, 0.85, 0.90} 

\begin{tikzpicture}

\begin{axis}[%
width=2in,
height=2in,
scale only axis,
xmin=0,
xmax=3500,
xlabel style={font=\color{white!15!black}},
xlabel={Number of Users},
ymin=0,
ymax=1100,
ylabel style={font=\color{white!15!black}, align=center},
ylabel={Two Norm of Excess Demand},
legend style={
legend cell align=left, align=left,
  fill opacity=0.8,
  draw opacity=1,
  text opacity=1,
  anchor=north west,
  font = \scriptsize,
  at = {(0.15, 0.27)},
  draw=white!80!black
},
axis background/.style={fill=white},
]
\addplot [color=mycolor4, line width=1.5pt]
  table[row sep=crcr]{%
100 101.4381036346848\\
250 200.08786934296324\\
500 300.3544965253534\\
750 380.9892609141406\\
1000 440.9724404516301\\
1250 501.43679913568826\\
1500 507.72336180261146\\
2000 634.0002077622074\\
2500 686.0930727235386\\
3000 768.9781766490134\\
};

\addplot [color=mycolor3, line width=1.5pt]
  table[row sep=crcr]{%
100 128.3986220426612\\
250 257.0085051541275\\
500 388.0827220732487\\
750 491.1849197529599\\
1000 570.0229065962436\\
1250 646.6003225530784\\
1500 670.0634584215586\\
2000 820.1780910213171\\
2500 895.1277896275203\\
3000 999.4842442033568\\
};
\end{axis} 

\begin{axis}[%
width=0in,
height=0in,
at={(0in,0in)},
scale only axis,
xmin=0,
xmax=1,
ymin=0,
ymax=1,
axis line style={draw=none},
ticks=none,
axis x line*=bottom,
axis y line*=left
]
\end{axis}

\end{tikzpicture}%
    \end{subfigure}
    \vspace{-25pt}
    \caption{{\small \sf  Comparison between Algorithm~\ref{alg:PrivacyPreserving} that has an additive price update step to a corresponding algorithm with a multiplicative price update step on regret and constraint violation metrics. 
    }} 
    \label{fig:comparisonPlot3}
\end{figure}

\subsection{Numerical Validation of Positivity of Prices in Algorithm~\ref{alg:PrivacyPreserving}} \label{apdx:numericalValidationStrictPositivity}

In this section, we present the results of a numerical experiment to validate that the prices remain strictly positive throughout the operation of Algorithm~\ref{alg:PrivacyPreserving} with a fixed step size. To this end, we consider two market settings: (i) the setting described in the counterexample in the proof of Theorem~\ref{thm:lbStatic}, and (ii) instance two described in Section~\ref{sec:experimental-setup-details}. For the experiments, we let the number of users $n$ range between $500$ to $10,000$ users, consider a step-size of the price updates as $\gamma = \frac{1}{200 \sqrt{n}}$, and compute the minimum prices across all goods for 300 instances. In particular, Figure~\ref{fig:positiveValidation} depicts the minimum prices of all goods across 300 instances, which validates the positivity of the prices during the operation of Algorithm~\ref{alg:PrivacyPreserving}.

\begin{figure}[tbh!]
    \centering
    \begin{subfigure}[t]{0.45\columnwidth}
        \centering
%
%
\definecolor{mycolor3}{rgb}{0.00000,0.44700,0.74100}%
\definecolor{mycolor2}{rgb}{1.0, 0.88, 0.21}
\definecolor{mycolor1}{rgb}{0.89, 0.0, 0.13}
\begin{tikzpicture}

\begin{axis}[%
width=2in,
height=2in,
legend style={
legend cell align=left, align=left,
  fill opacity=0.8,
  draw opacity=1,
  text opacity=1,
  at={(0.02,0.95)},
  anchor=north west,
  draw=white!80!black
},
scale only axis,
xmin=0,
xmax=11000,
xlabel style={font=\color{white!15!black}},
xlabel={Number of Users (Thousands)},
ymin=0,
ymax=1.2,
yticklabels={0.0, 0.0, 0.2, 0.4, 0.6, 0.8, 1.0, 1.2},
xtick=\empty,
    extra x ticks={0,2000, 4000, 6000, 8000, 10000},
    extra x tick labels={0, 2, 4, 6, 8, 10},
ylabel style={font=\color{white!15!black}, align=center},
ylabel={Minimum Price},
axis background/.style={fill=white}
]
\addplot [color=mycolor3]
  table{%
500 0.614721220586477
750 0.705628499906843
1000 0.766830101328316
1250 0.807872680525438
1500 0.839044365820337
2000 0.885037142159742
2500 0.920676305295961
3000 0.949493876064716
5000 0.99947768616033
7500 0.999826794919243
10000 0.9999
};

\addplot [color=mycolor3,only marks]
  table{%
500 0.614721220586477
750 0.705628499906843
1000 0.766830101328316
1250 0.807872680525438
1500 0.839044365820337
2000 0.885037142159742
2500 0.920676305295961
3000 0.949493876064716
5000 0.99947768616033
7500 0.999826794919243
10000 0.9999
};
\end{axis}

\begin{axis}[%
width=0in,
height=0in,
at={(0in,0in)},
scale only axis,
xmin=0,
xmax=1,
ymin=0,
ymax=1,
axis line style={draw=none},
ticks=none,
axis x line*=bottom,
axis y line*=left
]
\end{axis}

\end{tikzpicture}%
    \end{subfigure} \hspace{5pt}
    \begin{subfigure}[t]{0.45\columnwidth}
        \centering
%
%
\definecolor{mycolor3}{rgb}{0.00000,0.44700,0.74100}%
\definecolor{mycolor2}{rgb}{1.0, 0.88, 0.21}
\definecolor{mycolor1}{rgb}{0.89, 0.0, 0.13}
\begin{tikzpicture}

\begin{axis}[%
width=2in,
height=2in,
legend style={
legend cell align=left, align=left,
  fill opacity=0.8,
  draw opacity=1,
  text opacity=1,
  at={(0.02,0.95)},
  anchor=north west,
  draw=white!80!black
},
scale only axis,
xmin=0,
xmax=11000,
xlabel style={font=\color{white!15!black}},
xlabel={Number of Users (Thousands)},
ymin=0,
ymax=1.2,
yticklabels={0.0, 0.0, 0.2, 0.4, 0.6, 0.8, 1.0, 1.2},
xtick=\empty,
    extra x ticks={0,2000, 4000, 6000, 8000, 10000},
    extra x tick labels={0, 2, 4, 6, 8, 10},
ylabel style={font=\color{white!15!black}, align=center},
ylabel={Minimum Price},
axis background/.style={fill=white}
]
\addplot [color=mycolor3]
  table{%
500 0.0653112485080349
750 0.122822208080793
1000 0.203139597811679
1250 0.29725375665961
1500 0.386309482172456
2000 0.541219419311885
2500 0.662642397952839
3000 0.76473451164479
5000 0.985615736564613
7500 0.996120642004941
10000 0.99805
};

\addplot [color=mycolor3,only marks]
  table{%
500 0.0653112485080349
750 0.122822208080793
1000 0.203139597811679
1250 0.29725375665961
1500 0.386309482172456
2000 0.541219419311885
2500 0.662642397952839
3000 0.76473451164479
5000 0.985615736564613
7500 0.996120642004941
10000 0.99805
};
\end{axis}

\begin{axis}[%
width=0in,
height=0in,
at={(0in,0in)},
scale only axis,
xmin=0,
xmax=1,
ymin=0,
ymax=1,
axis line style={draw=none},
ticks=none,
axis x line*=bottom,
axis y line*=left
]
\end{axis}

\end{tikzpicture}%
    \end{subfigure}
    \vspace{-25pt}
    \caption{{\small \sf  Numerical validation of the positivity of prices during the operation of Algorithm~\ref{alg:PrivacyPreserving} in two market settings: (i) the market instance in the proof of Theorem~\ref{thm:lbStatic} (left), and (ii) instance two described in Section~\ref{sec:experimental-setup-details} (right). The y-axis denotes the minimum price across all goods across 300 problem instances, i.e., 300 runs of Algorithm~\ref{alg:PrivacyPreserving} on different instances drawn from the specified distribution corresponding to each market setting.
    }} 
    \label{fig:positiveValidation}
\end{figure}

\section{Relation to Approximate Equilibria} \label{sec:apx-equilibria}

In this section, we present the connection between our studied performance metrics and market equilibria. To this end, we first note that our regret and constraint violation metrics approximate the optimal offline Eisenberg Gale aggregated social objective and constraint satisfaction, respectively, which, as beautifully proven, corresponds to perfect Pareto efficiency and envy-freeness under complete information of the utility and budget parameters of users. As a result, obtaining sub-linear guarantees for our regret and constraint violation metrics serves as a proxy for a solution corresponding to an approximate market equilibrium, as the distance to the optimal offline objective and constraint satisfaction of an algorithm indicate its proximity to the optimal offline equilibrium solution. For instance, we note that the per-period regret of Algorithm~\ref{alg:PrivacyPreserving} with a fixed step size of $O(\frac{1}{\sqrt{n}})$ is $\frac{1}{\sqrt{n}}$, which decays and approaches zero as the number of users becomes large. This fact suggests that, on average, the allocations made by the online algorithm approach that of the optimal offline solution as the number of users becomes large, further suggesting that the price-iterates approach the market equilibrium in expectation. Furthermore, we reiterate that achieving low regret corresponding to the Eisenberg Gale objective implies that no user can suffer too much, i.e., receive very low utilities, as the objective is a (weighted) product of all users’ utilities.

Our studied problem setting and corresponding performance metrics directly relate to notions of Pareto efficiency and envy-freeness. To this end, we first note that our constraint violation metric can serve as a measure of Pareto inefficiency, which is typically related to the extent to which the capacity constraints are not satisfied (e.g., see~\cite{sinclair2021sequential}), i.e., the number of unsold goods, as when certain goods are unsold some users can become better off without making others worse off. Noting that our theoretical guarantees for constraint violation hold for both the settings of over or under-consumption of resources, our constraint violation bounds thus serve as a measure of the degree of Pareto inefficiency of our obtained solution, which is sub-linear in the number of users for our proposed algorithms. 

As for envy-freeness, we first note that our proposed algorithms correspond to posted-price mechanisms, wherein users observe the posted prices and freely (and truthfully) choose which goods to purchase to obtain their most favored bundle of goods given the set prices. In this regard, our proposed algorithms are envy-free by design as all users obtain their most favored bundle of goods, given the set prices upon their arrival. Furthermore, even though under our algorithms, users typically observe different prices, we note that most users observe prices that are similar to other users implying an envy-freeness with regards to the prices faced by users, i.e., there is little that users can gain by swapping their observed prices with that faced by most other users. To elucidate this point, we conducted numerical experiments of Algorithm~\ref{alg:PrivacyPreserving} with a fixed step size under the two market settings: (i) the setting described in the counterexample in the proof of Theorem~\ref{thm:lbStatic}, and (ii) instance two described in Section~\ref{sec:experimental-setup-details}. For the experiments, we let the number of users $n$ be $5,000$ and consider a step-size of the price updates as $\gamma = \frac{1}{5 \sqrt{n}}$. Figure~\ref{fig:PriceConvValidation} depicts the evolution of the prices of the goods under both our market instances and demonstrates that a majority of (about 90\% of) the users observe prices within a small price band under both market instances. In general, we note that given the $O(\frac{1}{\sqrt{n}})$ step-size of the price updates in Algorithm~\ref{alg:PrivacyPreserving}, one can expect that it will take about $O(\sqrt{n})$ steps for the price to move from the initial price vector to a new price vector that, from that point on, stabilizes in a particular band. Thus, Algorithm~\ref{alg:PrivacyPreserving} (with a fixed step size) can be interpreted as achieving approximate envy-freeness where only $O(\sqrt{n})$ users observe arbitrary prices, while the remaining users observe prices within a small price band.




\begin{figure}[tbh!]
    \centering
    \begin{subfigure}[t]{0.45\columnwidth}
        \centering
            \include{Figures/priceconvergence_may2024_5goods}
    \end{subfigure} \hspace{5pt}
    \begin{subfigure}[t]{0.45\columnwidth}
        \centering
        \vspace{-6pt}
%
%
\definecolor{mycolor1}{rgb}{0.00000,0.44700,0.74100}%
\definecolor{mycolor2}{rgb}{0.85000,0.32500,0.09800}%
\begin{tikzpicture}

\begin{axis}[%
width=2.5in,
height=2in,
at={(0in,0in)},
scale only axis,
xmin=0,
xmax=5000,
xlabel style={font=\color{white!15!black}},
xlabel={Number of Users},
ymin=0.4,
ymax=0.8,
ylabel style={font=\color{white!15!black}},
ylabel={Price of Good},
axis background/.style={fill=white},
legend style={legend cell align=left, align=left, draw=white!15!black}
]
\addplot [color=mycolor1, semithick]
  table{%
0 0.698585786437627
50.5050505050505 0.68901871259852
101.010101010101 0.674390432457489
151.515151515152 0.657084626892038
202.020202020202 0.630108930704439
252.525252525253 0.612086973249856
303.030303030303 0.60622945932972
353.535353535354 0.609760944623155
404.040404040404 0.602458998002294
454.545454545455 0.595340921536348
505.050505050505 0.582768274990617
555.555555555556 0.563841350541686
606.060606060606 0.558283869574413
656.565656565657 0.553642559064595
707.070707070707 0.554189952439737
757.575757575758 0.546992392550698
808.080808080808 0.535974161913482
858.585858585859 0.537076222846553
909.090909090909 0.537633722820244
959.59595959596 0.528224316271261
1010.10101010101 0.501714586692529
1060.60606060606 0.491053078378629
1111.11111111111 0.489566076395045
1161.61616161616 0.488486687271797
1212.12121212121 0.493409587602616
1262.62626262626 0.489242108141442
1313.13131313131 0.498114990141167
1363.63636363636 0.506318826888155
1414.14141414141 0.521021387414502
1464.64646464646 0.530963514203679
1515.15151515152 0.52700737666761
1565.65656565657 0.528846759848338
1616.16161616162 0.52335422833297
1666.66666666667 0.525246761177438
1717.17171717172 0.519481672593092
1767.67676767677 0.51711455263769
1818.18181818182 0.512630941756396
1868.68686868687 0.500163109881486
1919.19191919192 0.511419131389904
1969.69696969697 0.510207384383918
2020.20202020202 0.519347181315413
2070.70707070707 0.51201754285172
2121.21212121212 0.502767400277064
2171.71717171717 0.502060246010507
2222.22222222222 0.502133998468613
2272.72727272727 0.502776169891336
2323.23232323232 0.502703255765587
2373.73737373737 0.523317090689429
2424.24242424242 0.524943458757384
2474.74747474747 0.521323764787099
2525.25252525253 0.518816476989266
2575.75757575758 0.523327305570292
2626.26262626263 0.51042632081056
2676.76767676768 0.520011421655585
2727.27272727273 0.53034230810636
2777.77777777778 0.526938038110067
2828.28282828283 0.523560840176371
2878.78787878788 0.523110929755549
2929.29292929293 0.520066943799039
2979.79797979798 0.535171197110132
3030.30303030303 0.539049387030476
3080.80808080808 0.524315325229294
3131.31313131313 0.529076747313474
3181.81818181818 0.537693019853301
3232.32323232323 0.537917635391869
3282.82828282828 0.530740080303533
3333.33333333333 0.516539647416492
3383.83838383838 0.506195407589112
3434.34343434343 0.505507853110111
3484.84848484848 0.508275438448235
3535.35353535354 0.517728031333221
3585.85858585859 0.500109899044425
3636.36363636364 0.494363513966035
3686.86868686869 0.490068060925829
3737.37373737374 0.501576921266991
3787.87878787879 0.506840729269863
3838.38383838384 0.503470265461108
3888.88888888889 0.513907786788574
3939.39393939394 0.502004290504405
3989.89898989899 0.501859274710507
4040.40404040404 0.499268332842188
4090.90909090909 0.499656929949348
4141.41414141414 0.504418075992529
4191.91919191919 0.504255005515134
4242.42424242424 0.47927816853005
4292.92929292929 0.493192936144824
4343.43434343434 0.504893333554762
4393.93939393939 0.523035613515861
4444.44444444444 0.530241486858905
4494.94949494949 0.536164296069165
4545.45454545455 0.541400346311462
4595.9595959596 0.533694968938012
4646.46464646465 0.509388787322385
4696.9696969697 0.50562747019607
4747.47474747475 0.494931522097084
4797.9797979798 0.495497407187621
4848.48484848485 0.504368963768297
4898.9898989899 0.492880911697823
4949.49494949495 0.486598201501017
5000 0.496836568155159
};
\addlegendentry{Good 1}

\addplot [color=mycolor2, semithick]
  table{%
0 0.700606091526731
50.5050505050505 0.671208906130447
101.010101010101 0.649752780508864
151.515151515152 0.633899527395284
202.020202020202 0.630606238925733
252.525252525253 0.620713521043474
303.030303030303 0.601186709471394
353.535353535354 0.574115909058968
404.040404040404 0.560412431326815
454.545454545455 0.548690745185318
505.050505050505 0.544784946708506
555.555555555556 0.549216950627626
606.060606060606 0.541320529508542
656.565656565657 0.53397206681255
707.070707070707 0.522577594335577
757.575757575758 0.520616843272539
808.080808080808 0.523622043795077
858.585858585859 0.515040560565555
909.090909090909 0.507879148153289
959.59595959596 0.511947785009112
1010.10101010101 0.533152184631159
1060.60606060606 0.538929632495883
1111.11111111111 0.536755057085903
1161.61616161616 0.534665187845614
1212.12121212121 0.527417884770202
1262.62626262626 0.528887805552996
1313.13131313131 0.51833929312414
1363.63636363636 0.508354882865413
1414.14141414141 0.491510562230578
1464.64646464646 0.479283978410052
1515.15151515152 0.482416108157116
1565.65656565657 0.479298947803696
1616.16161616162 0.484292407189225
1666.66666666667 0.481427192001056
1717.17171717172 0.486908400758367
1767.67676767677 0.488741129363738
1818.18181818182 0.492824111918105
1868.68686868687 0.504934487831873
1919.19191919192 0.493041693717839
1969.69696969697 0.493880723388397
2020.20202020202 0.483918153928804
2070.70707070707 0.491377591349164
2121.21212121212 0.500558226735208
2171.71717171717 0.501021577449529
2222.22222222222 0.500729011086285
2272.72727272727 0.499903744102979
2323.23232323232 0.49981095065547
2373.73737373737 0.478073636101877
2424.24242424242 0.476374224424294
2474.74747474747 0.480388669971028
2525.25252525253 0.483032455441546
2575.75757575758 0.478197475370801
2626.26262626263 0.491858190838675
2676.76767676768 0.481601653669119
2727.27272727273 0.470231014866926
2777.77777777778 0.474095169182509
2828.28282828283 0.477859329941711
2878.78787878788 0.478358400347615
2929.29292929293 0.481665331244425
2979.79797979798 0.464900244897451
3030.30303030303 0.460514852026895
3080.80808080808 0.477329705093155
3131.31313131313 0.472034696751289
3181.81818181818 0.462370114320333
3232.32323232323 0.462333722556948
3282.82828282828 0.470657635625116
3333.33333333333 0.486132989708955
3383.83838383838 0.496727865246612
3434.34343434343 0.497229936446541
3484.84848484848 0.494221477078202
3535.35353535354 0.484184303221044
3585.85858585859 0.502237993772329
3636.36363636364 0.507777108445756
3686.86868686869 0.511847858596563
3737.37373737374 0.500481665106926
3787.87878787879 0.495069316404782
3838.38383838384 0.498426500043132
3888.88888888889 0.487603143173106
3939.39393939394 0.499805071609841
3989.89898989899 0.499900249464409
4040.40404040404 0.502444375060611
4090.90909090909 0.502023561248835
4141.41414141414 0.497218090464462
4191.91919191919 0.497352236161593
4242.42424242424 0.521458619372274
4292.92929292929 0.508362271640622
4343.43434343434 0.496718123574586
4393.93939393939 0.477570081377034
4444.44444444444 0.469732793265054
4494.94949494949 0.463285988817963
4545.45454545455 0.457552677752053
4595.9595959596 0.466817821752412
4646.46464646465 0.493162686411997
4696.9696969697 0.496861750283628
4747.47474747475 0.507383653922015
4797.9797979798 0.506709038432777
4848.48484848485 0.497769870827692
4898.9898989899 0.509059788801369
4949.49494949495 0.515059157954843
5000 0.505144229550828
};
\addlegendentry{Good 2}

\end{axis}

\begin{axis}[%
width=0in,
height=0in,
at={(0in,0in)},
scale only axis,
xmin=0,
xmax=1,
ymin=0,
ymax=1,
axis line style={draw=none},
ticks=none,
axis x line*=bottom,
axis y line*=left,
legend style={legend cell align=left, align=left, draw=white!15!black}
]
\end{axis}
\end{tikzpicture}%
    \end{subfigure}
    \vspace{-25pt}
    \caption{{\small \sf  Evolution of the price vector of Algorithm~\ref{alg:PrivacyPreserving} (with a fixed step size) with the number of users on the instance two described in Section~\ref{sec:experimental-setup-details} (left) and the setting described in the counterexample in the proof of Theorem~\ref{thm:lbStatic} (right). In both market instances, the price vectors stabilize in a small band after the arrival of the first few users.
    }} 
    \label{fig:PriceConvValidation}
\end{figure}    

\end{document}